\documentclass[preprint,review,1p,12pt,sort&compress]{elsarticle}
\usepackage{epic,eepic}
\pdfoutput=1
\usepackage{rotating}
\usepackage[normalsize]{subfigure}
\usepackage{graphicx}
\usepackage{amsmath,amssymb}
\usepackage{multicol}
\usepackage{multirow}
\usepackage{array}
\usepackage{verbatim}
\usepackage[section]{placeins}
\usepackage{appendix}
\usepackage{numcompress}
\usepackage{caption}
\usepackage{epstopdf}
\usepackage{lineno}
\usepackage{commath}

\begin{document}
\begin{frontmatter}
\title{Acoustical characteristics of segmented plates with contact interfaces}
\author[PuMe]{Srinivas Varanasi}
\ead{v.s.s.srinivas@gmail.com}
\author[PuMe]{Thomas Siegmund\corref{cor1}}
\ead{siegmund@purdue.edu}
\author[PuHe]{J. Stuart Bolton}
\ead{bolton@purdue.edu}

\address[PuMe]{School of Mechanical Engineering, 585 Purdue Mall, West Lafayette, IN 47907-2088, Phone: (765) 494-9766, U.S.A.}
\address[PuHe]{Ray W. Herrick Laboratories, School of Mechanical Engineering, Purdue University, 177 S. Russell St., West Lafayette, IN 47907-2099, Phone: (765) 494-2132, U.S.A.}
\cortext[cor1]{Corresponding author}
\begin{abstract}
The possibility of shifting sound energy from lower to higher frequency bands is investigated. The system configuration considered is a segmented structure having non-linear stiffness characteristics. It is proposed here that such a frequency-shifting mechanism could complement metamaterial concepts for mass-efficient sound barriers. The acoustical behavior of the material system was studied through a representative two-dimensional model consisting of a segmented plate with a contact interface. Multiple harmonic peaks were observed in response to a purely single frequency excitation, and the strength of the response was found to depend on the degree of non-linearity introduced. The lower and closer an excitation frequency was to the characteristic resonance frequencies of the base system, the stronger was the predicted higher harmonic response. The broadband sound transmission loss of these systems has also been calculated and the low frequency sound transmission loss was found to increase as the level of the broadband incident sound field increased. The present findings support the feasibility of designing material systems that transfer energy from lower frequency bands, where a sound barrier is less efficient, to higher bands where energy is more readily dissipated.
\end{abstract}
\end{frontmatter}

\section*{Nomenclature}

\noindent $c:$ Sound speed in air\\
$\rho_{0}:$ Mass density of air\\
$d_{cl}:$ Clearance between segment interfaces\\
$D_{i}:$ Positive-traveling wave amplitude in the downstream duct\\
$D_{r}:$ Negative-traveling wave amplitude in the downstream duct\\
$\delta:$ Contact overclosure\\
$f:$ Frequency of sound waves\\
$f_{ex}:$ Excitation frequency of a single frequency sound source\\
$f_{s}:$ Sampling frequency for time domain data\\
$f_m:$ Characteristic modal frequencies\\
$I_{0}:$ Intensity of the sound field on the incident side\\
$I_{0+}:$ Net $+ve$ Intensity of the sound field on the incident side\\
$I_{t}:$ Intensity of the sound field on the transmitted side\\
$I_{t+}:$ Net $+ve$ Intensity of the sound field on the transmitted side\\
$I_{s}$: Sound source Intensity\\
IL: Sound Intensity Level\\
$k:$ Wavenumber in the ambient fluid\\
$L_{0}:$ Edge length of the panel overall\\
$L_{p}:$ Distance between the outer ends of the segments\\
$\omega:$ Circular frequency in radians/s\\
$P_{1}, P_{2}, P_{3}$ and $P_{4}:$ Complex pressure amplitudes at the 'virtual' microphone \indent locations 1, 2, 3 and 4, respectively\\
$P\vert_{x_{1}=0}:$ Pressure at the incident side of the plate\\
$P\vert_{x_{1}=t_{p}+t_{s}}:$ Pressure on the transmitted side of the plate\\
$P_{i_{rms}}:$ Root mean square sound pressure acting on the incident face of the plate\\
$p_{c}:$ Contact pressure\\
$s:$ Slope relating the contact pressure with overclosure\\
$\sigma:$ Mass per unit area\\
$P_{w}:$ Lowpass Gaussian white noise sound source\\
SPL: Sound Pressure Level\\ 
STL: Sound Transmission Loss\\
$\mathrm{STL}_{\mathrm{limp}}:$ Analytical sound transmission loss of a limp panel for a given mass per unit area $\sigma$\\
powerShiftFrac: Fraction of net incident power shifted to frequencies other than the excitation frequency on the transmitted side in \%\\
$t_{p}:$ Thickness of the segments\\
$t_{s}:$ Thickness of the skin\\
$u_{1}$ and $u_{2}:$ Displacements of the two-dimensional model in $x_{1}$ and $x_{2}$\\
$u_{rot}$: Rotational displacement of the two-dimensional model in $x_{3}$ direction\\
$U_{i}:$ Positive-traveling wave amplitude in the upstream duct\\
$U_{r}:$ Negative-traveling wave amplitude in the upstream duct\\
$V\vert_{x_{1}=0}:$ Acoustic particle velocity at the incident side of the plate\\
$V\vert_{x_{1}=t_{p}+t_{s}}:$ Acoustic particle velocity at the transmitted side of the plate\\
$x_{1}$ and $x_{2}:$ Model coordinates\\

\section{Introduction}\label{sect:introduction}

Metamaterial-based sound barrier solutions~\cite{liuScience2000,yangPRL08,naifyJAP2010,varanasiAPAC2013, varanasi2017experiments} hold promise as light weight material alternatives for low frequency noise control. Yet, such approaches generally suffer from poor performance at their resonance frequencies which is an important issue that needs to be addressed. Potential solutions which mitigate that drawback include: incorporation of damping mechanisms, stacking multiple panels in sequence~\cite{yangAPL10} to expand the benefit-region frequency-span, use of multi-celled arrays~\cite{naify2JAP2011} mutually compensating for the poor performance of individual cells, disrupting the in-phase motion of the cells at their resonance frequencies, thereby reducing the corresponding radiation efficiency, or transferring the energy from the problematic frequency regions to higher frequencies where the barrier materials are more efficient~\cite{fahy2007sound}. 

The latter approach, i.e., the shift of energy to higher frequencies, can potentially be realized by using material systems having non-linear response characteristics. Lazarov and Jensen~\cite{lazarov2007} have shown that the presence of non-linear oscillators allows for spreading the local resonances of a periodic structure over a wider frequency range when compared to structures having locally-attached linear oscillators. Forced-excitation studies of materials having a bilinear stiffness character revealed their rich sub- and higher harmonic response~\cite{ugoIJNM2007,chuAIAA1992,pengIJMS2007}. Work has been performed to understand the response of such material systems when subject to forced harmonic excitation loads~\cite{ugoIJNM2007,chuAIAA1992,pengIJMS2007,maejawaBJSME1980,shawJSV1983} for fault diagnostic applications. 

The idea of shifting energy to higher frequencies where energy absorption is more efficient was studied by Denis et al.~\cite{denisIJNM2017} in the context of vibration mitigation. The authors studied plates with tapered power-law wedges possessing a flexural wave trapping effect, also referred to as the acoustic black hole (ABH) effect, and examined the geometric non-linearity aspect of the tapered wedges in shifting energy to higher frequencies. The authors showed through numerical simulations and experiments that the response characteristics of such plates have a significant non-linear signature and energy is transferred to higher frequencies. Also, the character of the response power spectral density was dependent on excitation amplitude. Energy absorbing visco-elastic layers were used in the tapered regions to dissipate the trapped energy, thereby resulting in vibration mitigation. Touz{\'e} et al.~\cite{liJSV2019} demonstrated numerically that effective vibration mitigation can be achieved by using a combination of ABH power-law wedges and contact non-linearity. The mechanism of vibration mitigation in the low frequency band was attributed to transfer of energy to higher frequencies by contact non-linearity and the more efficient damping of energy achieved at the higher frequencies through the use of ABH wedges combined with viscoelastic damping material.

However, the prospect of using material systems with non-linear response characteristics for sound barrier applications was investigated only recently by Rekhy et al.~\cite{rekhy2019}. The authors studied both numerically and experimentally the characteristics of plate structures with a mass-loaded membrane and an impactor; thus creating a non-linear system, and the feasibility of shifting energy from lower frequencies to higher frequencies was demonstrated. The Sound Transmission Loss (STL) was measured across a frequency band of white noise by using a four-microphone standing wave tube procedure~\cite{songJASA2000}. An improvement in the STL at the lower frequencies was attributed to the up-frequency shift of the incident energy. However, the measurement of the inter-microphone transfer functions in that procedure requires that the upstream and downstream signals be linearly related at each frequency, which is unlikely to be the case when the incident broadband random signal passes through a nonlinear system. Further, the authors used a one-load measurement, which is not appropriate for asymmetrical samples like theirs. Thus, their results are interesting but require further verification.

Here, we consider a planar cellular panel with its unit cells as shown in Fig.~\ref{fig:sgmntn_embodiments}. Unit cells can either comprise a cell frame and a cell-filling monolithic plate~\cite{varanasiAPAC2013}, Figs.~\ref{fig:sgmntn_embodiments}(b) and (d) or a cell-filling segmented plate, Figs.~\ref{fig:sgmntn_embodiments}(c) and (e). A planar cellular panel with its unit cell comprising segmented plates with contact interfaces can possess linear, bilinear or nonlinear stiffnesses. Although there are a number of variations in possible contact behavior, the qualitative nature of their harmonic responses to a pure tone excitation was expected to be similar~\cite{dimarogonasEFM1996,andreausJSV2011, chuAIAA1992, pengIJMS2007}. The behavior has been studied extensively in the context of fault diagnostics, but the main objective of the present study was to explore the feasibility of employing contact non-linearity in the design of efficient sound barriers. Therefore, a simple contact scenario of frictionless hard contact was studied in this work: i.e., the examination of the sound transmission characteristics of segmented plates possessing bilinear or non-linear stiffness behavior due to the presence of contact interfaces. Here, numerical models were employed to study the characteristics of representative two-dimensional models comprising a segmented plate having a contact interface, Fig.~\ref{fig:material}(b),  and connected to each other with an adhesive skin having a very small stiffness. 

\section{Model Definition}\label{sect:modelDef}
 Two-dimensional models, Fig.~\ref{fig:material}, based on the three-dimensional embodiment of segmented metamaterial panels, Fig.~\ref{fig:sgmntn_embodiments}, were studied. The two-dimensional models consisted of two plate segments attached to a skin that provides an integrating support to the structure. The distance between the outer ends of the segments, $L_{p}$, Fig.~\ref{fig:material}(a), was chosen to be $63.5$~mm and the thickness of the skin, $t_{s}$, was taken to be $50$~$\mu$m. The thickness of the plate segments was set to $t_{p}=1.27$~mm, equivalent in stiffness at its center to a three-dimensional plate with its in-plane dimensions aspect ratio being equal to $1$ and having a thickness of $1.00$~mm (assuming the same elastic modulus for the three-dimensional and two-dimensional cases). The choice of $L_{p}$ was inspired by the  duct dimensions of an available standing wave tube apparatus~\cite{ASTMimpedtube} with a square cross-section of $63.5\times63.5$~mm~\cite{techrevBrK}. The plate thickness was chosen based on the specimens used in experimental investigations previously conducted by the authors~\cite{varanasiAPAC2013,varanasi2017experiments}. Monolithic specimens whose thickness was $1$~mm were previously studied by the authors~\cite{varanasiAPAC2013} using a standing wave tube set up, as were application-scale cellular panels whose unit cell interior thickness was on the order of $1$~mm~\cite{varanasi2017experiments}. Such a choice of thickness would allow for incorporation of the present idea into the concept of cellular panels for a future experimental study. The thickness of the skin,  $t_{s}$,  was representative of a typical adhesive tape. The choice of the clearance, $d_{cl}$, magnitudes was determined by the possible interface fineness that can be achieved with machined acrylic plates.
 
 Two models based on the configurations shown in Fig.~\ref{fig:material}(b) are referred to here as M$1$ and M$2$; they differed from each other in the clearance separating the segments. While M$1$ had a clearance ($d_{cl}$) of  $10$~$\mu$m, M$2$ was given a clearance of $100$~$\mu$m. A third segmented plate model, M$3$, with a skin thickness $t_{s}$ of $500$~$\mu$m and zero clearance ($d_{cl}$) between the segments was also studied. All the other dimensions characterizing the model remained the same as specified above. An unsegmented equivalent of models M$1$ and M$2$, Fig.~\ref{fig:material}(a), representing a structure with linear response, was also used in this study for comparative analysis, and is referred to as model M$0$. Model M$0$ includes the segments and the skin with all of them bonded to each other. The  dimensions of the various models  are summarized in Table~\ref{tab:M1M2M4dimensions}. Typical acrylic material properties, i.e., $E=3.7$~GPa, Poisson's ratio $\nu=0.35$ and a mass density of $\rho=1160$~kg/m$^3$, were applied to both the skin and the plate segments in all models. 
 
 The contact interaction between the two plate segments was modeled using a solid-to-solid contact described by a linear pressure over-closure relationship, as illustrated in Fig.~\ref{fig:presOverclosure}. Such a model is available in the FE-code ABAQUS~\cite{abaqusdocs}. The contact pressure ($p_{c}$) - overclosure ($\delta$) relationship is $p_{c}=s\delta$, with $s=10^{14}$~Pa/m for M$1$ and M$2$, and $s=10^{13}$~Pa/m for M$3$. The contact constraint was enforced using a Direct method~\cite{abaqusdocs}. The Direct method strictly enforces a given pressure-overclosure behavior for each constraint, without approximation or use of augmentation iterations: this is akin to enforcing a hard contact with a penalty constraint. The displacements $u_{1}$ and $u_{2}$ in the $x_{1}$ and $x_{2}$ directions were monitored at two nodes located on the far right end on each of the contact interfaces.

\section{Analysis Methods}\label{sect:methods}

\subsection{Numerical Analysis Methods}\label{subsect:meth_static_analysis}

Two-dimensional finite element (FE) models were built for a static analysis to evaluate the stiffness characteristics of the plate systems. The analysis was performed with FE-code ABAQUS~\cite{abaqusdocs}. The present model employs a combination of solid elements for the plate components as well as beam elements for the skin component. The use of plate elements was not possible in this study since the face-edge-to-face-edge contact conditions between the segmented plates cannot be enforced in a model built with plate elements. The plate segments were discretized using two-dimensional plane strain quadratic elements (CPE8R in ABAQUS). The segments were modeled by using a biased mesh with finer resolution towards the contact interface: the smallest element size was ${\sim}1\%$ of $L_{p}$ and the largest element size was ${\sim}3\%$ of $L_{p}$. The thickness dimension of the segments was discretized uniformly with an element size of $5\%$ of  $t_{p}$. The skin was modeled by using one-dimensional linear structural beam elements (B21 in ABAQUS) which are based on Timoshenko beam theory with a uniform element size of $0.5$~mm: i.e., ${\sim}0.8\%L_{p}$. The choice of analytical beam elements over continuum solid elements to model the skin greatly reduced the computational expense. The out-of-plane depth of the plate segments and the skin were taken to be unity. 

Figures~\ref{fig:methods_static_analysis}(a) and (b) show the loading and boundary conditions. A uniform static pressure load of $100$~Pa was applied consecutively to the the skin in the $+ve$ $x_{1}$ direction, Fig.~\ref{fig:methods_static_analysis}(a), and the segment faces in the $-ve$ $x_{1}$ direction, Fig.~\ref{fig:methods_static_analysis}(b). The load applied to the skin was released while the segment faces were loaded. The ends of the beam extending beyond the plate segments were fully constrained: i.e., $u_{1} = u_{2}$ = $u_{rot}$ = 0, where $u_{1}$ and $u_{2}$ are the displacements in the $x_{1}$ and $x_{2}$ directions, respectively, and $u_{rot}$ was the rotation about the $x_{3}$ axis.

\subsection{Modal Analysis}\label{subsect:meth_modal_analysis}

Two-dimensional FE models were created to evaluate the natural frequencies of the models M$0$, M$1$, M$2$ and M$3$. The natural frequency extraction methods in the FE-code ABAQUS were employed with the details of this procedures described in~\cite{newman1973fast,ramaswamyMITthesis, parlett1981Book}. Characteristic eigenmodes of the models were determined  by modal analyses performed by using the Lanczos eigensolver method~\cite{abaqusdocs}. The domain discretization and the boundary conditions remained the same as in the static analysis. Note that a linear procedure cannot account for contact non-linearity. Thus, the characteristic modal frequencies ($f_m$) for M$1$, M$2$ and M$3$ were determined for the two following configurations: (1) the \textit{open} configuration, in which the plate segments were coupled only by the thin skin so that the plate segments interacted with each other without any contact, i.e., representing a configuration with a wide gap and small amplitude vibrations; and (2) the \textit{closed} configuration in which the plate segments were fused to each other to form a conventional monolithic plate. The characteristics of bilinear materials~\cite{chatiJSV1997} are determined by their bilinear frequencies, which are the harmonic means of the corresponding modal frequencies in the open and closed configurations. By using those modal frequencies along with those of M$1$ and M$2$ in their open configuration, the corresponding first bilinear frequencies were determined. Similarly, for M$3$, its first bilinear frequency was determined by using its modal frequencies in the \textit{open} and \textit{closed} (interface tied) configurations. The various natural frequencies are listed in Table~\ref{tab:M0M1M2M3freq} The bilinear frequencies of the models were used when choosing the excitation frequencies of the sound sources in the subsequent acoustical analyses.   

\subsection{Acoustical Analysis}\label{meth_acoustical_analysis}
\subsubsection{FE model of Standing Wave Tube}



A two-dimensional FE model was also built to study the acoustical properties of the models, Fig.~\ref{fig:methods_acoustical_analysis}. Again, the FE-code ABAQUS ~\cite{abaqusdocs} was employed. The model was built to reproduce the standing wave tube setup that is typically used for characterizing barrier materials~\cite{songJASA2000, ASTMimpedtube,techrevBrK}. The numerical model involved a coupled structural-acoustic procedure based on implicit dynamic analysis using implicit time integration methods~\cite{abaqusdocs}. That procedure allows for simulating the non-linear behavior of contact interactions. The contact interface was modeled following the same rules described above in the context of the static analysis. The model consisted of two acoustic domains (upstream and downstream) and a solid domain consisting of the segmented plate structure separating the two acoustic domains. The lengths of the up- and downstream acoustic domains were taken to be $180$ mm each with their width equal to $63.5$~mm. Acoustic elements (4-node bilinear) were used to model the acoustic domains. The two acoustic domains were discretized with an element size of $6$~mm along the length and $3$~mm along the width dimensions. The wavelength at the maximum frequency in the analysis, i.e., $5000$~Hz, contains $11$ elements, thus, well-exceeding the usual requirement of at least $6$ elements per wavelength. The discretization of the structural domain remained the same as described for the static analysis models.

\subsubsection{Harmonic Sound Source}

A sound source radiating a single frequency $\left(f_{ex}\right)$ sound pressure was simulated at the upstream end of the acoustic domain by prescribing the normal derivative of the acoustic pressure per unit density. In addition, the upstream termination was modeled as non-reflecting by imposing the specific acoustic impedance of air $\rho_{0}c$ ($\rho_{0}$ is the density of air and $c$ is the speed of sound in air). Similarly, the downstream end of the acoustic domain was also modeled as an anechoic (non-reflecting) termination. Each of the models M$0$, M$1$, M$2$ and M$3$ was studied for their behavior at three excitation frequencies, which were selected based on their corresponding first characteristic bilinear frequencies. 

The predicted sound pressures were recorded by using virtual 'microphones' at two axial locations both in the upstream and downstream acoustical domains at  time intervals of $1\times10^{-4}$~sec: i.e., at a sampling rate, $f_{s}$, of $10$~kHz, the Nyquist frequency thus being $5$~kHz. The upstream 'microphones' were located at $x_{1}=-150.0$ and $-138.0$~mm, and the downstream 'microphones' at $x_{1}=151.3$ and $163.3$~mm. The maximum frequency of interest here was nominally $2000$~Hz, thus ensuring that the sampling frequency, $f_{s}$, was at least $5$~times higher than the maximum frequency of interest. Note that higher order acoustic modes would begin to propagate at approximately $2700$~Hz in a standing wave tube having the cross-sectional dimension considered here. The pressure measurement locations were chosen so that they were at least a distance $L_{p}$ away from the unit cell on either side to avoid near-field effects.  The results are presented in terms of sound intensity spectra which were calculated as described in Section~\ref{subsubsecMicDataAnSSSection} below. Apart from recording the sound pressures at the virtual 'microphone' locations, the displacements $u_{1}$ and $u_{2}$ were monitored at the far right ends of the segments' contact interfaces. The latter information was used to ascertain the occurrence of contact and to correlate the observations of displacements and sound pressure.

\subsubsection{Analysis of the Microphone Data}\label{subsubsecMicDataAnSSSection}

An acoustic layer having linear behavior can be characterized by a $2\times2$ constant coefficient transfer matrix which can be determined by using the sound pressure data taken from the four microphones in a standing wave tube~\cite{songJASA2000}. That matrix can further  be used to determine the  STL or absorption characteristics of the target material. However, in the case of a material having nonlinear characteristics, a linear transfer matrix cannot be assumed because of the multi-frequency response resulting from a single frequency sound source excitation. However, although the structural response might be nonlinear, the wave propagation in the fluid domains (ducts) can remain linear and is still governed by the linear wave equation if sound levels are not too high. Hence, the pressure data (converted to the frequency domain from the time domain) from each of the microphone pairs was used separately to resolve the waves traveling in the $+ve$ $x_{1}$ and $-ve$ $x_{1}$ directions in the upstream and downstream ducts. That information was further used to determine the pressures and acoustic particle velocities on both sides of the test specimen~\cite{songJASA2000}: i.e., on the incident and transmitted sides, Fig.~\ref{fig:methods_acoustical_analysis}. From that data, the time-averaged sound intensity spectra on both sides of the test specimen were computed following the procedure described in \ref{sec:soundIntensityApp}. The time-averaged sound intensity spectra were further resolved into $+ve$ $x_{1}$ and $-ve$ $x_{1}$ traveling components based on their algebraic sign. The spectrum with $+ve$ valued intensity on the upstream side of the specimen was the net incident intensity, here referred to as $I_{0+}$, and the  $+ve$ valued intensity on the downstream side corresponded to the net transmitted intensity, here referred to as $I_{t+}$. The fraction of sound power shifted to frequencies other than the incident frequency was quantified by summing the corresponding intensities in $I_{t+}$, dividing the result by the sum of the net transmitted intensities $I_{t+}$, and expressing that ratio in percentage terms, i.e., $\dfrac{\sum\limits_{f{\neq}{f_{ex}}}^{} I_{t+}}{\sum\limits_{}^{} I_{t+}}{\times}100$, referred to subsequently as powerShiftFrac. The sound pressure level, SPL, acting on the incident face of the plates at the source frequency was also calculated.


\subsubsection{Lowpass Gaussian White Noise as Sound Source}

While the harmonic sound source excitation studies helped to illustrate the qualitative response characteristics, a broadband sound source excitation study was conducted to quantitatively study the STL characteristics of the segmented plates. A lowpass Gaussian white noise time-domain signal ($P_{w}$) was generated by applying a low-pass FIR filter with a cutoff frequency of $1000$~Hz to a Gaussian white noise signal with a mean value of $0$~Pa and a standard deviation of $1$~Pa. The signal was generated for a total time length of $10$~sec with a sampling frequency of $10000$~Hz. Signals with standard deviations of $10$, $25$, $50$ and $100$~Pa were then created by scaling the standard deviation to $10$, $25$, $50$ and $100$~Pa, respectively. The signal with a standard deviation of $10$~Pa was used for the comparative study of models M$0-$M$3$ while all the signals were used to study the effect of sound source amplitude on M$1$. The modeling of the sound source end as non-reflecting helped to improve the signal-to-noise ratio by minimizing reflections at the source end. The predicted sound pressures were recorded by using virtual 'microphones' at two axial locations in both the upstream and downstream acoustic domains at time intervals of $1\times10^{-4}$~sec, as previously noted. Sound pressures were also recorded at two additional locations; one closer to the sound source in the up-stream duct,  and the second in the down-stream duct nearer to the sample compared to the other two microphones to serve as reference signals, respectively. The upstream reference 'microphone' was located at $x_{1}=-162.0$~mm and the downstream 'microphone' was located at $x_{1}=139.3$~mm. Those signals were used in the post-processing and analysis of the microphone data as detailed in the next section. The virtual 'microphone' locations were chosen to be sufficiently far away from the sample in the standing wave tube so that near-field effects were avoided.

\subsubsection{Analysis of the Microphone Data for White Noise Excitation}\label{subsubsubsecMicDataAnSSSection2}

The time-domain non-deterministic signals from the six microphones described above were processed using the procedure laid out for the standard test method to measure normally incident sound transmission of acoustical materials~\cite{ASTMimpedtube}. The complex acoustic transfer functions between the reference and the other two microphones in each of the up-stream and down-stream ducts were determined by taking the ratio of the one-sided cross-spectral density of each of the microphone signals with the reference microphone one-sided power spectral density. The physical process was assumed to be stationary and ergodic in nature and time-domain averaging was used to evaluate the cross-spectral densities. The complex transfer functions were in-turn used to resolve the incident and  reflected sound waves as a function of frequency in the up-stream and down-stream ducts, separately. Sound intensities on the incident~($I_{0}$) and transmitted~($I_{t}$) side  of the sample were evaluated in the frequency domain by using Eqs.~(\ref{eq:PndV}) and~(\ref{eq:I0It}) given in \ref{sec:soundIntensityApp}. The $\mathrm{STL}$ in the frequency band of excitation ($0-1000$~Hz) was computed by taking the ratio of net transmitted sound intensity ($I_{t+}$) to sound source intensity ($I_{s}$) and converting the result to decibels: i.e.,
\begin{equation}\label{eq:stlFRMTrintensity}
\mathrm{STL}=-10\log_{10}(\|{I_{t+}/{I_{s}}}\|).
\end{equation}
Note that the transfer matrix method described in~\cite{ASTMimpedtube} could not be used to compute $\mathrm{STL}$ since the plate response was non-linear in nature. Instead, the $\mathrm{STL}$ measure defined above is in-line with the method typically used to characterize a sound barrier in a  reverberation room test setup~\cite{varanasi2017experiments}. The predictions were compared against an analytical limp panel STL. For a limp panel with a mass per unit area of $\sigma$, the STL for a normal-incidence sound field $\left(\mathrm{STL}_{\mathrm{limp}}\right)$ was calculated as~\cite{fahy2007sound}:
\newcommand{\ud}{\,\mathrm{d}}
\begin{equation}\label{eq:limppanel}
\mathrm{STL}_{\mathrm{limp}}=10\log_{10}{\frac{4{\rho_{0}}^{2}c^2+{\omega}^2{\sigma}^{2}}{4{\rho_{0}}^{2}c^2}}.
\end{equation}
Here, $\rho_{0}$ is the density of air, $c$ is the speed of sound in air, $\omega=2{\pi}f$ where $f$ is the frequency in Hz, and $\sigma$ is the areal mass.

\section{Results}\label{sect:resNDdiscuss}
\subsection{Model stiffness}\label{subsect:res_stiffness}

The skin and the plate-plate-face-edge contact are an integral part of the segmented plate configuration of interest. Contact and skin deformation are the key factors in the deformation response of the system. Figure~\ref{fig:results_static_analysis}(a) shows the resultant force versus $u_{1}$ (deflection in the $x_{1}$ direction) for models M$0$, M$1$ and M$2$, all of which have the same thickness for the beam representing the skin. The resultant force was calculated by multiplying the applied uniform pressure by the area of the plate segments normal to the $x_{1}$ direction. 

First, studying the stiffness behavior when a load was applied to the segment faces in the $-ve$ $x_{1}$ direction, it can be seen that M$1$, $d_{cl}=10$~$\mu$m clearance, shows a sharp change in slope due to the closing of the contact interface at an axial displacement of $u_{1}\approx$~125~$\mu$m. The displacement $u_{1}$ at which the plates come in contact was analytically estimated to be equal to $\frac{L_pd_{cl}}{4t_p}$ (when it is assumed that the skin kinks at its center point and that the curvature at the remaining locations is negligible - see~\ref{sec:u1Analytical}). The latter value was $125$~$\mu$m in the present case, the same as the numerically predicted value. The ratio of the high to low stiffnesses in this phase was $127$. The estimated displacement $u_{1}$ after which M$2$ with $d_{cl}=100$~$\mu$m comes in contact was 1250~$\mu$m. The contact does not close even for the maximum static load applied, but its behavior becomes nonlinear due to the geometric nonlinearity of the unsupported segment of the skin at its center portion. Here, it can also be seen that the deflection curve for M$2$ traces that of M$1$ until the point where the contact occurs for M$1$ but then continues in the same trajectory since contact did not occur in the M$2$ case. For M$0$, the deflection behavior is linear with its slope being equal to that of M$1$ in the closed configuration. 

Next, studying the characteristics of loading on the skin in the $+ve$ $x_{1}$ direction, it can be seen that both M$1$ and M$2$ exhibit geometric nonlinearity, and that there is no sharp change in the slope, since no contact occurs. There is a minor quantitative difference in the part of the curve in the first quadrant for M$2$ compared to its counterpart in the third quadrant apart from the change in sign of the slope from $-ve$ to $+ve$. The force magnitude for the same deflection magnitude was slightly above $3$~N for loading on the skin in the $+ve$ $x_{1}$ direction when compared to the corresponding force magnitude for loading on the segment faces in the  $-ve$ $x_{1}$ direction, which was slightly below $3$~N. In the case of M$0$, the stiffness remains linear and was the same as seen in the $-ve$ $x_{1}$ loading direction. 

Figure~\ref{fig:results_static_analysis}(b) shows the force-deflection behavior for M$3$ which has a thicker beam representing the skin ($10$ times thicker than that of M$0$, M$1$ and M$2$) and with a zero thickness clearance. Here, it can be seen that the curve segments are linear in the $-ve$ and $+ve$ loading directions with a slope change occurring at $x_{1}=0$ with the curve in the third quadrant having a higher slope. The ratio of the high and low stiffness values in this case was $1.8$. 

\subsection{Characteristic Frequencies}\label{subsect:res_frequencies}
 The first mode of the equivalent unsegmented plate M$0$, Fig.~\ref{fig:results_modal_analysis}(a), occurs at $429$~Hz. Figures~\ref{fig:results_modal_analysis}(b,c) show the first mode shapes for the models M$1$, M$2$ and M$3$, qualitatively, with the details of their occurrence depending on the specific model stiffness and mass characteristics. The occurrences of the modes for M$1$ and M$2$ were very close since the only difference between the two was the contact interface clearance. Hence, their equivalent in the closed configuration can be taken to be the same: i.e., M$0$. The first modal frequencies in the contact-open, contact-closed configurations, and the corresponding bilinear frequencies are listed in Table~\ref{tab:M0M1M2M3freq}. With the idea of having excitation frequencies for the acoustical analysis that bracket the first bilinear frequency for each case, excitation frequencies of $20, 80$ and $320$~Hz were chosen for M$1$ and M$2$. The same excitation frequencies were chosen for the unsegmented M$0$ case to allow for a comparative analysis. The excitation frequencies were chosen to be $80, 320$ and $640$~Hz for the case of M$3$ since its first bilinear frequency is $451.5$~Hz.

\subsection{Acoustical Response}\label{subsect:res_ac_analysis}
\subsubsection{Effect of Stiffness Non-linearity}



Figures~\ref{fig:100Pa_M0}(a-f) show the net $+ve$ sound intensity spectra ($I_{0+}$ and $I_{t+}$) on the upstream and downstream sides of the unsegmented plate (M$0$) for excitation ($f_{ex}$) frequencies of $20, 80$ and $320$~Hz, respectively, at an amplitude of $100$~Pa (strength of the incident sound pressure wave at the source end of the up-stream duct). In these plots, the intensity is expressed in a decibel scale with a reference intensity of $10^{-12}$~W/m$^2$, and in that form is further referred to as the Sound Intensity Level (IL). From Figs.~\ref{fig:100Pa_M0}(a-f), it can be observed that harmonics were predicted in the transmitted spectra ($I_{t+}$) with decreasing orders of magnitude compared to $I_{0+}$. The last significant harmonic was around $20$~dB while $I_{0+}$ was around $80$ and $100$~dB for $f_{ex}=20$ and $80$~Hz, respectively. The magnitude of the lowest level harmonic in $I_{t+}$ was around $20$~dB compared to $120$~db of $I_{0+}$ for $f_{ex}=320$~Hz. An insignificant response was also predicted for $320$~Hz excitation at non-integral multiples of the corresponding excitation frequency. Table~\ref{tab:powShiftFrac} summarizes powerShiftFrac for all the cases studied in this work and Table~\ref{tab:pressureAtxZero} lists the corresponding ${\mathrm{SPL}_{x_{1}=0}}$ on the incident side of the panel. For $20$, $80$ and $320$~Hz excitations, powerShiftFrac for M$0$ was $0.08\%$, $0.11\%$ and $0.19\%$, respectively. The shift in power was not significant with an increasing trend as the excitation frequency approached the first flexural resonance frequency of  $429$~Hz. Considering the model's linear stiffness behavior, the occurrence of upper-harmonic was not expected. The reasons for the occurrence of the low order upper-harmonics are discussed in  Section~\ref{res_discuss}. 
%


Figures~\ref{fig:100Pa_M1}(a-f) show $I_{0+}$ and $I_{t+}$ of the segmented plate model M$1$ for $f_{ex}=20, 80$ and $320$~Hz, respectively, at a sound source amplitude of $100$~Pa. Figures~\ref{fig:100Pa_M1_zoomed}(a-b) show the zoomed-in views of $I_{t+}$ for $f_{ex}=20, 80$~Hz excitation frequencies. The occurrence of upper-harmonics for excitations of $20$ and $80$~Hz extended to $3900$ and $5000$~Hz, respectively, with the lowest level harmonic of magnitude $0$~db and their corresponding $I_{0+}$ at $120$~dB. For $f_{ex}=320$~Hz, there were only three upper-harmonics in $I_{t+}$ with the farthest at $960$~Hz and around $30$~dB in magnitude against an $I_{0+}$ of $120$~dB. No subsequent upper-harmonics were predicted. There were some upper-harmonics at approximately the same order of magnitude as $I_{0+}$ for $f_{ex}=20$ and $80$~Hz. The magnitudes of the upper-harmonics decreased steadily as the frequency increased for all the cases of $f_{ex}$. No sub-harmonics were predicted for any of the three excitation frequencies, Figs.~\ref{fig:100Pa_M1_zoomed}(a-b), in $I_{t+}$. Some sub-harmonics were predicted for the excitation frequency of $80$~Hz in $I_{0+}$ with an average magnitude of $20$~dB. The powerShiftFrac for $20$, $80$ and $320$~Hz excitations were $66.8\%$, $51.7\%$ and $0.00\%$, respectively. The shift in power was very significant for $f_{ex}=20$ and $80$~Hz with the fraction being higher for $20$~Hz. The influence of the proximity of the excitation frequency to the first flexural frequency of $55.5$~Hz can once again be seen here. Considering the response for $f_{ex}=320$~Hz, unlike the other excitation frequencies, the powerShiftFrac was zero suggesting that contact did not occur at the interface. Figures~\ref{fig:100Pa_M1_disp}(a-c) show the transverse displacement, $u_{2}$, versus time of the far right end of segment~$1$, Fig.~\ref{fig:material}(b), for $f_{ex}=20, 80$ and $320$~Hz, respectively. It can be seen that contact does occur in the cases of $20$ and $80$~Hz, but not in the case of the $320$~Hz excitation. While the displacement $u_{2}$ was limited to $5$~microns in the former cases (the clearance was $10$~microns for M$1$), the maximum displacement reached for $320$~Hz excitation was below $5$~microns. 



Figure~\ref{fig:100Pa_M2}(a-f) depicts results for model M$2$ which only experienced a geometric nonlinearity, unlike M$1$ which experienced both geometric and contact non-linearities (Section~\ref{subsect:res_stiffness}). The occurrence of upper-harmonics in $I_{t+}$ extended to $840$~Hz for the $f_{ex}=20$~Hz excitation with its $I_{0+}$ at around $120$~dB. The drop in magnitude of $I_{t+}$ was linear in the dB scale and almost monotonically decreasing. The frequency range for non-zero $I_{t+}$ extended to $1280$ and $960$~Hz, respectively, for the $80$ and $320$~Hz excitations. No sub-harmonics were predicted for any of the excitation frequencies either in $I_{0+}$ or $I_{t+}$. All the observations made in the context of M$1$ apply to this case as well. The main difference was that the occurrence of upper-harmonics was less pronounced, with their strength dropping much more rapidly with increasing frequency which was reflected in powerShiftFrac dropping to $46.2\%$ and $7.46\%$ for $f_{ex}=20$ and $80$~Hz, respectively. There was no change in the powerShiftFrac for $f_{ex}=320$~Hz from $0.00\%$. The absence of contact closure in the case of M$2$ was confirmed by observing the transverse displacement of the far right end of segment~$1$, Fig.~\ref{fig:100Pa_M2_disp}(a-c). It can be seen that the maximum amplitude of the transverse displacement was lower than half of the clearance: i.e., $50$~$\mu$m at all the excitation frequencies. Note that the transverse displacement for the excitation frequencies of $20$ and $80$~Hz does not show a pure harmonic behavior although it is periodic. 


Figures~\ref{fig:100Pa_M3}(a-f) show the acoustic response characteristics of model M$3$ at an excitation amplitude of $100$~Pa for excitations of $80$, $320$ and $640$~Hz, respectively. Higher excitation frequencies were considered for this case to bracket the higher flexural bilinear frequency of $451.5$~Hz. Harmonics occurred up to $3900$~Hz in $I_{t+}$ for all three excitation frequencies with the farthest significant harmonic magnitudes of $10$, $20$ and $40$~dB, respectively. The corresponding $I_{0+}$ were $100$, $120$, and $120$~dB. In contrast to the models M$1$ and M$2$ which experienced geometric non-linearity, M$3$ showed only a contact non-linearity. Also, M$3$ had a lower bilinear stiffness ratio of $1.8$ as compared to the bilinear stiffness ratio of $127$ for M$1$, owing to the thicker beam used for representing the skin. The upper harmonic response was weaker than that for M$1$ and M$2$, as reflected in powerShiftFrac (Table~\ref{tab:powShiftFrac}) with the highest powerShiftFrac of $13.94\%$ for $80$~Hz excitation. The trend of decreasing powerShiftFrac with increasing excitation frequency continued for M$3$. The presence of a flexural resonance frequency of $451.5$~Hz lying in between $320$ and $640$~Hz apparently did not have much effect in reversing this trend. Unlike the response characteristics for M$0$, M$1$ and M$2$, sub-harmonics were predicted in the $I_{t+}$ spectra for $320$ and $640$~Hz excitations. The upper harmonics were caused by the contact closure, as reflected in the transverse displacement plots at $f_{ex}=80$, $320$ and $640$~Hz shown in Fig.~\ref{fig:100Pa_M3_disp}(a-c), respectively.






\subsubsection{Effect of Excitation Amplitude}
All the cases discussed in the previous section were based on the same source excitation strength of $100$~Pa. Figures~\ref{fig:10Pa_M1}(a-f) show $I_{0+}$ and $I_{t+}$ spectral characteristics of M$1$ for an excitation amplitude of $10$~Pa. By comparing the response for M$1$ at the two different amplitudes, it can be seen that upper-harmonic response was less pronounced at the reduced amplitude. Significant upper-harmonics were observed only for $f_{ex}=20$~Hz extending to $3840$~Hz. The drop in the strength of the upper-harmonics in $I_{t+}$ was steeper compared with the characteristics of M$1$ for the $100$~Pa excitation. This was reflected in the powerShiftFrac which was only $26.68\%$ as opposed to $66.80\%$ for the $100$~Pa excitation amplitude at $20$~Hz. Similarly, the powerShiftFrac dropped to $0.00\%$ at $f_{ex}=80$~Hz. Extended contact closure occurred for the $20$~Hz excitation, Fig.~\ref{fig:10Pa_M1_disp}(a-c), thus supporting the upper-harmonic acoustic response predictions. The powerShiftFrac for $10$~Pa amplitude excitation for M$2$ and M$3$ presented a similar trend in the drop of powerShiftFrac, as can be seen from Table~\ref{tab:powShiftFrac}. 


\subsubsection{Broadband Excitation Response Characteristics}

Figure~\ref{fig:stlBroadbandWhNoiseM0M1M2M3-meas2}(a) shows the $\mathrm{STL}$ characteristics of M$0$, M$1$, M$2$ for the same sound source excitation: i.e., the lowpass filtered ($0-1000$~Hz) Gaussian white noise excitation with zero mean and standard deviation of $10$~Pa. This STL was defined in Eq.~(\ref{eq:stlFRMTrintensity}) by using the ratio of the transmitted intensity to the incident sound intensity. The analytical STL characteristics of an equivalent limp panel with an areal mass of $1.53$~kg/m$^2$ is overlaid for comparison. Similarly, Figure~\ref{fig:stlBroadbandWhNoiseM0M1M2M3-meas2}(b) shows the same result for M$3$. The equivalent limp panel for M$3$ had an areal mass of $2.05$~kg/m$^2$. The STL characteristics of all the models primarily followed the trend of a classic edge-constrained plate~\cite{beranek1960, song_boltonJASA01, song_boltonNCEJ03} starting with a high STL at very low frequency, decreasing steeply until the first flexural resonance frequency, and then steadily increasing beyond it. The dip in the STL for M$1$ and M$2$ occurred around $50$~Hz, which corresponded to the first flexural resonance frequency of $55.5$~Hz (Table~\ref{tab:M0M1M2M3freq}). Similarly, the location of the dips in the M$0$ and M$3$ STLs corresponded to their first flexural resonance frequencies of $429$ and $451.5$~Hz, respectively. A notable improvement in STL was predicted for M$1$ and M$2$ beyond the dip frequency when compared to their equivalent limp panels with an average value of $2$~dB. Such an improvement was not predicted for M$3$ in the frequency range considered here.


Figure~\ref{fig:stlBroadbandWhNoiseM1AmpEffect}(a) shows the $\mathrm{STL}$ characteristics of M$1$ for the lowpass ($0-1000$~Hz) filtered Gaussian white noise excitation with zero mean and varying standard deviations from $10$ to $100$~Pa. The analytical STL characteristics of an equivalent limp panel with an areal mass of $1.53$~kg/m$^2$ was overlaid for comparison. It can be seen from this result that the improvement in STL becomes increasingly significant in the very low frequency range of $0-200$~Hz as the standard deviation of the white noise increases. A unifying feature for all the plots was that each of them followed the trend of a classic edge-constrained plate~\cite{beranek1960, song_boltonJASA01, song_boltonNCEJ03} but with a different frequency for the dip in STL: i.e., a lower excitation amplitude corresponded to a lower dip frequency. As the amplitude increased, a noticeable loss in STL was predicted in the frequency range of $200-600$~Hz. The loss in STL decreased beyond $600$~Hz eventually converging with the STL prediction for $10$~Pa amplitude. Figure~\ref{fig:stlBroadbandWhNoiseM1AmpEffect}(b) shows the coherence plots of downstream microphone m$3$ and the corresponding input sound source signal for the cases with standard deviations of $10$, $25$, $50$ and $100$~Pa, respectively. It can be seen from this result that the coherence value fell increasingly below unity in the low frequency range of $0-600$~Hz as the excitation amplitude was increased, suggesting an increasing non-linearity effect on the predictions. That observation correlated with the differences in the STL predictions seen in Fig.~\ref{fig:stlBroadbandWhNoiseM1AmpEffect}(a).

 
\section{Discussion}\label{res_discuss}

Upper-harmonics occur in the acoustical response of structures having geometric and contact non-linearities. The presence of contact interfaces particularly accentuates the non-linear effect due to the steep change in the stiffness. In fact, it was shown that a bilinear stiffness behavior can be expressed as a polynomial non-linearity of order $4$~\cite{pengIJMS2007} in the context of forced-excitation response studies. The predominant trend that can be noted from the powerShiftFrac for all the cases listed in Table~\ref{tab:powShiftFrac} was that the lower the pure tone excitation frequency, the larger was the amount of power shifted to higher frequencies. The only case where the trend was different was M$0$, which only exhibited a mild non-linearity discussed further below. In that case, the powerShiftFrac increased as the pure tone excitation frequency increased with a maximum powerShiftFrac of $0.19\%$. It has to be noted here that powerShiftFrac includes the power shifted to frequencies other than the excitation frequency but since the presence of sub-harmonics is not significant compared to the occurrence of upper-harmonics, it can be considered as a good quantifying indicator of the power shifted to frequencies higher than the excitation frequency. 

Potentially, there are multiple factors that determine the amount of power shifted to higher frequencies, with the proximity of the excitation frequency to the first flexural frequency being the second prominent one. The first flexural resonance frequency for M$1$ and M$2$ of $55.5$~Hz was located in between the $20$ and $80$~Hz excitation frequencies and was a bit closer to $80$~Hz. In the case of M$0$, the first flexural resonance frequency was $429$~Hz with $f_{ex}=320$~Hz being the closest to it, and that frequency had the highest powerShiftFrac of $0.19\%$. 

Two other factors affecting the shift of power to higher frequencies that stand out from the results are the effect of amplitude and the degree of non-linearity. The higher the amplitude of excitation, the stronger was the shift, as reflected in the powerShiftFrac for all the cases as listed in Table~\ref{tab:powShiftFrac}. Also, the overall shift in power was in the order of M$1 > $M$2 > $M$3 > $M$0$ which correlates with the bilinear contact stiffness ratio and geometric non-linearity. M$1$ had both the highest bilinear contact stiffness of $127$ and geometric nonlinearity reflected in the curved nature of the force-deflection curve in the first quadrant. M$2$ had a similar geometric non-linearity with a not-so-significant bilinear stiffness: the ratio of the tangent stiffness at $u_{1} = 300$ and $-300$~$\mu$m was ${\sim}1.1$. M$3$ represented a scenario primarily of contact non-linearity with a lower bilinear stiffness ratio of $1.8$ compared to M$1$. M$0$ represented a scenario closest to being linear with a bilinear stiffness ratio of $1$.

The reason for the occurrence of low order harmonics in the case of M$0$ which has linear static stiffness behavior was also examined. It was hypothesized that a potential geometric non-linearity arose from the magnitude of the deflections on the order of the skin slenderness. The analyses were repeated by turning off the terms accounting for the geometric non-linearity in the strain-displacement relationship. No significant upper-harmonics appeared in the corresponding response, thus confirming the hypothesis, Fig.~\ref{fig:100Pa_M0Linear}(a-f). It was concluded that the upper-harmonics were occurring due to the geometric non-linearity associated with the slender beam (representing the skin) connecting the two segments in combination with the applied boundary constraints.

It was found that the broadband response characteristics of the segmented plates followed the behavior of an edge-constrained plate~\cite{beranek1960, song_boltonJASA01, song_boltonNCEJ03}. A small but a notable averaged improvement of $2$~dB in STL was predicted for M$1$ and M$2$ beyond the dip frequency compared to their corresponding equivalent limp panel. Analysis with a deterministic signal sweep in the chosen frequency band might give more insight into the characteristics of the segmented plates with a stronger influence of the non-linearity on the STL characteristics. 

The amplitude of the lowpass filtered white noise had a significant effect on the predictions: i.e., the STL in the very low frequency range increased as the noise level increased. This further supports the idea of transferring the energy from lower to higher frequencies to increase the STL in the low frequency range and has potential applications for blocking high intensity sound sources at low frequencies. 

The idea of this study was to explore another mechanism for the hard-to-mitigate low frequency noise. Once the energy is shifted to higher frequencies, transmission can be mitigated more efficiently either by using a mass efficient barrier or dissipation of energy by an absorptive material. Both of those mechanisms are more effective at mitigating higher frequency noise. For example, energy absorbing visco-elastic material along with ABH wedges could be employed to trap and dissipate some power as demonstrated by Touz{\'e} et al.~\cite{liJSV2019}. The specific choices of dimensions of the models studied in this work would allow for incorporation of this idea into cellular panel concept for a future experimental study, which was beyond the scope of the present work. Materials with a higher degree of non-linearity will be required to demonstrate the shift in power to higher frequencies experimentally and eventually to realize the benefit of STL improvement.

\section*{Conclusions}
The sound barrier performance of planar cellular metamaterials is poor at their flexural resonance frequencies. The possibility of mitigating those limitations of cellular panels by incorporating additional dissipation mechanisms was studied here. One way to overcome those limitations is to transfer the energy from the frequency band having low STL to higher frequencies where it can be blocked more effectively. A material system having non-linear stiffness behavior introduced through contacts has the potential to realize the above mechanism. The effects of contact nonlinearity, geometric nonlinearity, and both acting together, were examined and compared with the corresponding linear case through two-dimensional numerical models of a segmented plate with a contact interface subjected to a normally incident sound field. Based on the numerical study, upper-harmonic response peaks were observed for single frequency sound excitations, the strength of which were found to strongly depend on the degree of non-linearity or bilinear stiffness ratio. The response was also found to depend on the relative location of the excitation frequency with respect to the eigenmodes of the structure, with stronger upper-harmonics occurring for excitation frequencies closer to the characteristic eigen-frequencies, thus supporting the idea of transferring energy from the deficit regions to higher frequencies. Broadband response of the materials was also examined and notable improvement in STL was predicted at low frequencies. Experimental validation of the observations made in this work would be the next logical step in taking the idea further. 

\noindent \textbf{Acknowledgment}: The authors gratefully acknowledge the financial support provided by
the United States Air Force Office of Scientific Research through the grant
FQ8671-090162. We also thank the reviewers for their detailed and helpful comments.


\newpage

\begin{table}[h!]
\caption{Clearance and the skin thickness of the two-dimensional segmented plate models M$1$, M$2$ and M$3$.}
\label{tab:M1M2M4dimensions}
\begin{center}
\begin{tabular}{ | l | l | l |l | }
\hline
Model & Clearance - $d_{cl}$ [$\mu$m] & Skin thickness - $t_{s}$ [$\mu$m] & Characteristics \\ \hline
M$0$ & NA & 50 & Unsegmented \\ \hline
M$1$ & $10$ & $50$ & Segmented \\ \hline
M$2$ & $100$ & $50$ & Segmented \\ \hline
M$3$ & $0$ & $500$ & Segmented \\ \hline
\end{tabular}
\end{center}
\end{table}

\begin{table}[h!]
\caption{Natural frequencies for the two-dimensional segmented plate models M$1$, M$2$ and M$3$ and the unsegmented model M$0$.}
\label{tab:M0M1M2M3freq}
\begin{center}
\begin{tabular}{ | l | l | l |l | }
\hline
Model & Contact-open & Contact-Closed & Bilinear Frequency \\ 
& first mode [Hz] & first mode [Hz] & [Hz]\\ \hline
M$0$ & NA & $429$ & NA \\ \hline
M$1$ & $29.6$ & $429$ & $55.5$ \\ \hline
M$2$ & $29.6$ & $429$ & $55.5$ \\ \hline
M$3$ & $391.3$ & $533.6$ & $451.5$ \\ \hline
\end{tabular}
\end{center}
\end{table}



\begin{table}[h!]
\caption{Power shifted to other frequencies on the transmitted side in \% of net incident power
 for the two-dimensional segmented plate models M$1$, M$2$ and M$3$ and the unsegmented model M$0$.}
\label{tab:powShiftFrac}
\begin{center}
\begin{tabular}{ | l | l | l | l | l | l | l |}
\hline
\multirow{2}{*}{Model} & \multirow{2}{*}{Source Amplitude} & \multirow{2}{*}{Source Amplitude} & \multicolumn{4}{l|}{powerShiftFrac in \% at $f_{ex}=$} \\ \cline{4-7} 
 & [Pa] & [dB] & $20$~Hz & $80$~Hz & 320~Hz & 640~Hz \\ \hline
M$0$ & $100$ & $131$ & $0.08$ & 0.11 & 0.19 & NA \\ \hline
M$1$ & $100$ & $131$ & $66.80$ & 51.70 & 0.00 & NA \\ \hline
M$2$ & $100$ & $131$ & $46.20$ & 7.46 & 0.00 & NA \\ \hline
M$3$ & $100$ & $131$ & NA & $13.94$ & $3.54$ & $0.16$ \\ \hline
M$1$ & $10$ & $111$ & $26.68$ & 0.00 & 0.00 & NA \\ \hline
M$2$ & $10$ & $111$ & $9.96$ & 0.00 & 0.00 & NA \\ \hline
M$3$ & $10$ & $111$ & NA & $11.63$ & $3.97$ & $0.15$ \\ \hline
\end{tabular}
\end{center}
\end{table}

\begin{table}[h!]
\caption{Sound pressure level on the incident side (${\mathrm{SPL}}_{x_{1}=0}$) for the two-dimensional segmented plate models M$1$, M$2$ and M$3$ and the unsegmented model M$0$.}
\label{tab:pressureAtxZero}
\begin{center}
\begin{tabular}{ | l | l | l | l | l | l | l |}
\hline
\multirow{2}{*}{Model} & \multirow{2}{*}{Source Amplitude} & \multirow{2}{*}{Source Amplitude} & \multicolumn{4}{l|}{${\mathrm{SPL}}_{x_{1}=0}$ on the incident side} \\ \cline{4-7} 
 & [Pa] & [dB] & $20$~Hz & $80$~Hz & 320~Hz & 640~Hz \\ \hline
M$0$ & $100$ & $131$ & $137.0$ & 137.0 & 136.6 & NA \\ \hline
M$1$ & $100$ & $131$ & $136.9$ & 135.7 & 136.7 & NA \\ \hline
M$2$ & $100$ & $131$ & $136.8$ & 135.1 & 136.7 & NA \\ \hline
M$3$ & $100$ & $131$ & NA & $137.0$ & $136.8$ & $136.7$ \\ \hline
M$1$ & $10$ & $111$ & $115.7$ & 115.0 & 116.7 & NA \\ \hline
M$2$ & $10$ & $111$ & $115.5$ & 115.0 & 116.7 & NA \\ \hline
M$3$ & $10$ & $111$ & NA & $117.0$ & $116.8$ & $116.7$ \\ \hline
\end{tabular}
\end{center}
\end{table}







 \addtocounter{figure}{+1}    
\newpage

\begin{figure}[htb!]
\setcounter{figure}{0}
\centering
\includegraphics[width=0.95\textwidth]{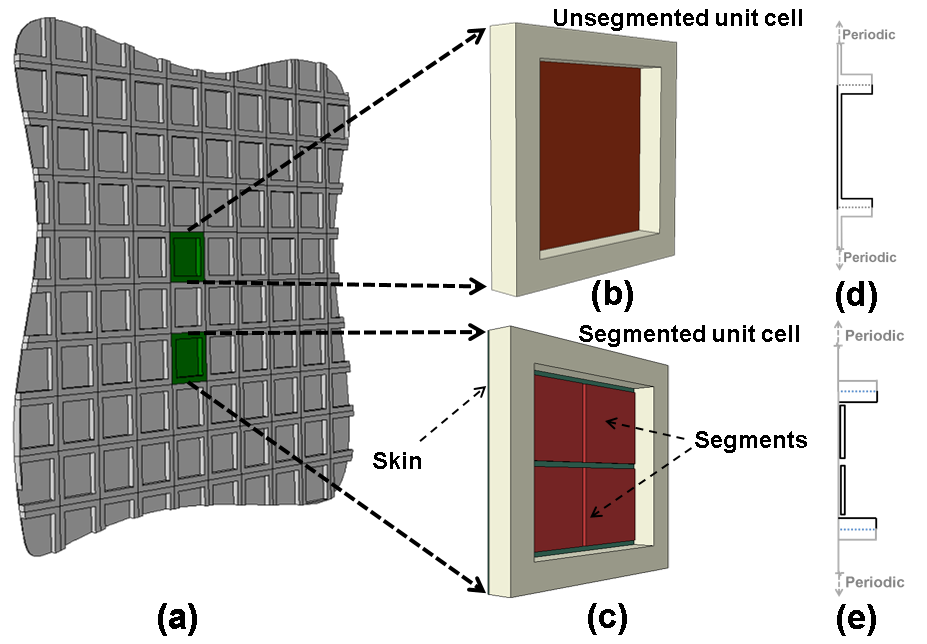}
\caption{(a) A planar cellular panel, (b) a unit cell with a monolithic plate filling the cell, (c) a unit cell with a segmented plate filling the cell, (d) and (e) side views of the unsegmented and segmented unit cells illustrating their periodicity.}
\label{fig:sgmntn_embodiments}
\end{figure}

\begin{figure}[htb!]
\centering
\subfigure[]{\includegraphics[width=0.23\textwidth]{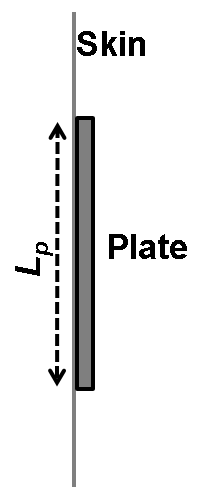}}
\subfigure[]{\includegraphics[width=0.45\textwidth]{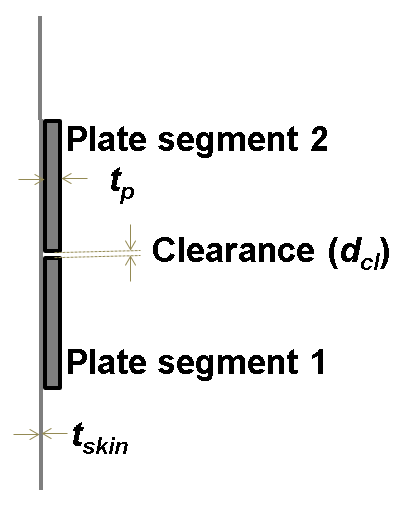}}
\caption{(a) Two-dimensional monolithic model, and (b) two-dimensional model of segmented plates with a contact interface supported by a skin.}
\label{fig:material}
\end{figure}

\begin{figure}[htb!]
\centering
\includegraphics[width=0.95\textwidth]{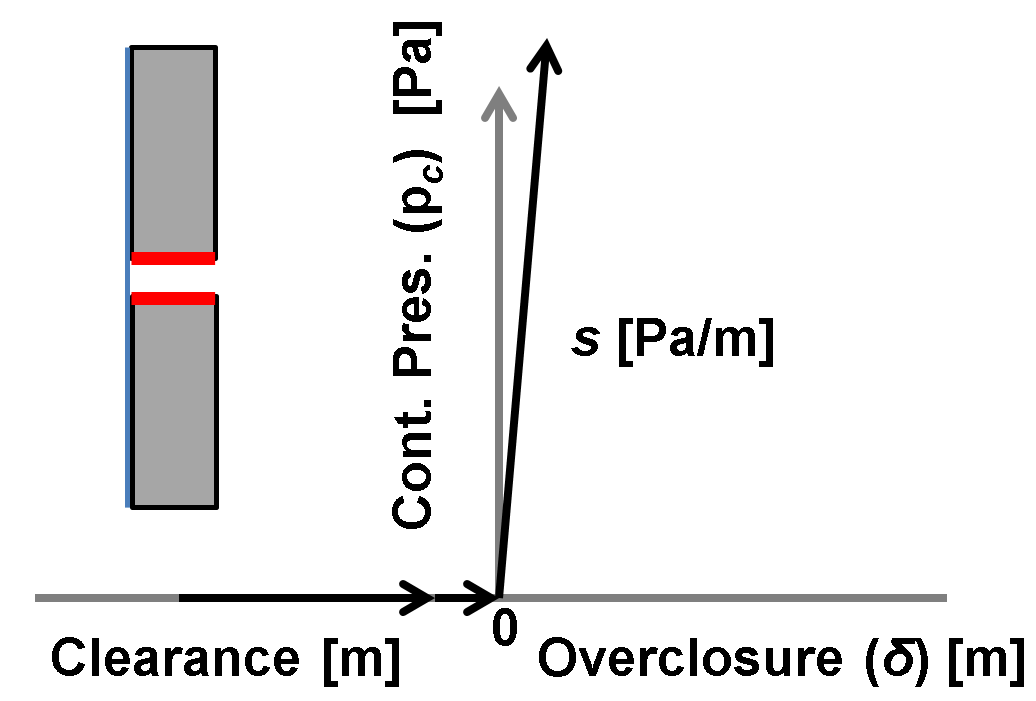}
\caption{The contact pressure ($p_{c}$) – over-closure ($\delta$) relationship describing the contact interaction.}
\label{fig:presOverclosure}
\end{figure}


\begin{figure}[htb!]
\centering
\includegraphics[width=0.9\textwidth]{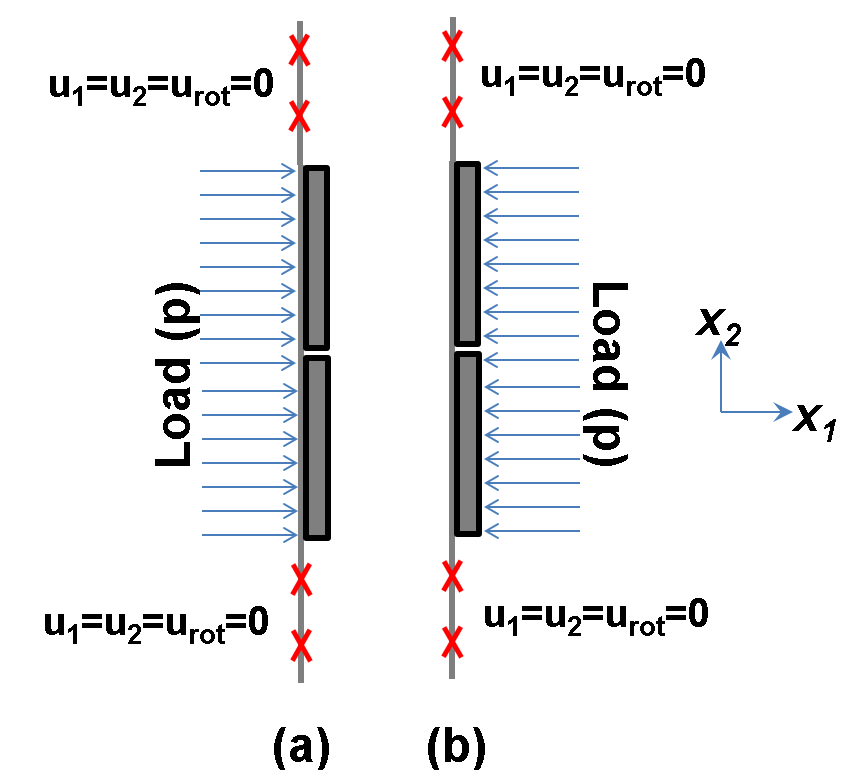}
\caption{(a) Static analysis models for obtaining the system stiffness for loading in (a) $+ve$~$x_{1}$ and (b) $-ve$~$x_{1}$ directions.}
\label{fig:methods_static_analysis}
\end{figure}

\begin{figure}[htb!]
\centering
 \includegraphics[width=0.95\textwidth]{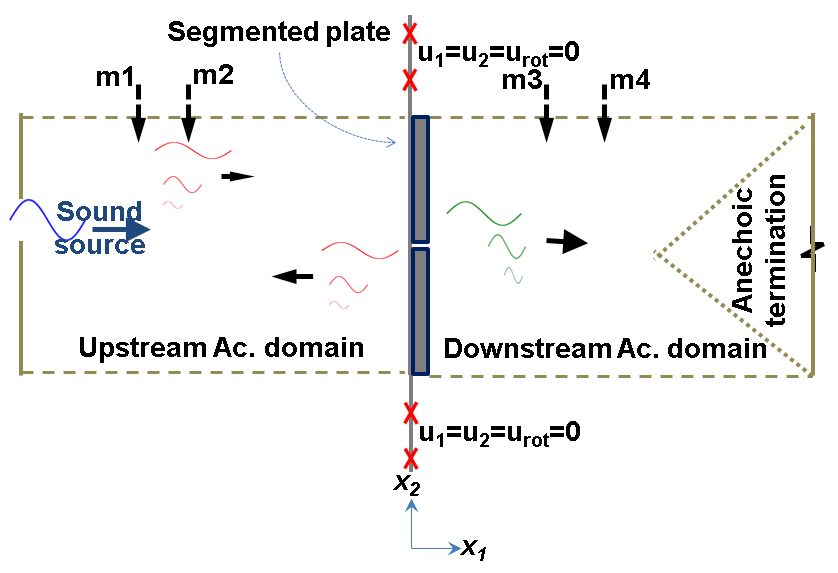}
 \caption{A schematic of the two-dimensional FE model used for acoustical characterization of the segmented plates models and the unsegmented plate model.}
\label{fig:methods_acoustical_analysis}
\end{figure}

\begin{figure}[htb!]
\centering
\subfigure[]{\includegraphics[width=0.6\textwidth]{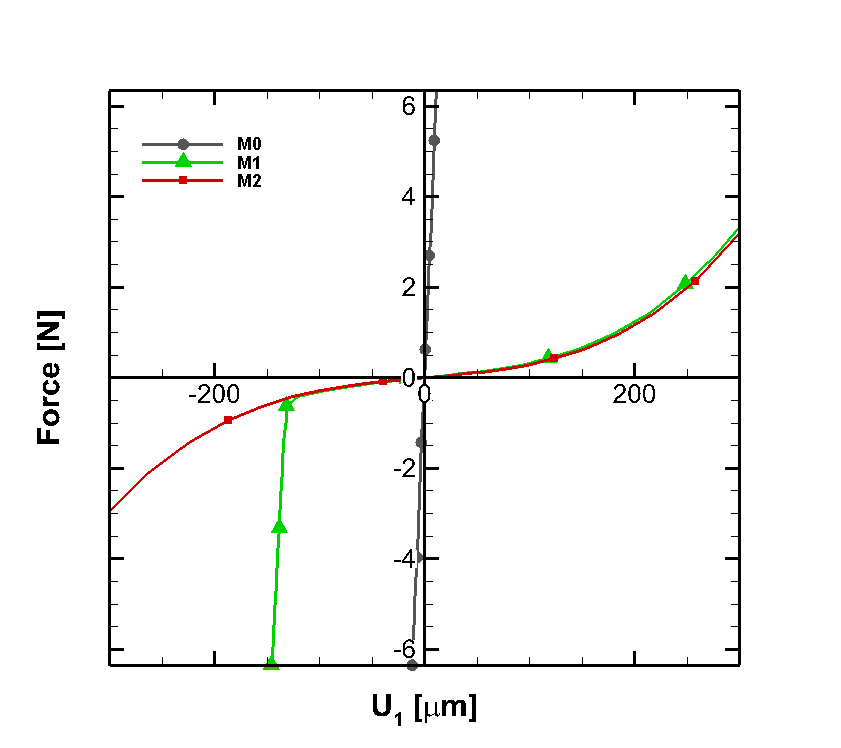}}
\subfigure[]{\includegraphics[width=0.6\textwidth]{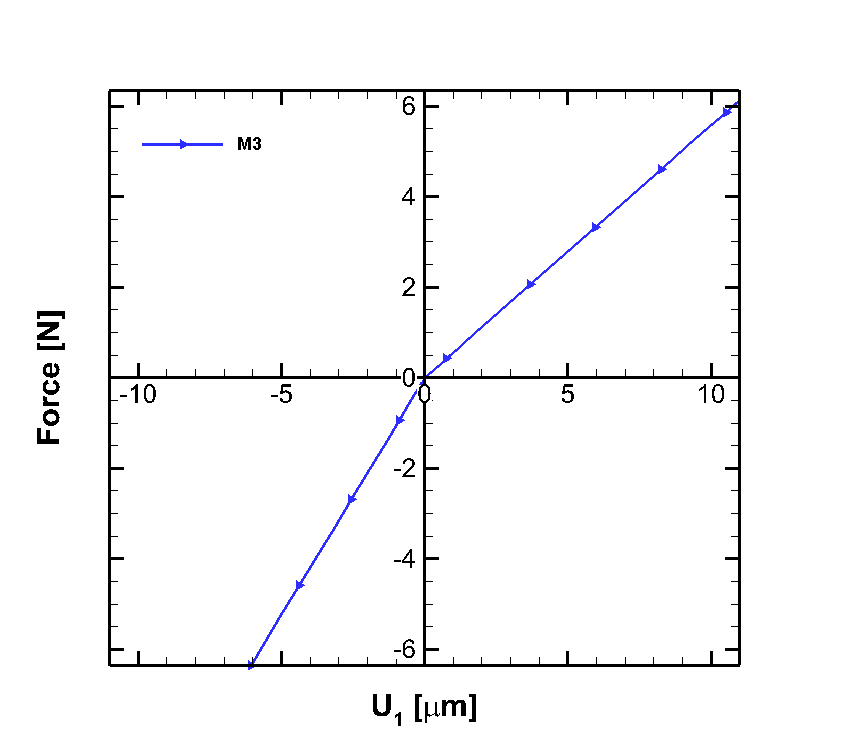}}
\caption{(a) Force versus deflection behavior for models M$0$, M$1$ and M$2$. (b)  Force versus deflection behavior for model M$3$.}
\label{fig:results_static_analysis}
\end{figure}

\begin{figure}[htb!]
\centering
\includegraphics[width=0.4\textwidth]{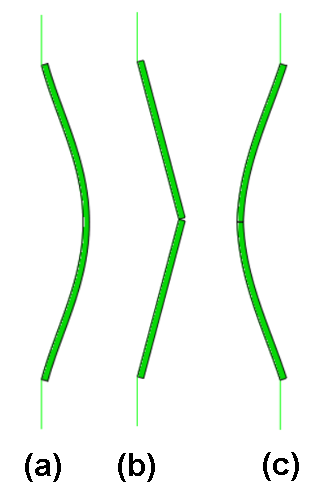}
\caption{(a) The first mode shape of the unsegmented plate, (b) the first mode shape of M$1$, M$2$ and  M$3$ with the contact open, and  (c) the first mode shape of M$1$, M$2$ and  M$3$ with the contact closed.}
\label{fig:results_modal_analysis}
\end{figure}

\FloatBarrier

\begin{figure}[htb!]
\centering
\subfigure[]{\includegraphics[width=0.48\textwidth]{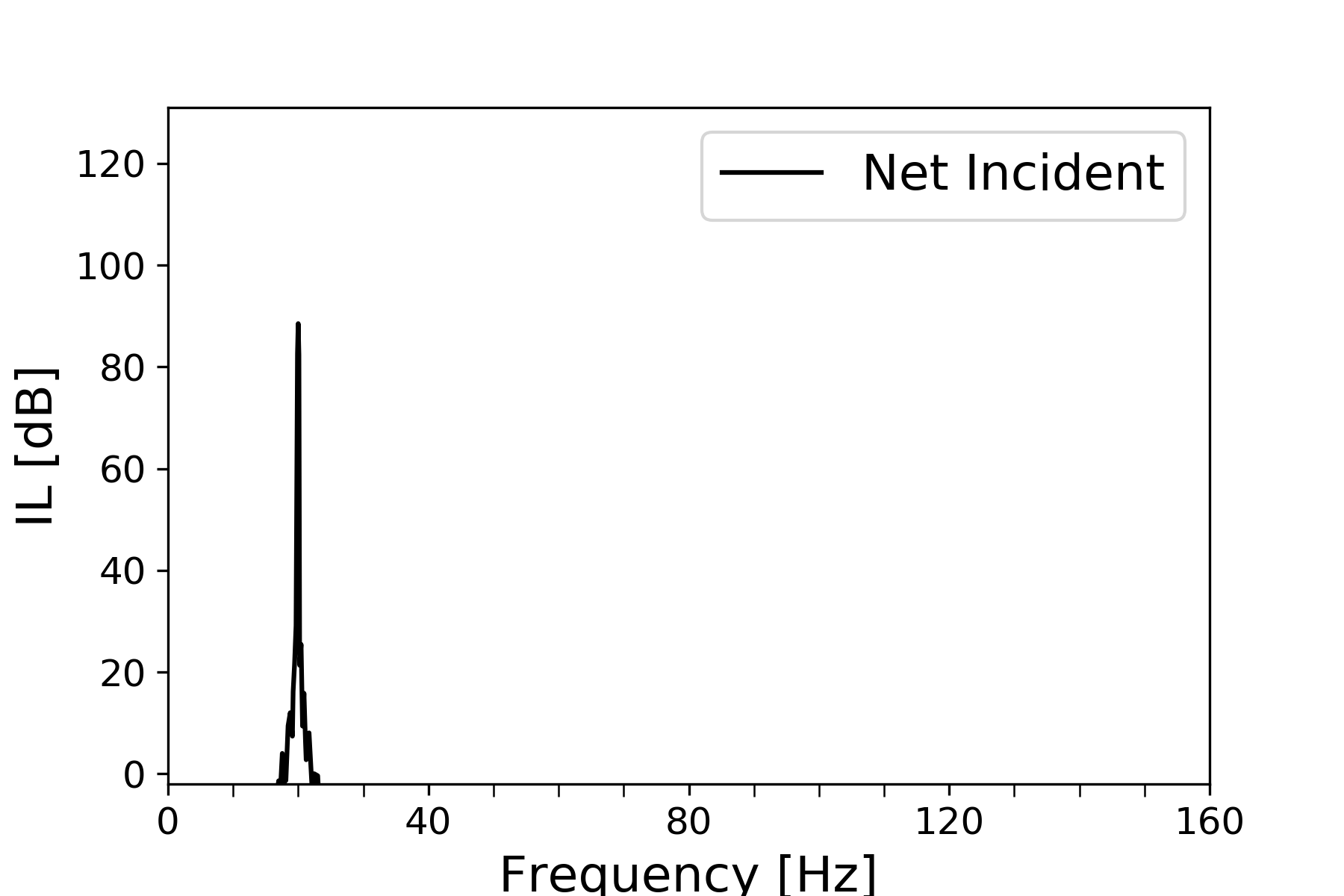}}
\subfigure[]{\includegraphics[width=0.48\textwidth]{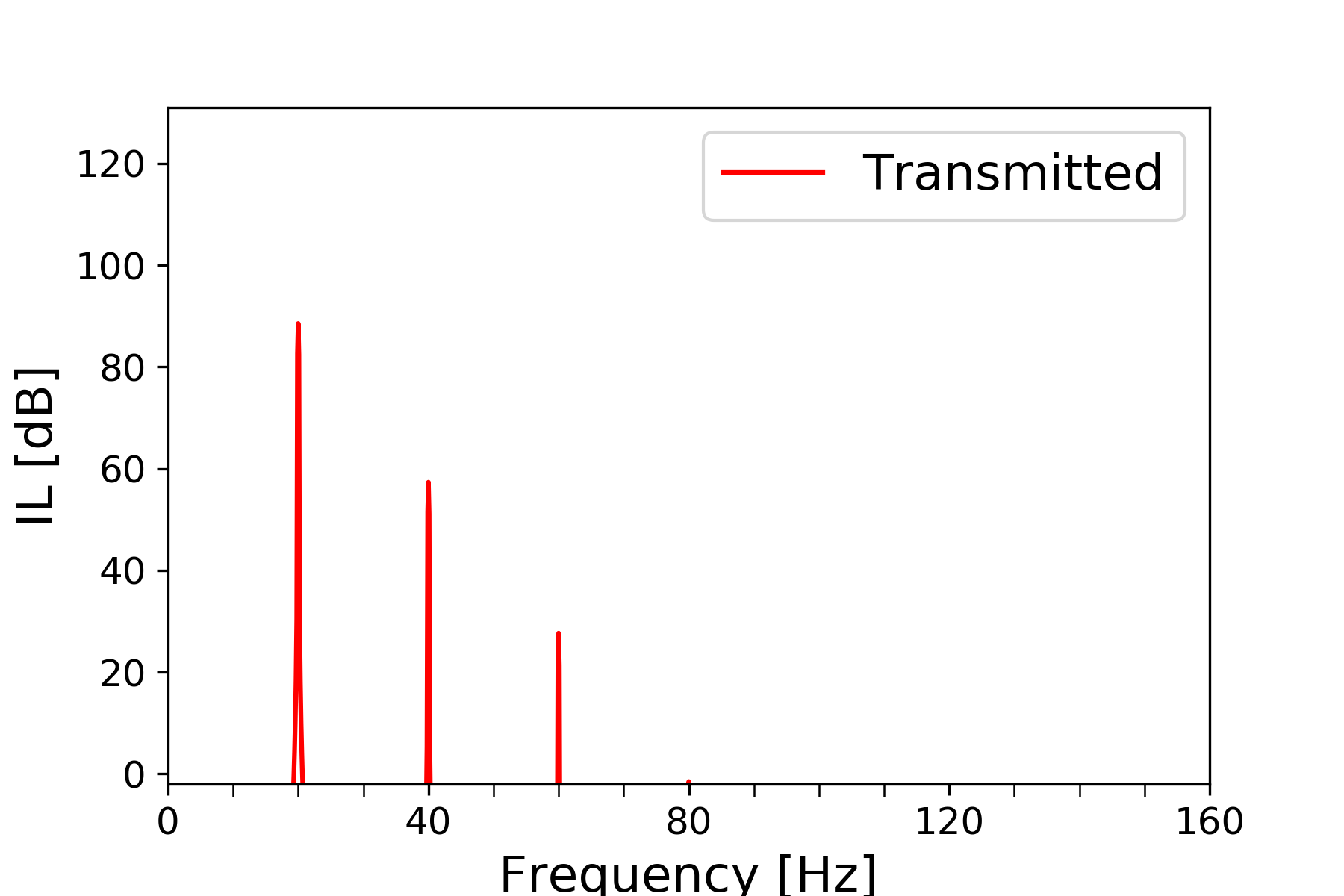}}\\
\subfigure[]{\includegraphics[width=0.48\textwidth]{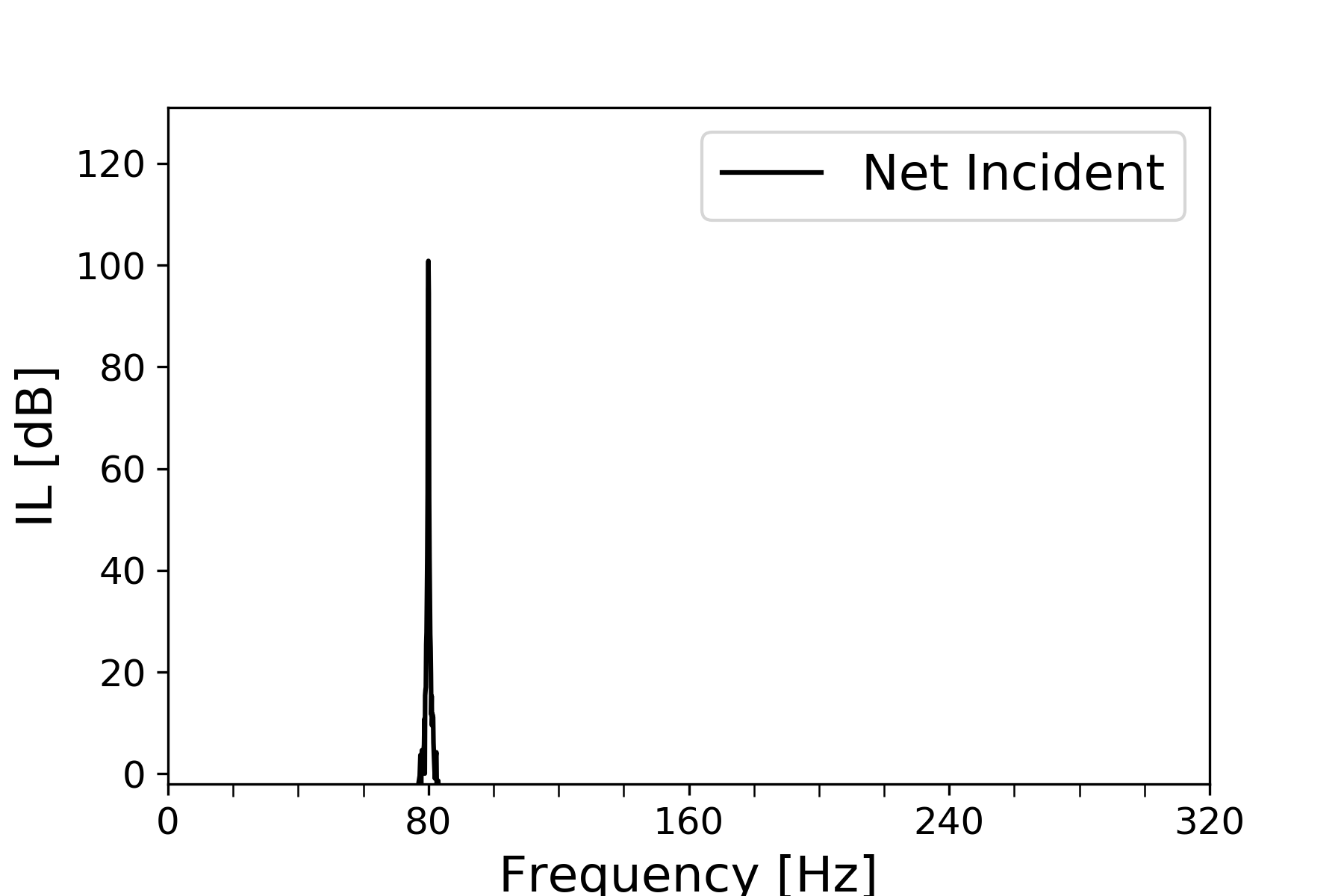}}
\subfigure[]{\includegraphics[width=0.48\textwidth]{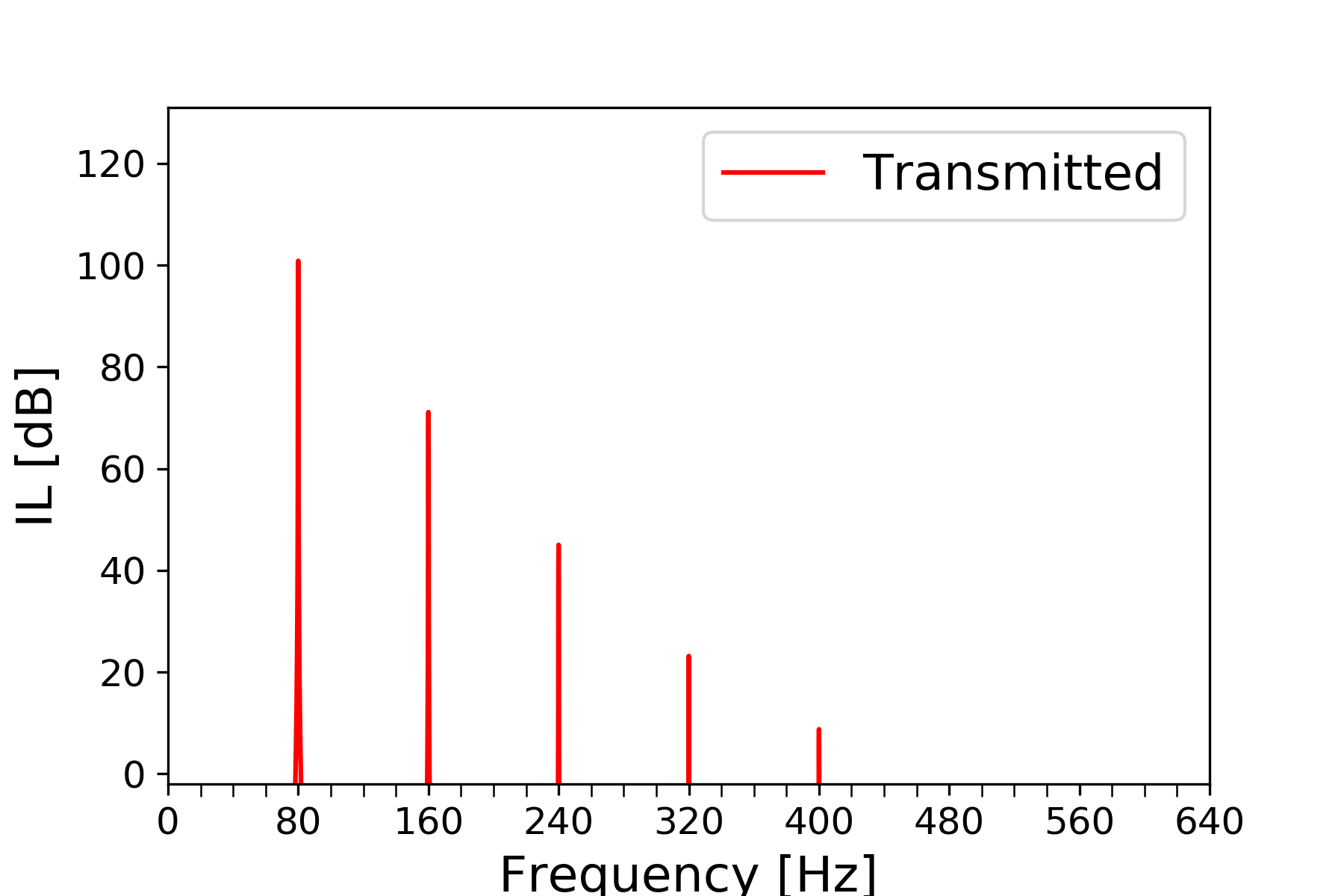}}\\
\subfigure[]{\includegraphics[width=0.48\textwidth]{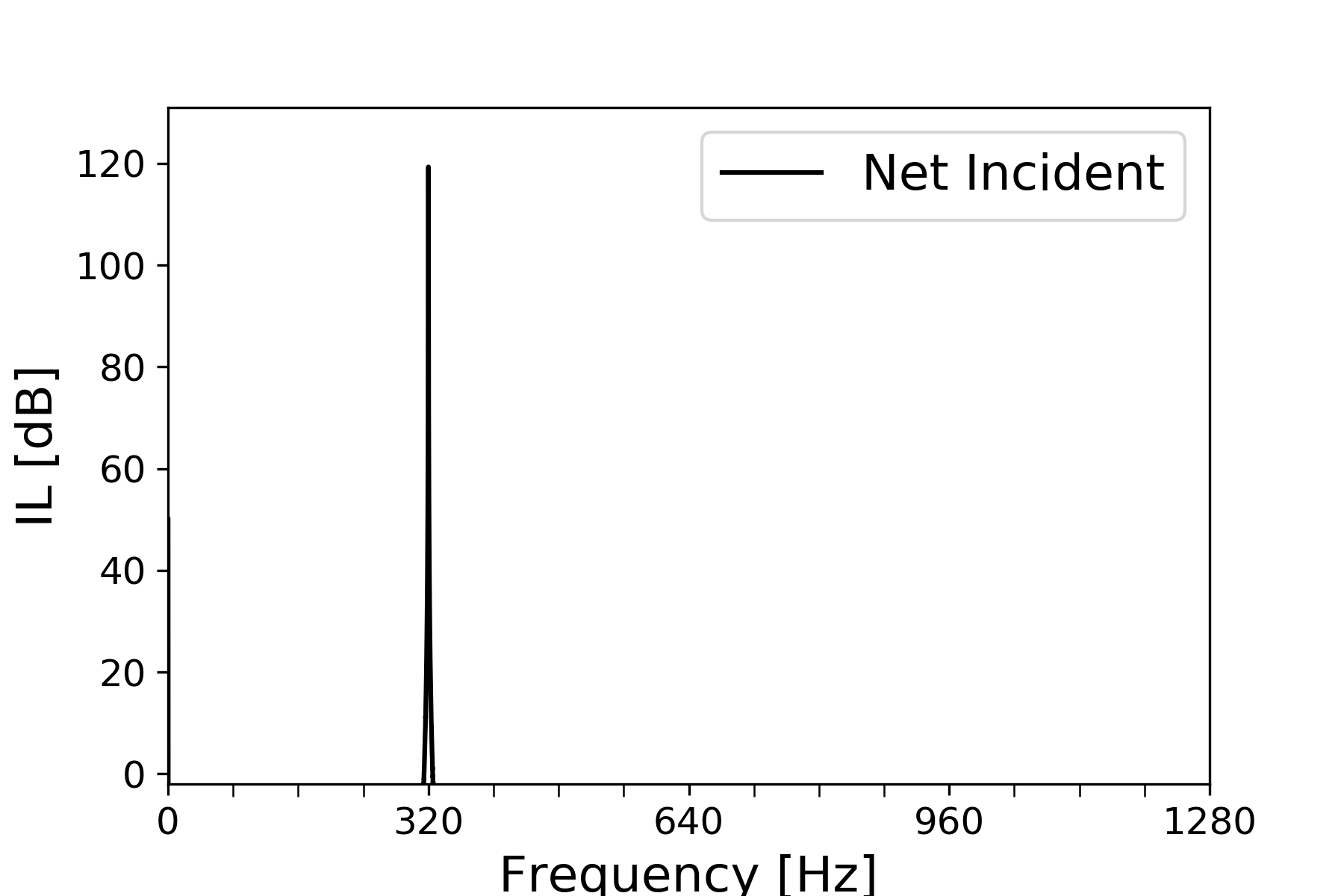}}
\subfigure[]{\includegraphics[width=0.48\textwidth]{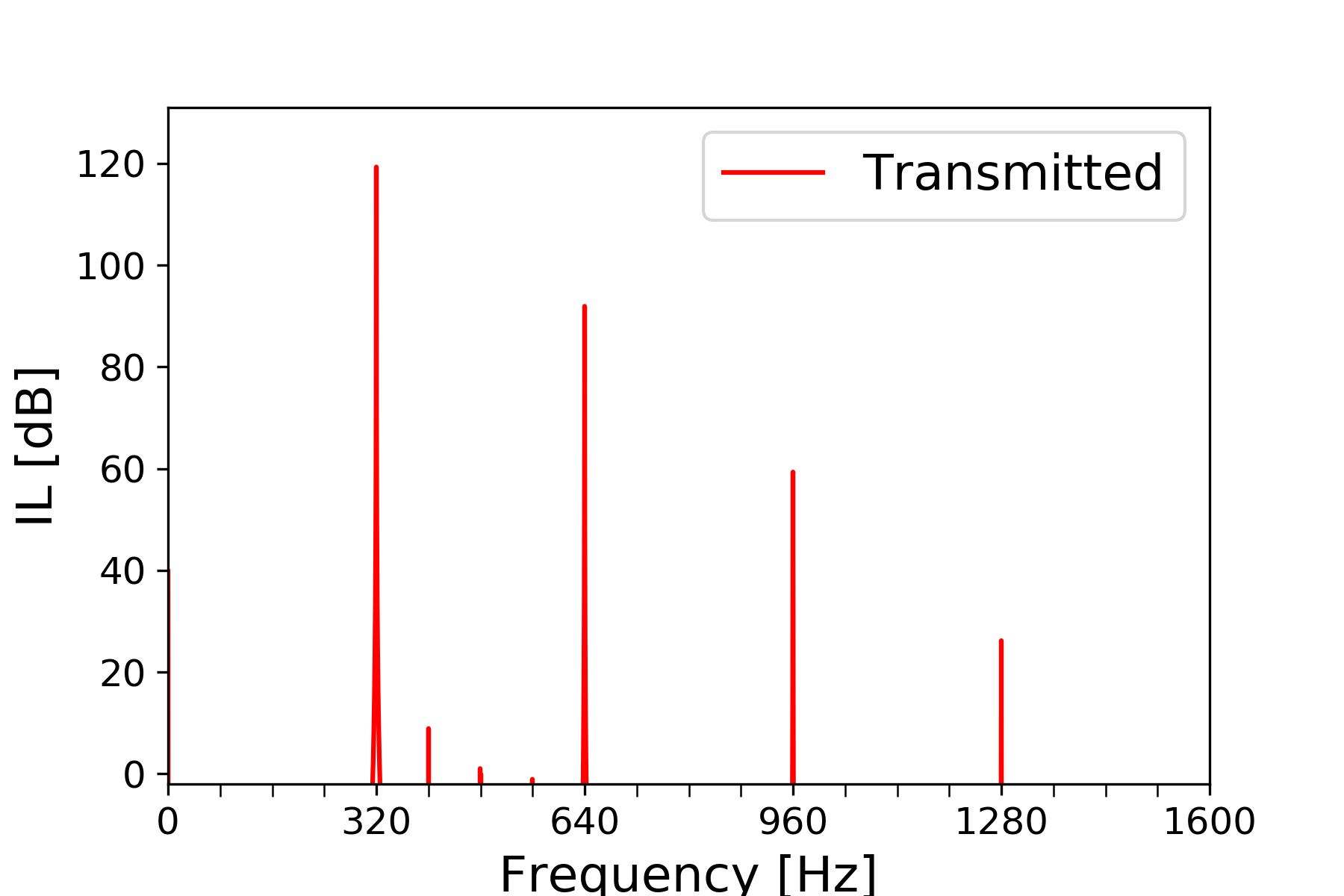}}
\caption{Net incident and transmitted sound intensity spectra of M$0$ for a sound source excitation of $100$~Pa amplitude at $20$~Hz, $80$~Hz and $320$~Hz, respectively.}
\label{fig:100Pa_M0}
\end{figure}


\begin{figure}[htb!]
\centering
\subfigure[]{\includegraphics[width=0.48\textwidth]{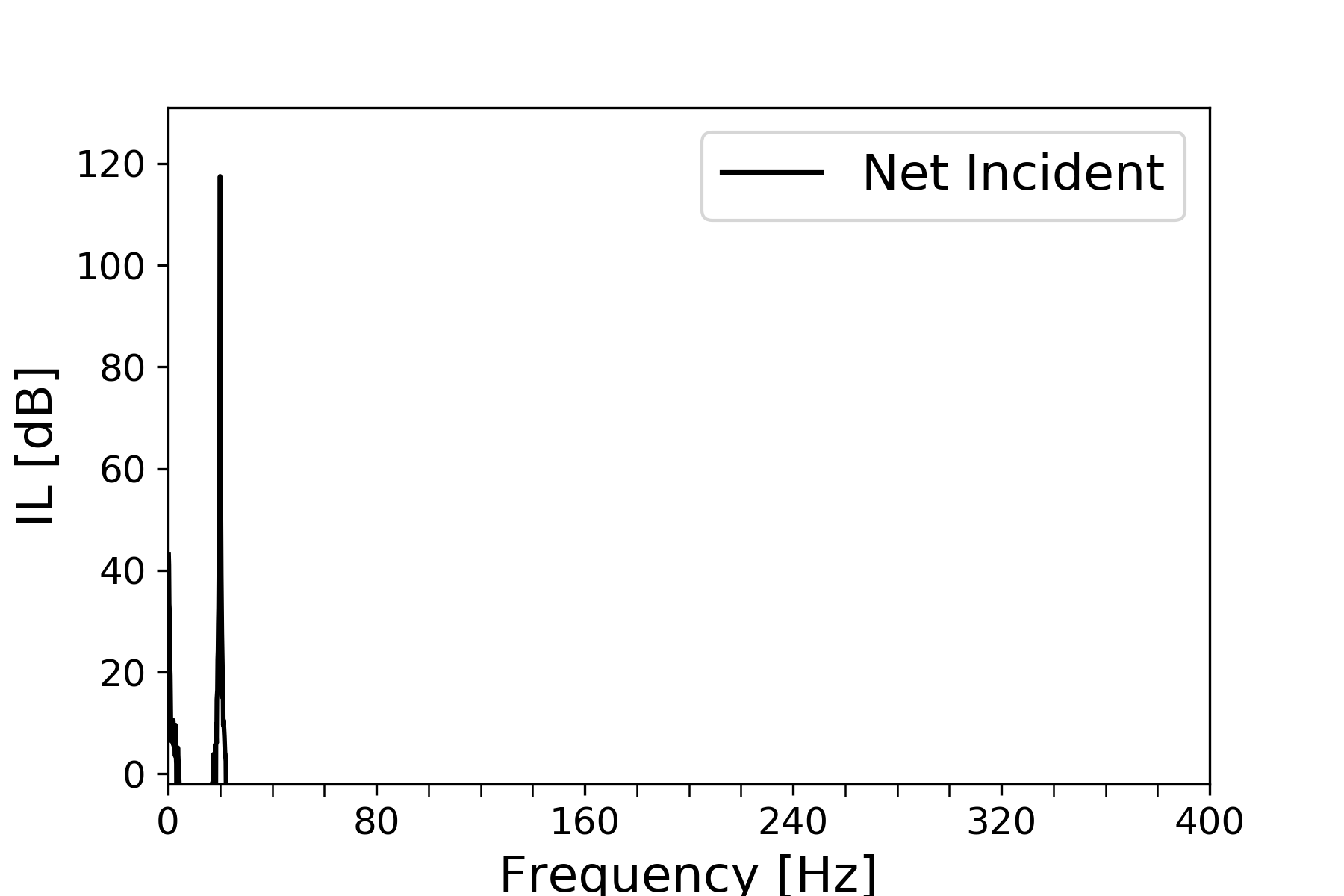}}
\subfigure[]{\includegraphics[width=0.48\textwidth]{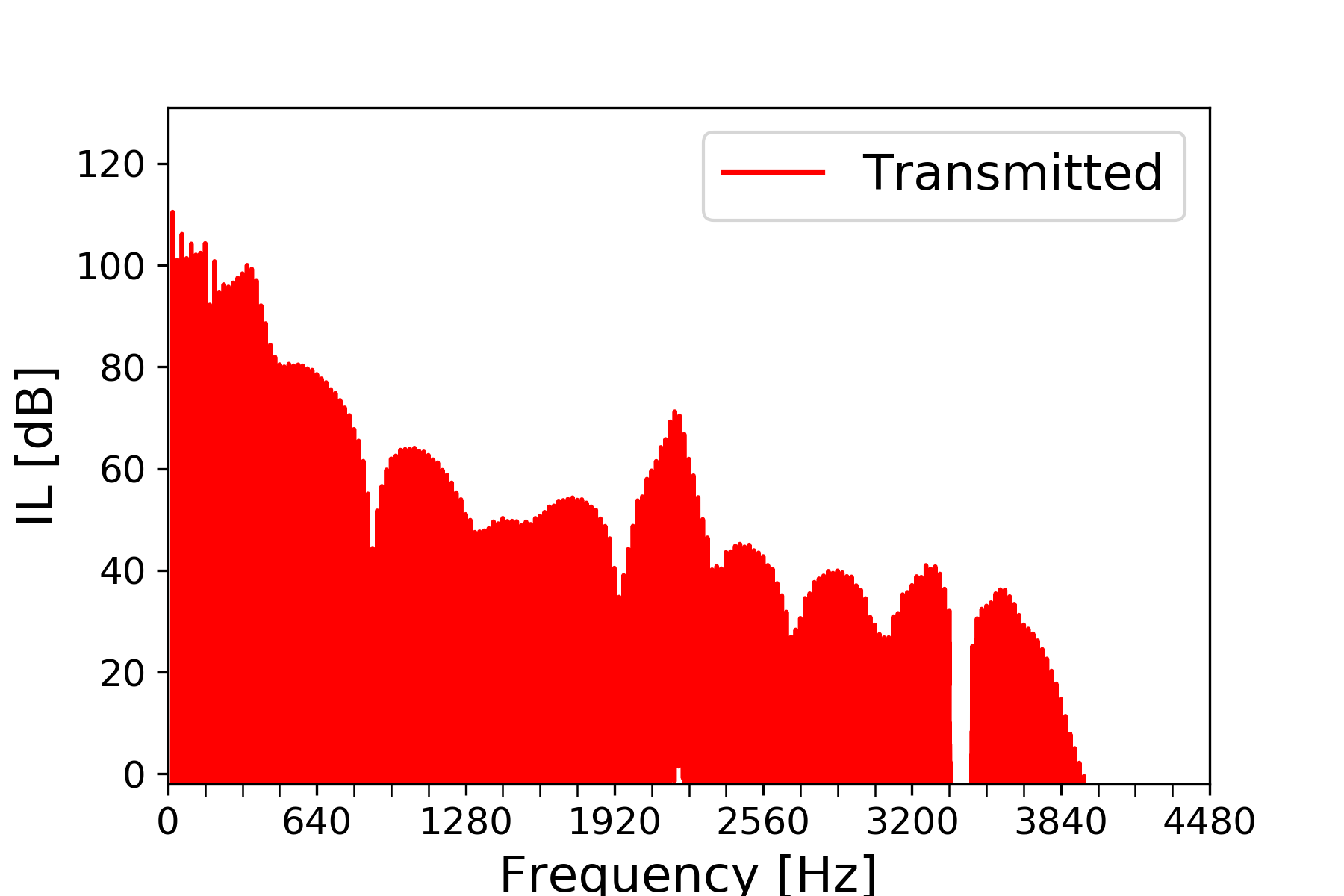}}\\
\subfigure[]{\includegraphics[width=0.48\textwidth]{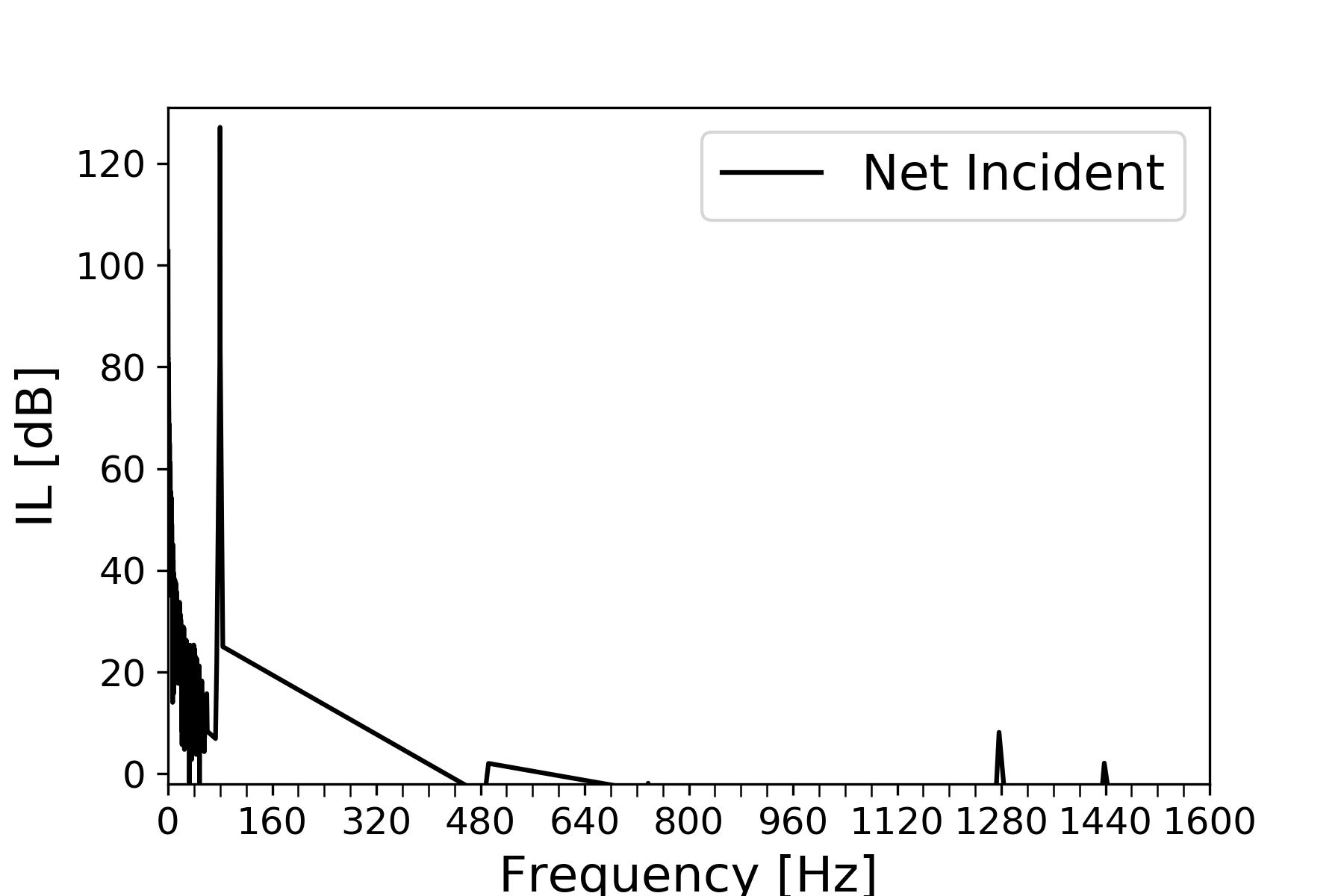}}
\subfigure[]{\includegraphics[width=0.48\textwidth]{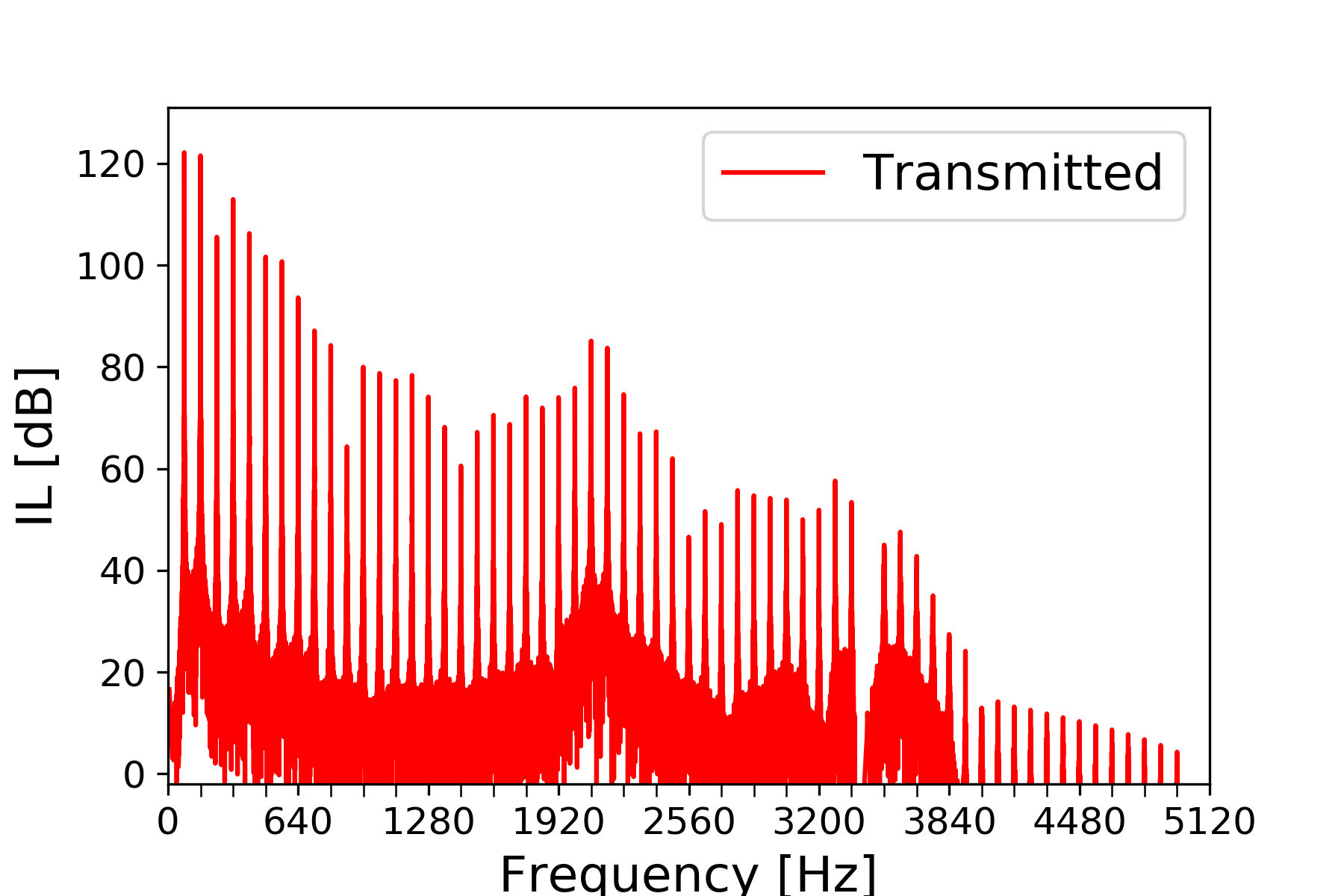}}\\
\subfigure[]{\includegraphics[width=0.48\textwidth]{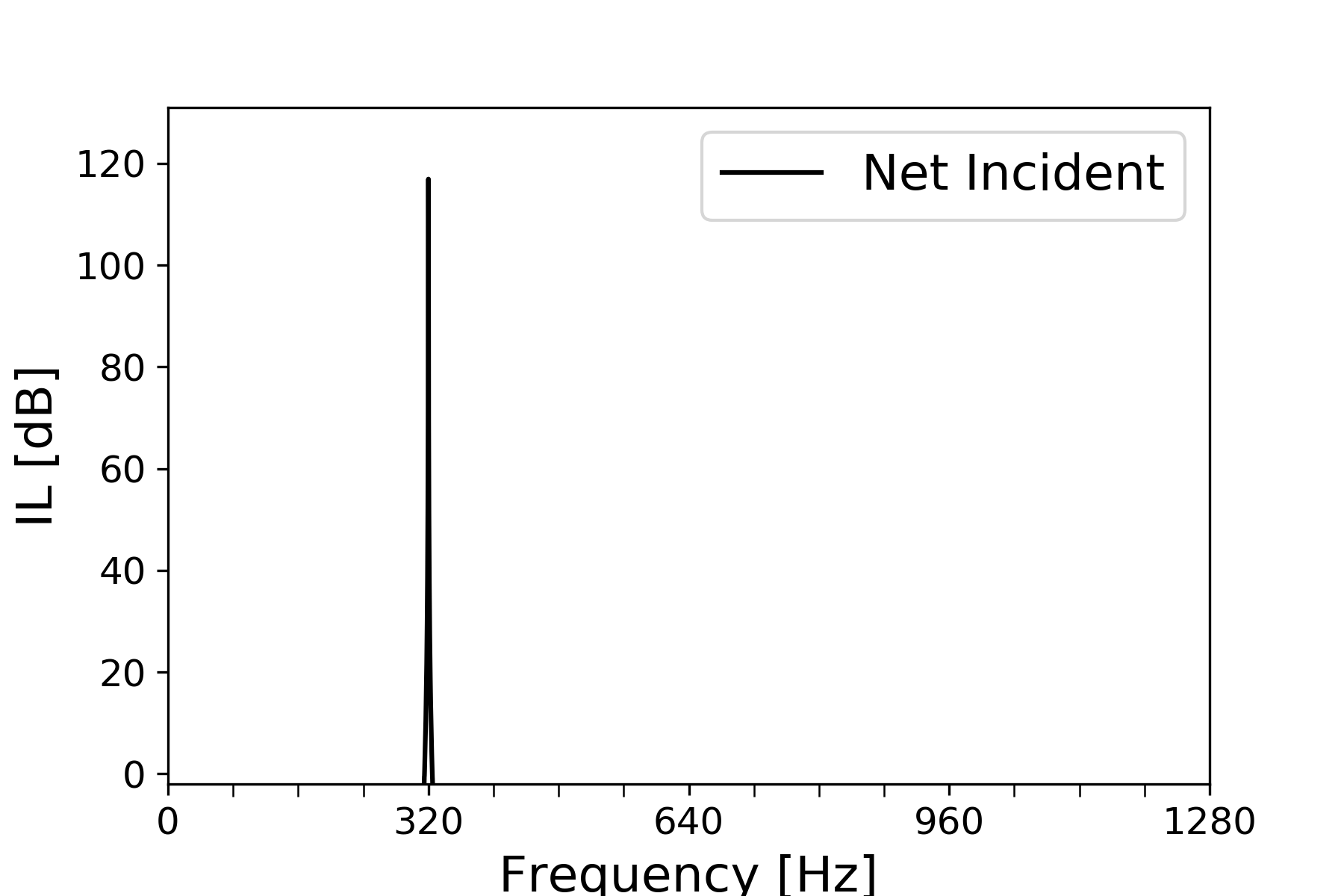}}
\subfigure[]{\includegraphics[width=0.48\textwidth]{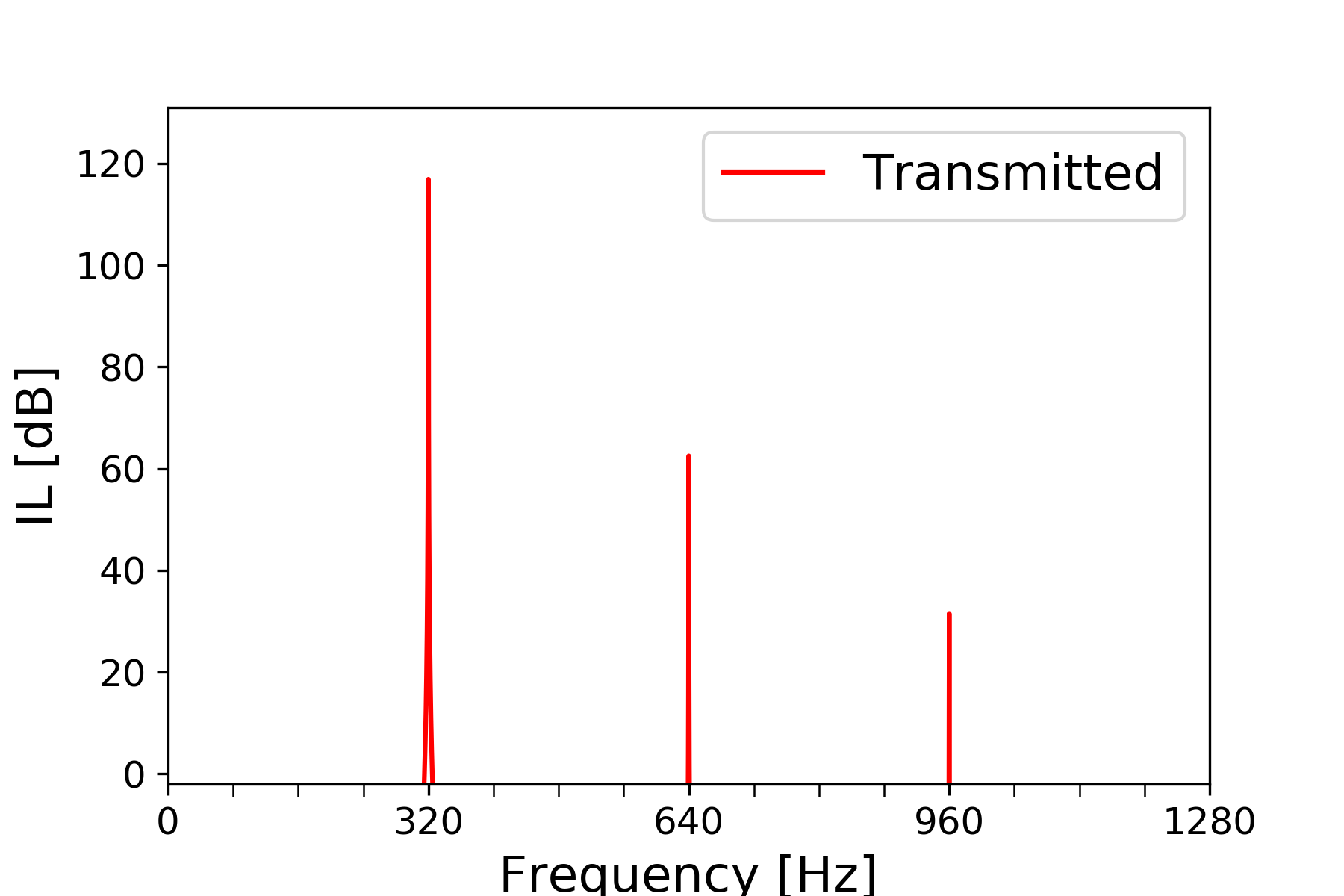}}\\
\caption{Net incident and transmitted sound intensity spectra of M$1$ for a sound source excitation of $100$~Pa amplitude at $20$~Hz, $80$~Hz and $320$~Hz, respectively.}
\label{fig:100Pa_M1}
\end{figure}

\begin{figure}[htb!]
\centering
\subfigure[]{\includegraphics[width=0.48\textwidth]{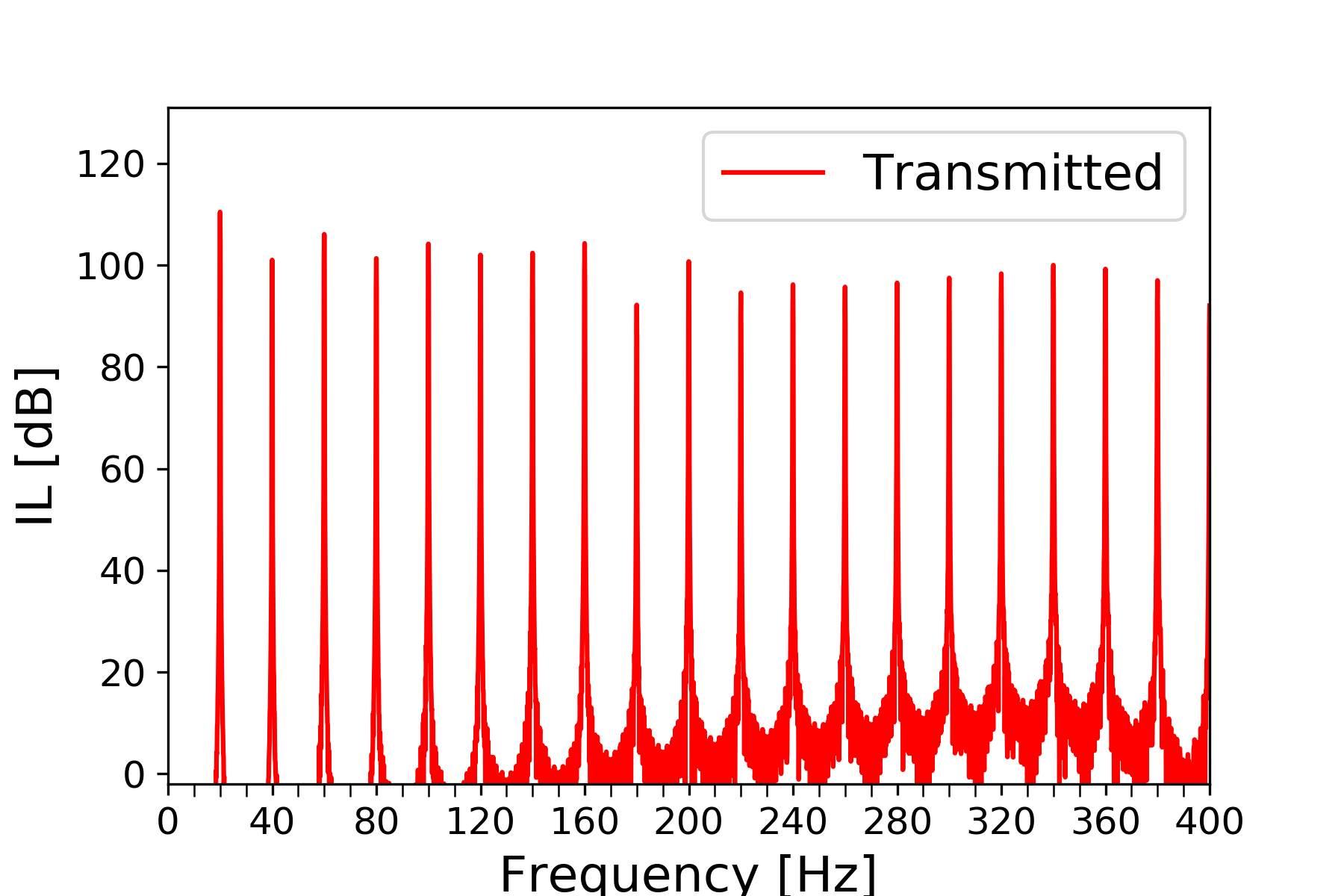}}
\subfigure[]{\includegraphics[width=0.48\textwidth]{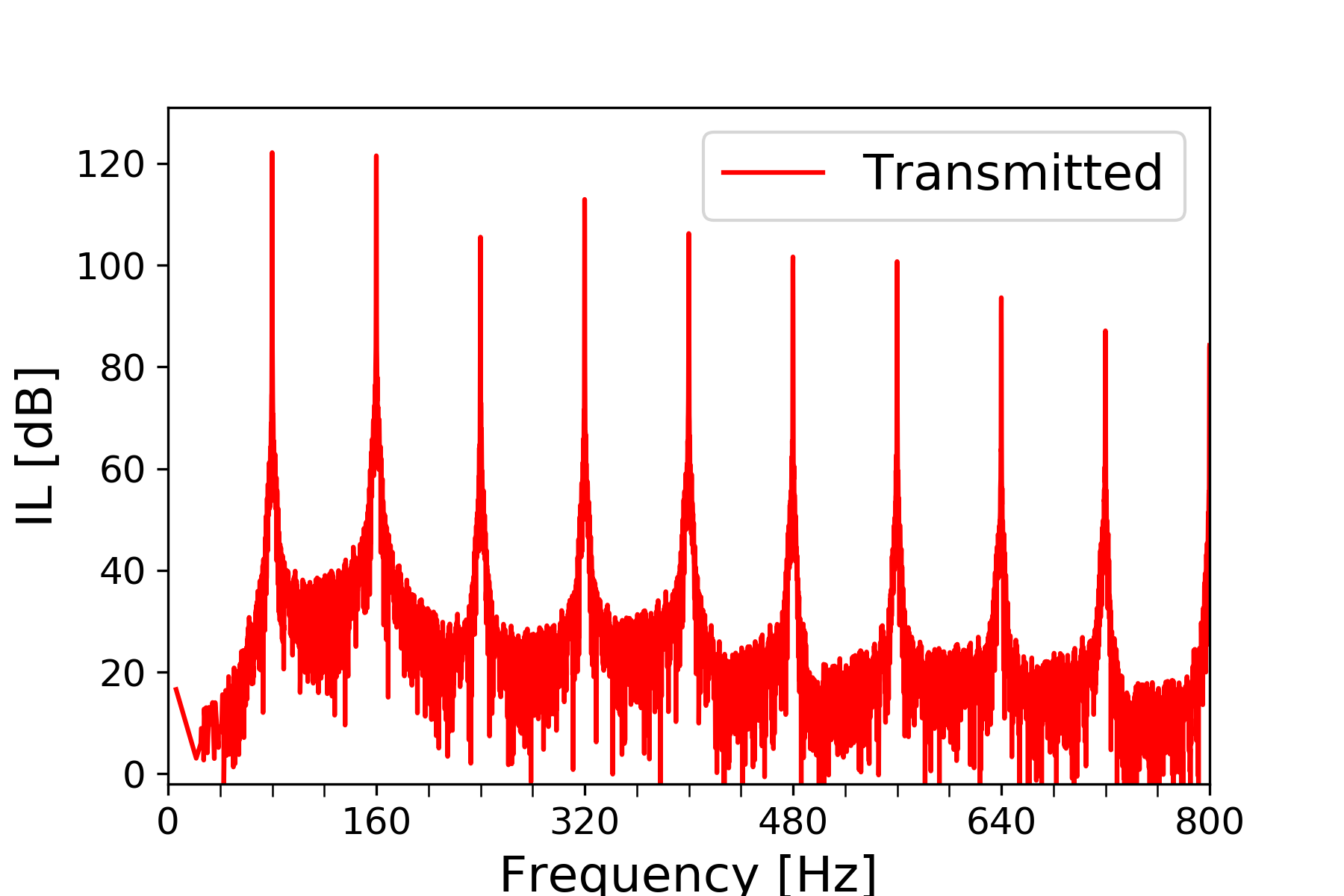}}
 \caption{Zoomed-in views of net transmitted sound intensity spectra of M$1$ for a sound source excitation of $100$~Pa amplitude at (a) $20$~Hz and (b) $80$~Hz.}
\label{fig:100Pa_M1_zoomed}
\end{figure}
\FloatBarrier

\begin{figure}[htb!]
\centering
\includegraphics[width=0.99\textwidth]{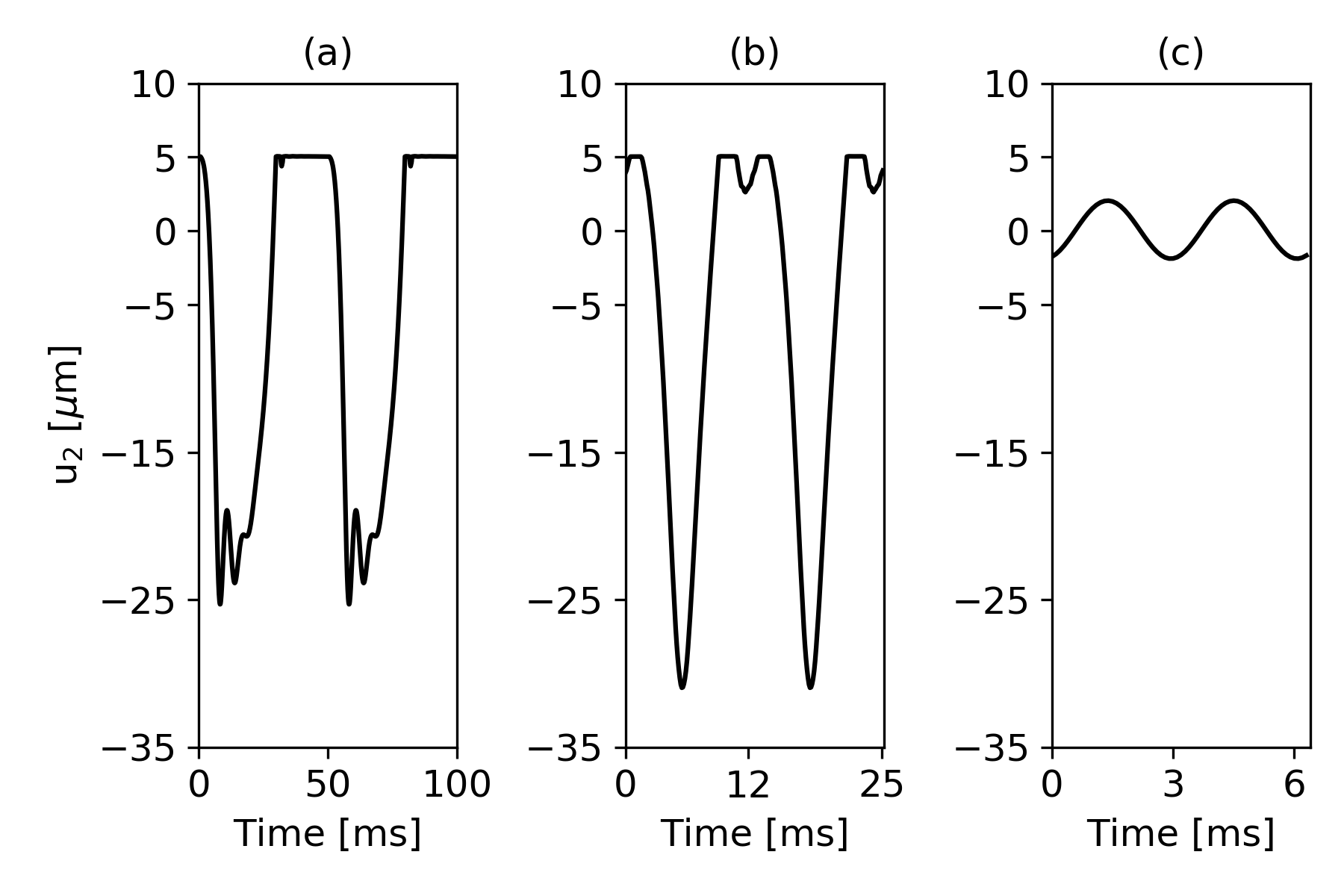}
\caption{Transverse displacement $\left(u_{2}\right)$ of M$1$ for $f_{ex}=$ $20$~Hz, $80$~Hz and $320$~Hz.}
\label{fig:100Pa_M1_disp}
\end{figure}

\begin{figure}[htb!]
\centering
\setcounter{subfigure}{0}
\subfigure[]{\includegraphics[width=0.48\textwidth]{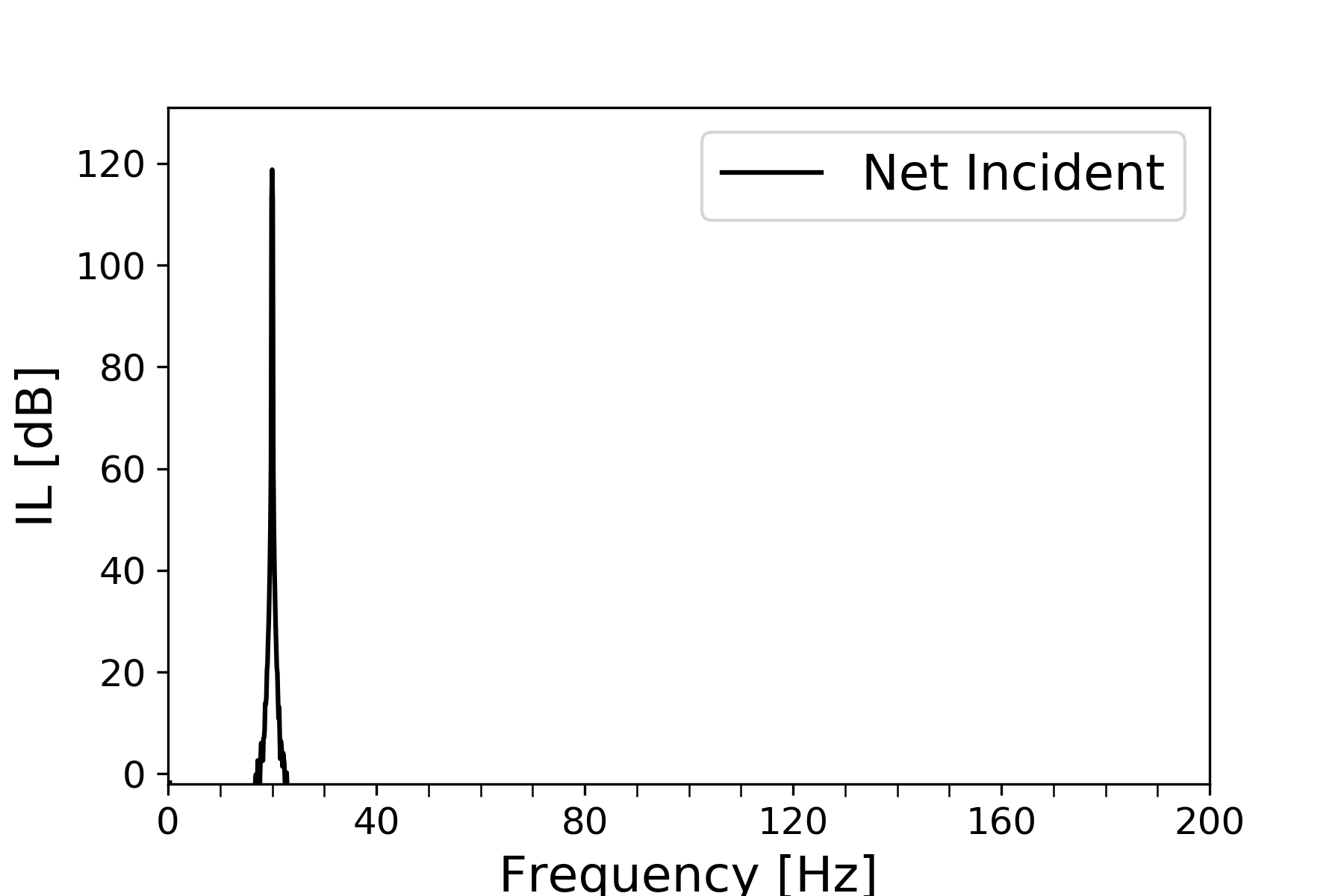}}
\subfigure[]{\includegraphics[width=0.48\textwidth]{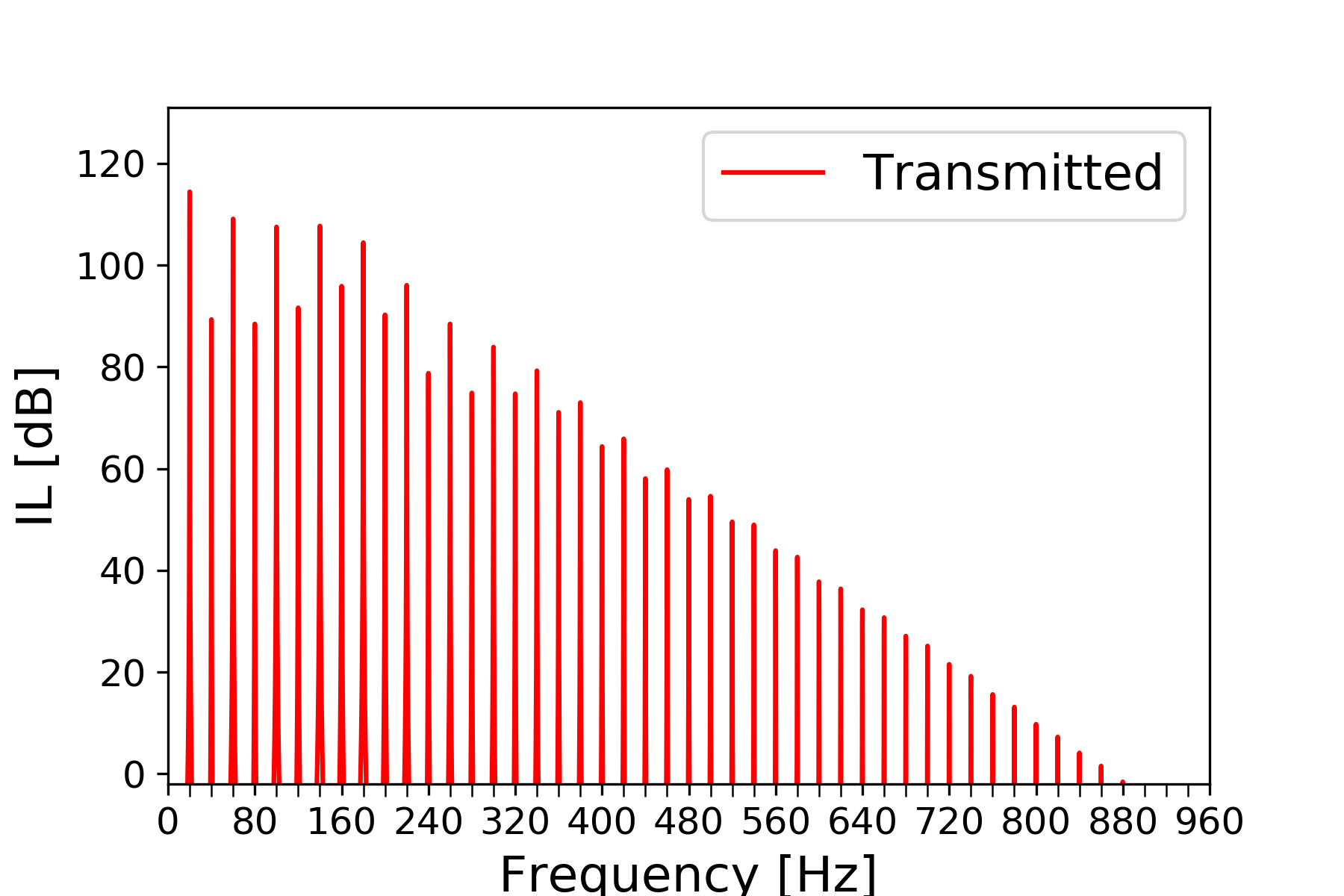}}\\
\subfigure[]{\includegraphics[width=0.48\textwidth]{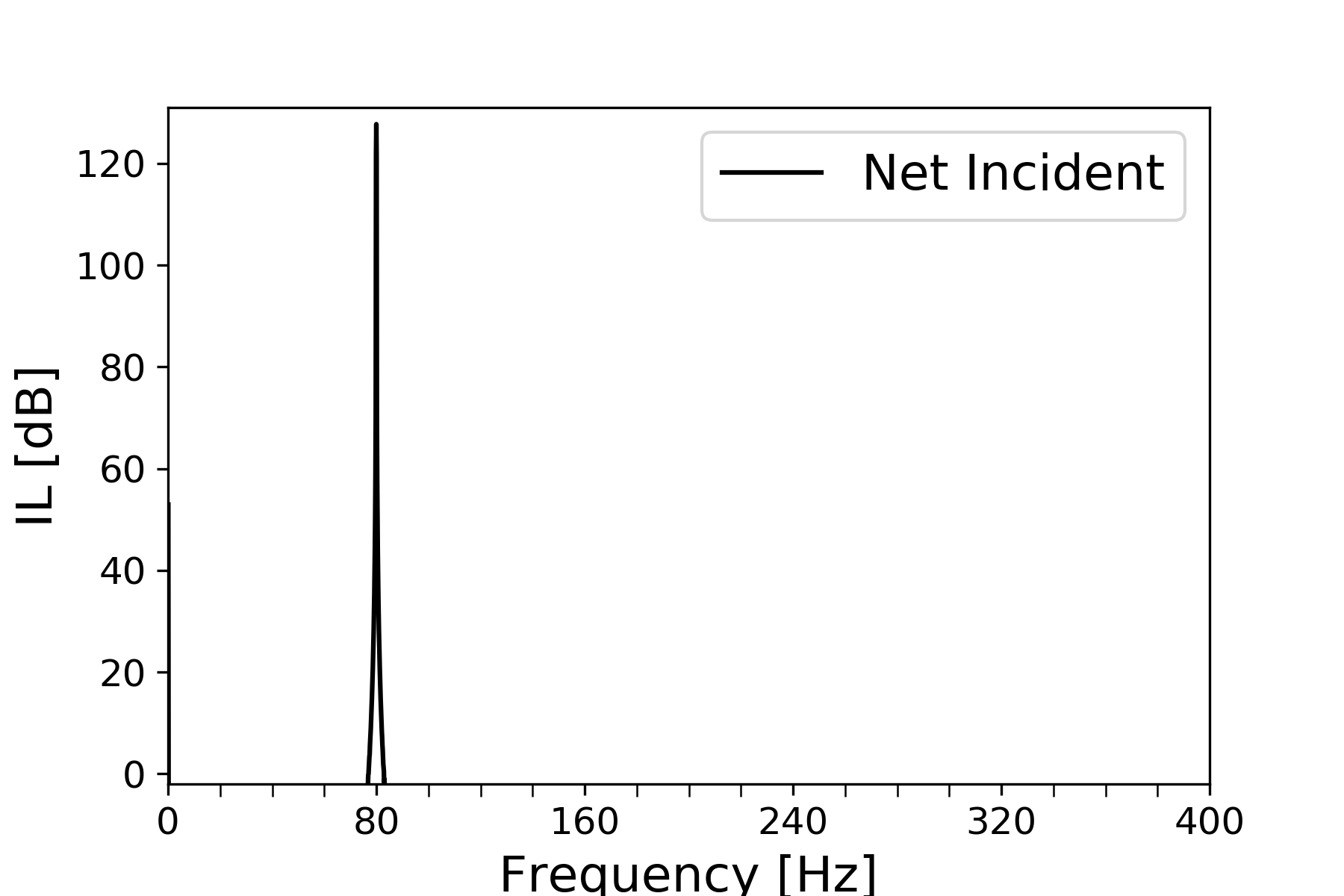}}
\subfigure[]{\includegraphics[width=0.48\textwidth]{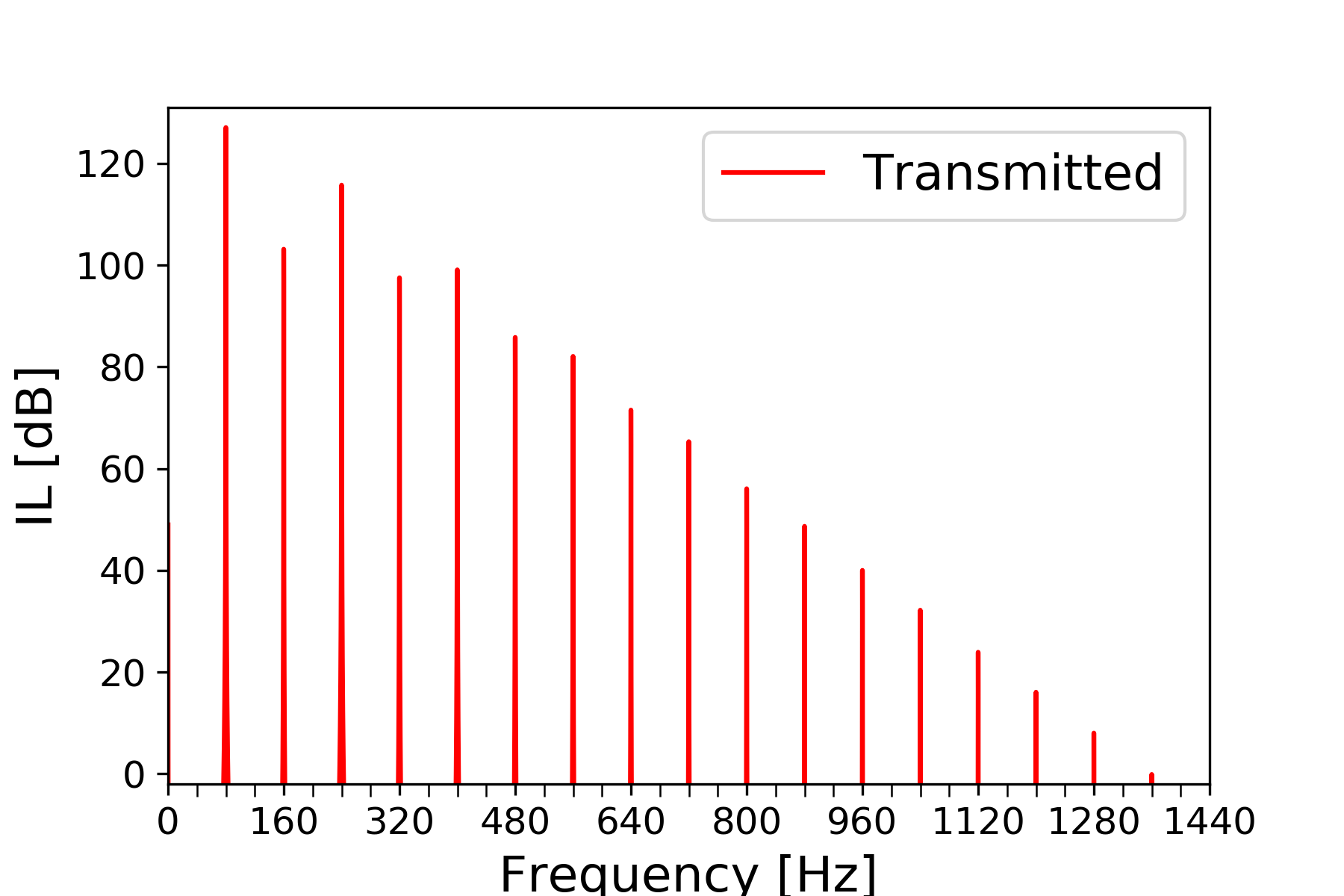}}\\
\subfigure[]{\includegraphics[width=0.48\textwidth]{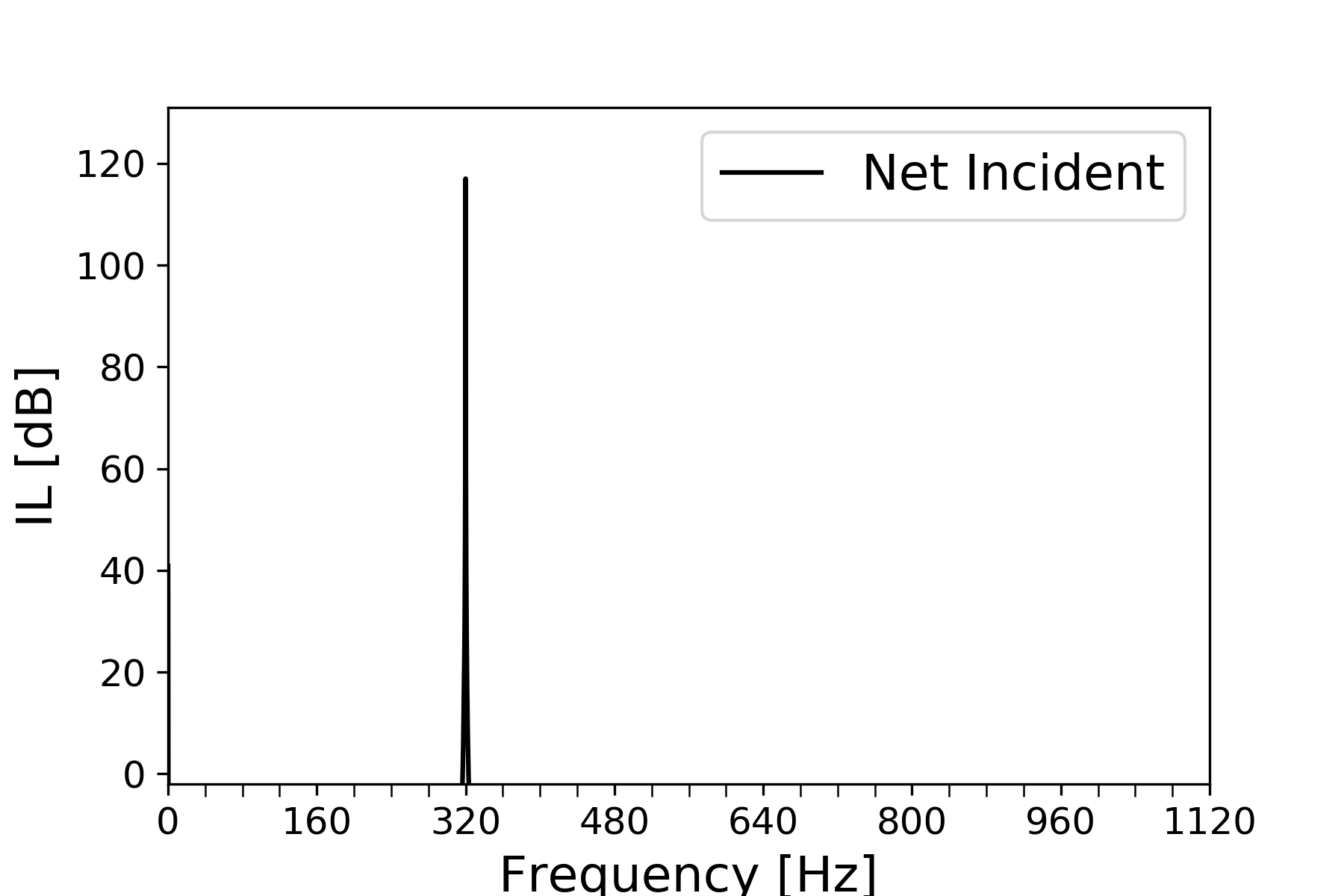}}
\subfigure[]{\includegraphics[width=0.48\textwidth]{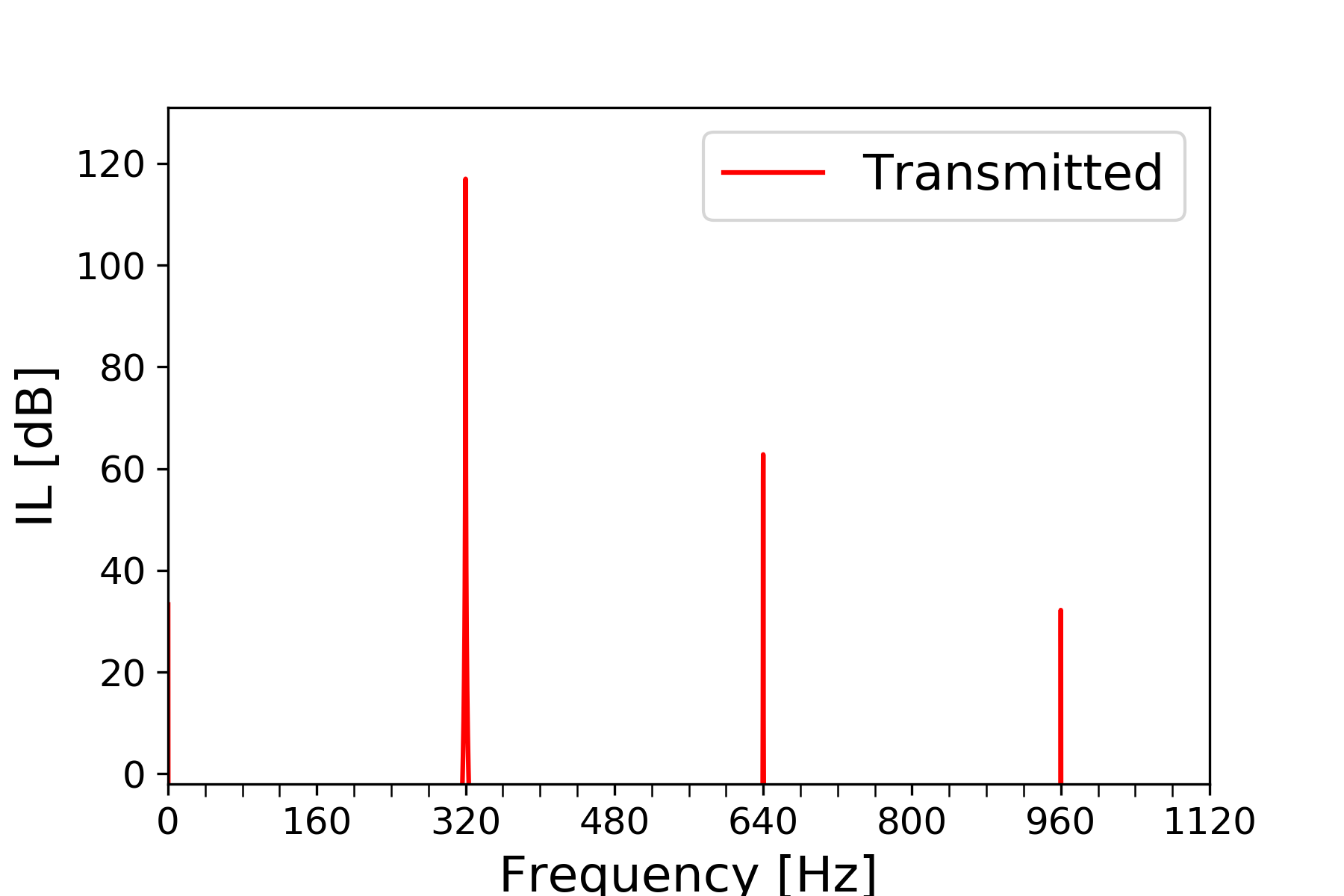}}\\
\caption{(a-f) Net incident and transmitted sound intensity spectra of M$2$ for a sound source excitation of $100$~Pa amplitude at $20$~Hz, $80$~Hz and $320$~Hz, respectively.}
\label{fig:100Pa_M2}
\end{figure}

\begin{figure}[htb!]
\centering
\includegraphics[width=0.99\textwidth]{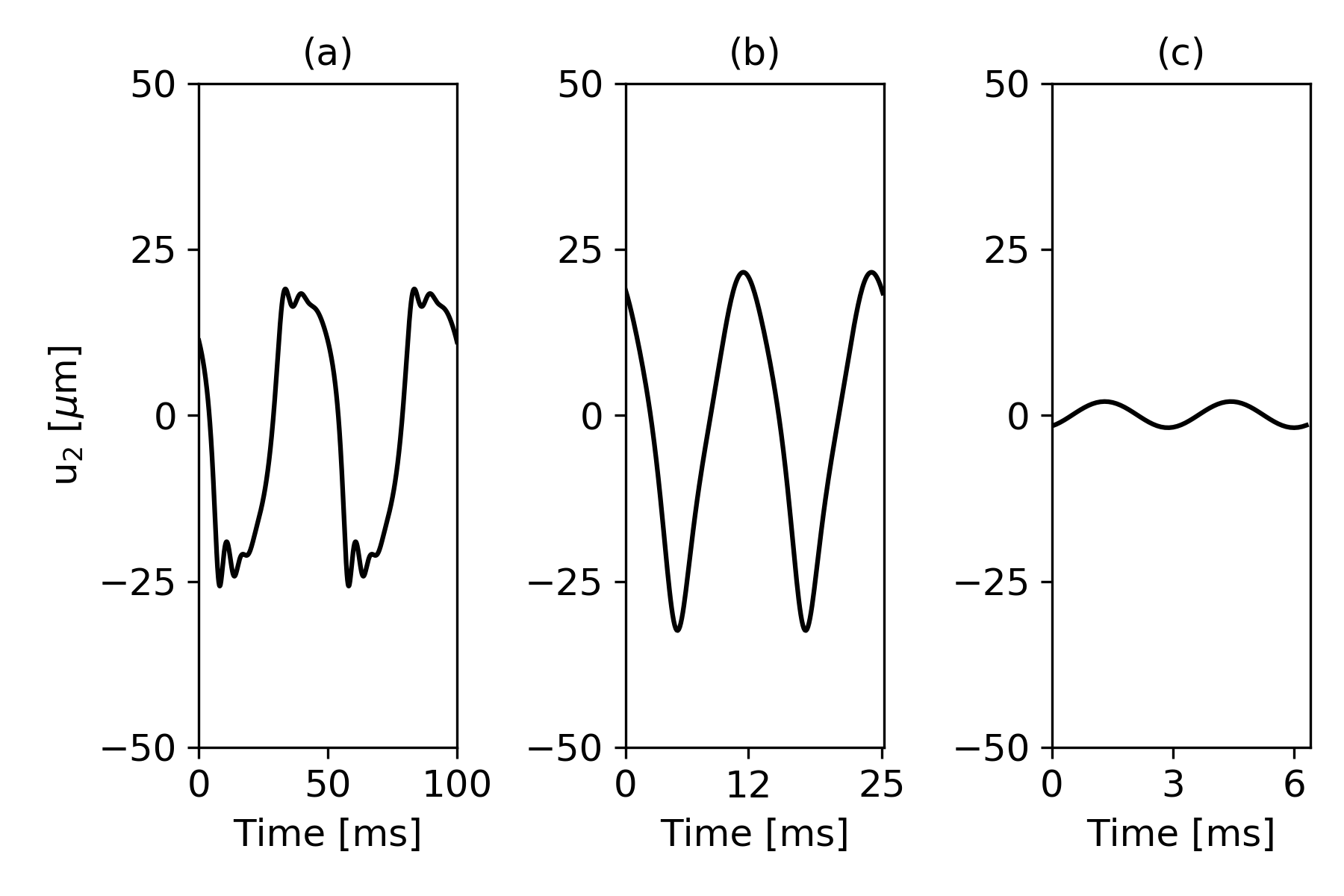}
\caption{Transverse displacement $\left(u_{2}\right)$  of M$2$ for $f_{ex}=20$~Hz, $80$~Hz and $320$~Hz.}
\label{fig:100Pa_M2_disp}
\end{figure}


\begin{figure}[htb!]
\centering
\setcounter{subfigure}{0}
\subfigure[]{\includegraphics[width=0.48\textwidth]{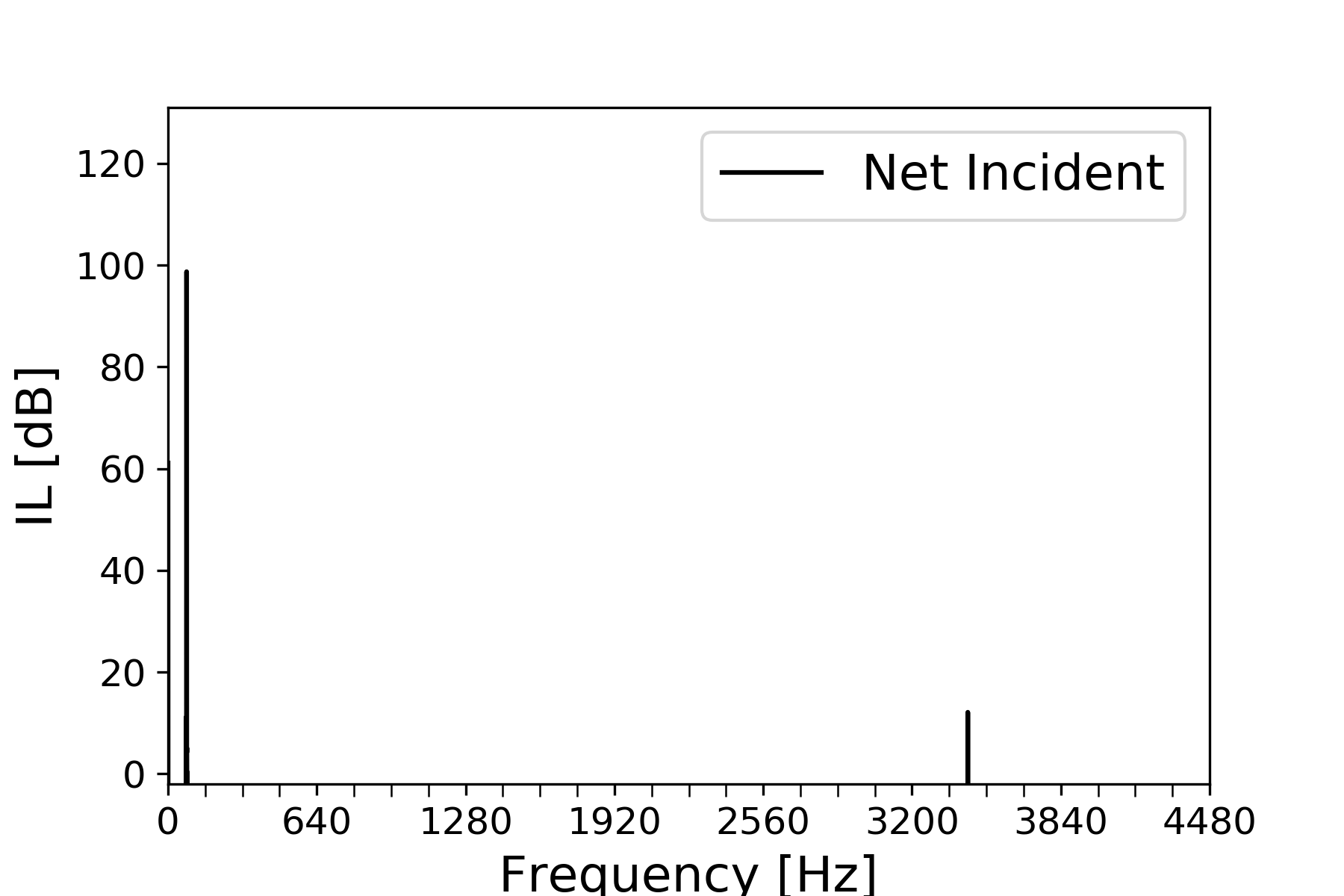}}
\subfigure[]{\includegraphics[width=0.48\textwidth]{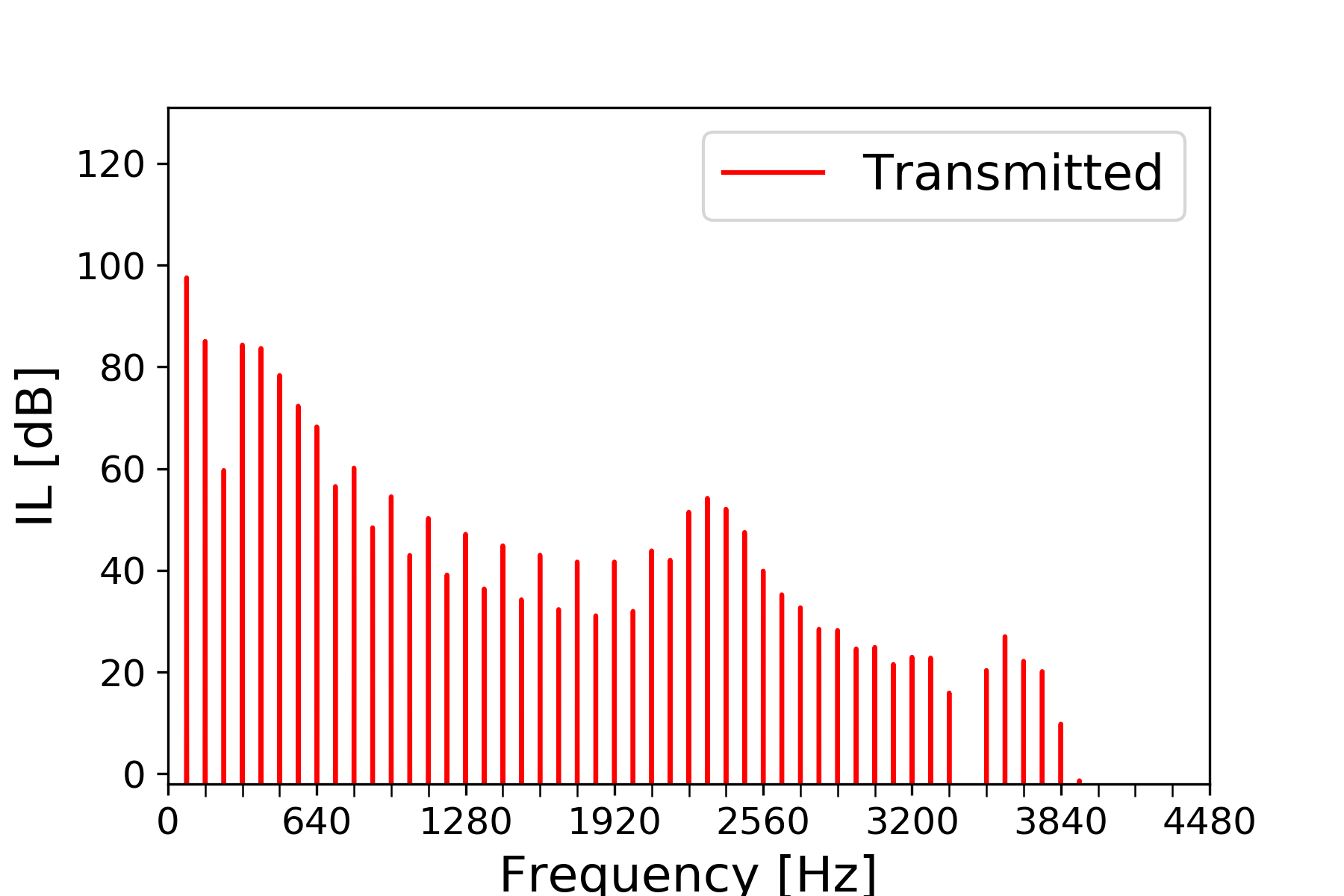}}\\
\subfigure[]{\includegraphics[width=0.48\textwidth]{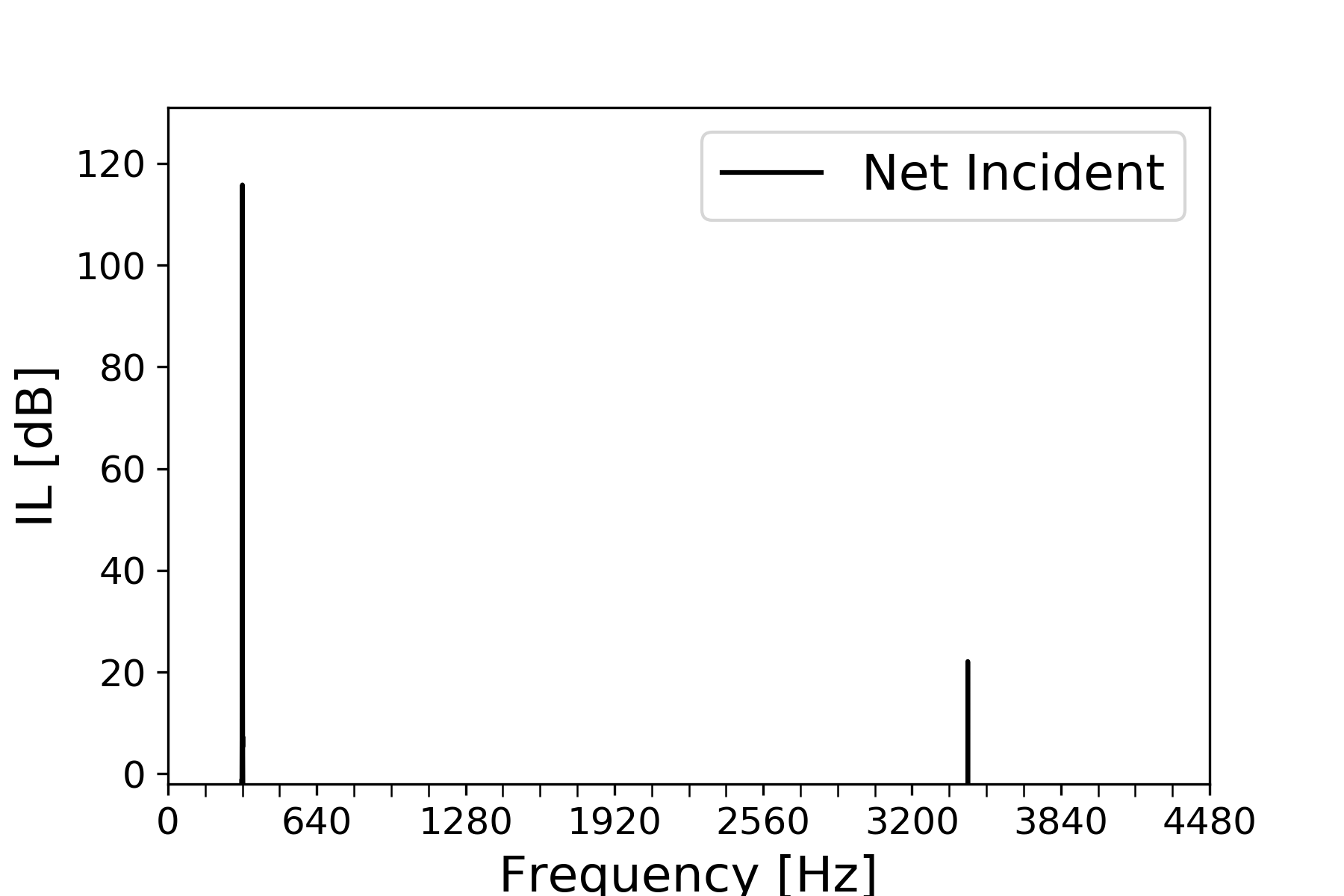}}
\subfigure[]{\includegraphics[width=0.48\textwidth]{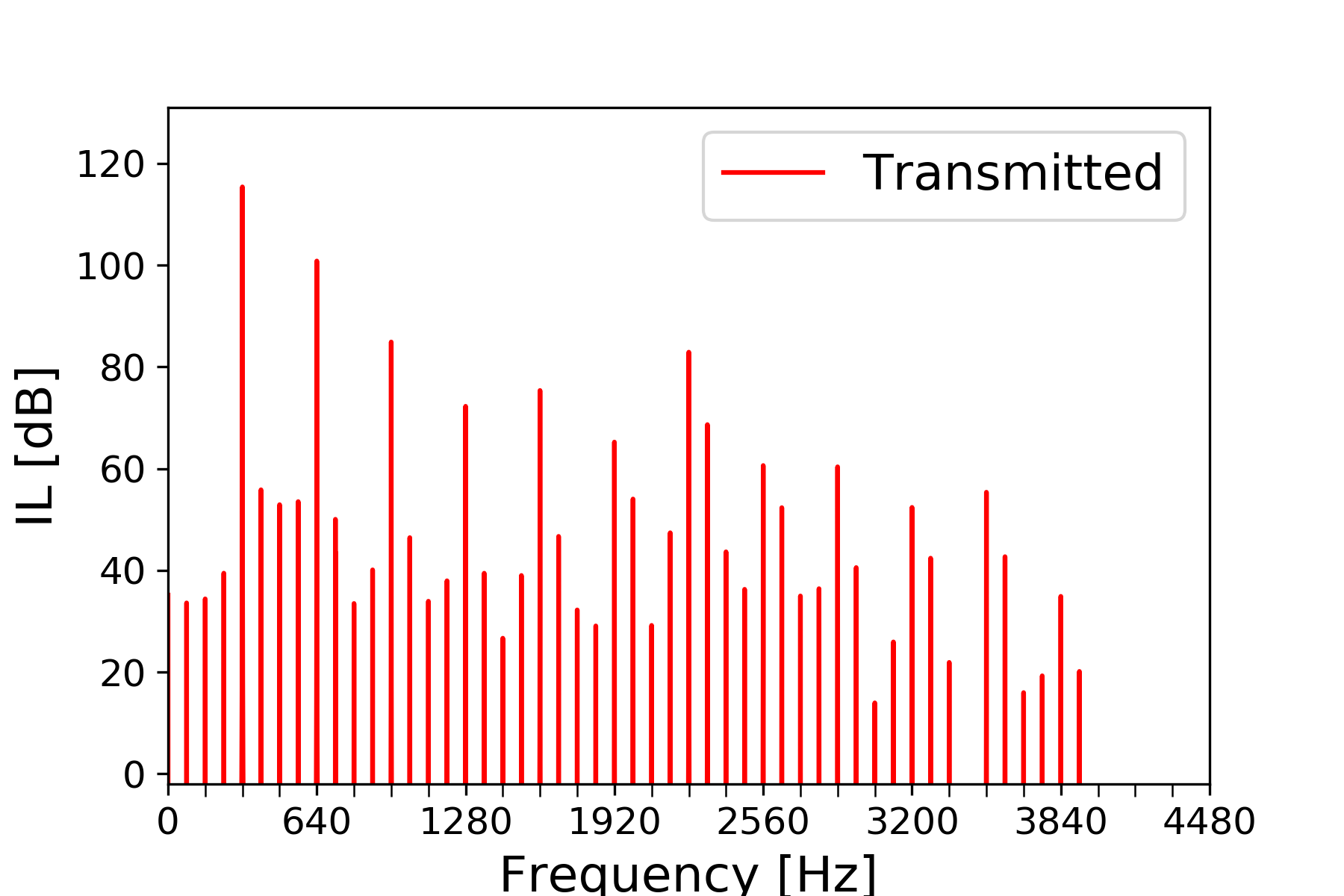}}\\
\subfigure[]{\includegraphics[width=0.48\textwidth]{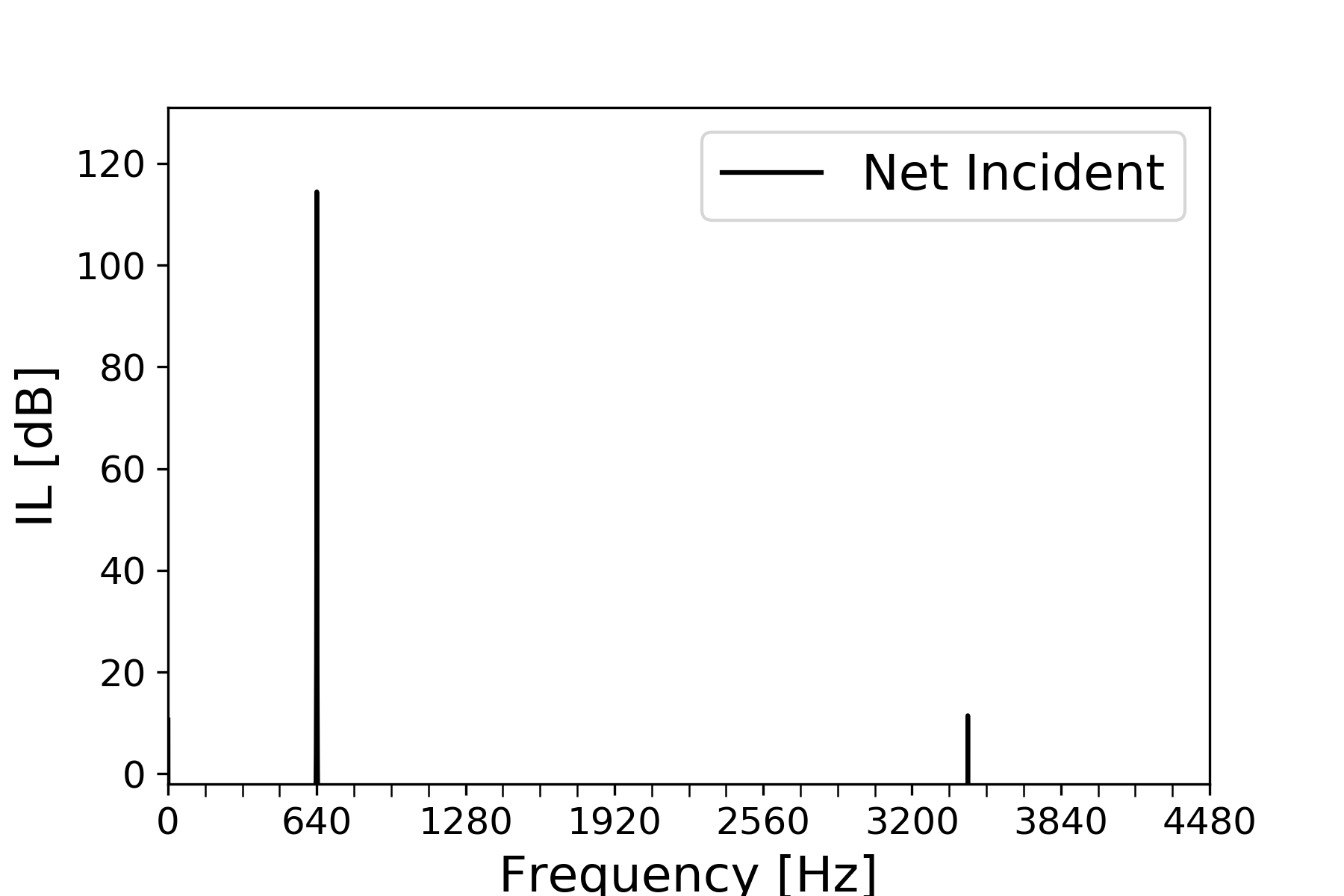}}
\subfigure[]{\includegraphics[width=0.48\textwidth]{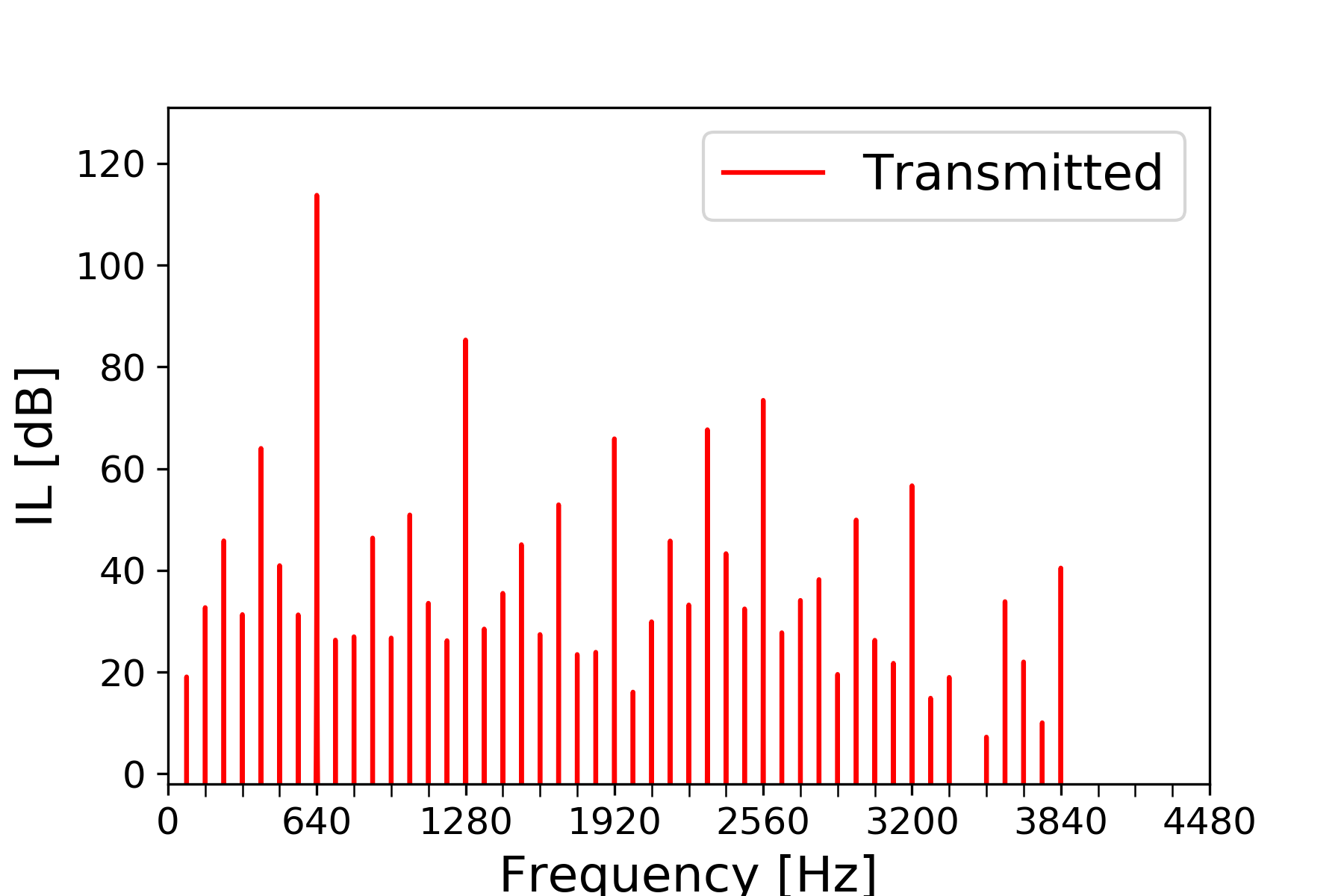}}\\
\caption{(a-f) Net incident and transmitted sound intensity spectra of M$3$ for a sound source excitation of $100$~Pa amplitude at $80$~Hz, $320$~Hz and $640$~Hz, respectively.}
\label{fig:100Pa_M3}
\end{figure}
\FloatBarrier

\begin{figure}[htb!]
\centering
\includegraphics[width=0.99\textwidth]{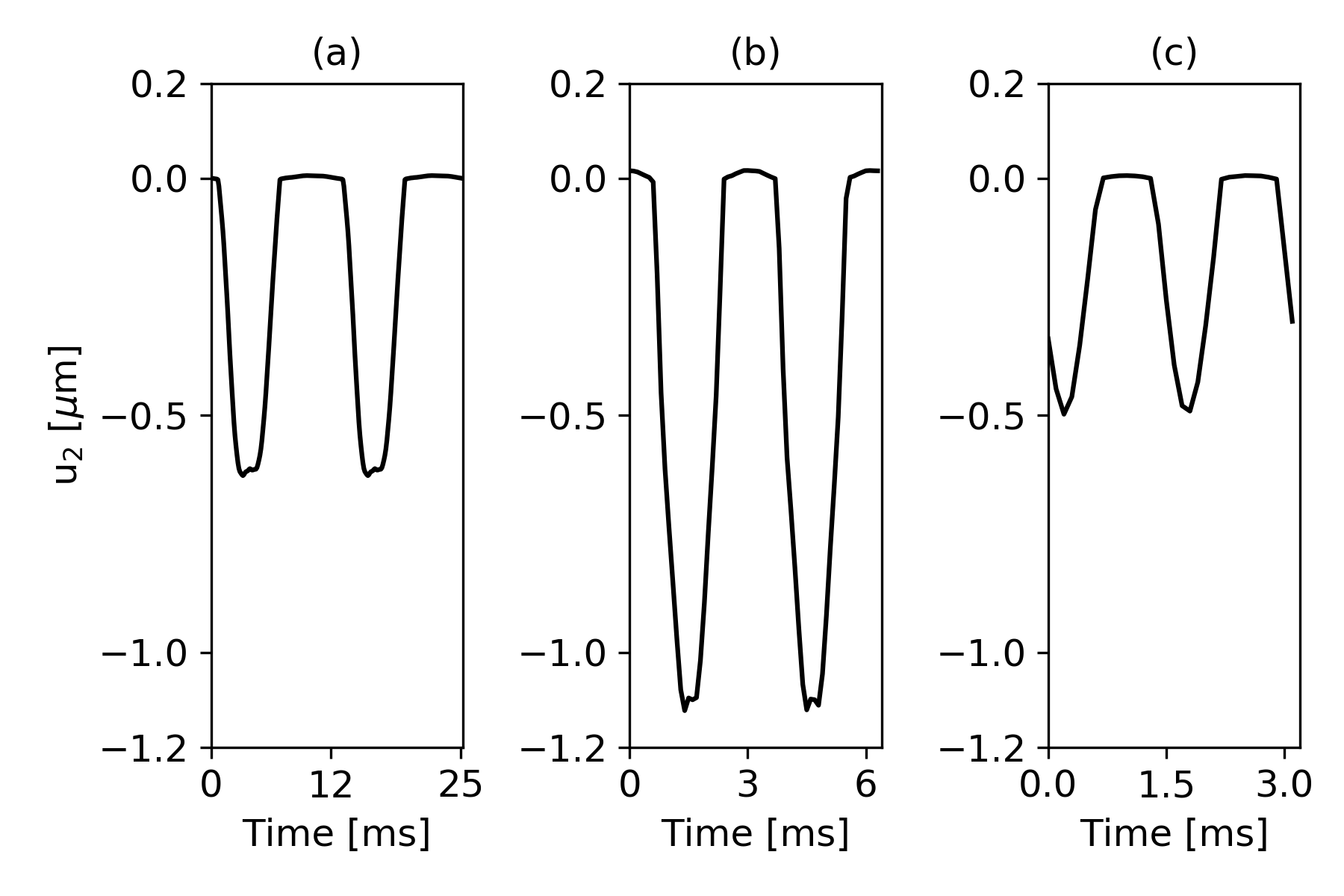}
\caption{Transverse displacement $\left(u_{2}\right)$  of M$3$ for $f_{ex}=80$~Hz, $320$~Hz and $640$~Hz.}
\label{fig:100Pa_M3_disp}
\end{figure}

\begin{figure}[htb!]
\centering
\setcounter{subfigure}{0}
\subfigure[]{\includegraphics[width=0.48\textwidth]{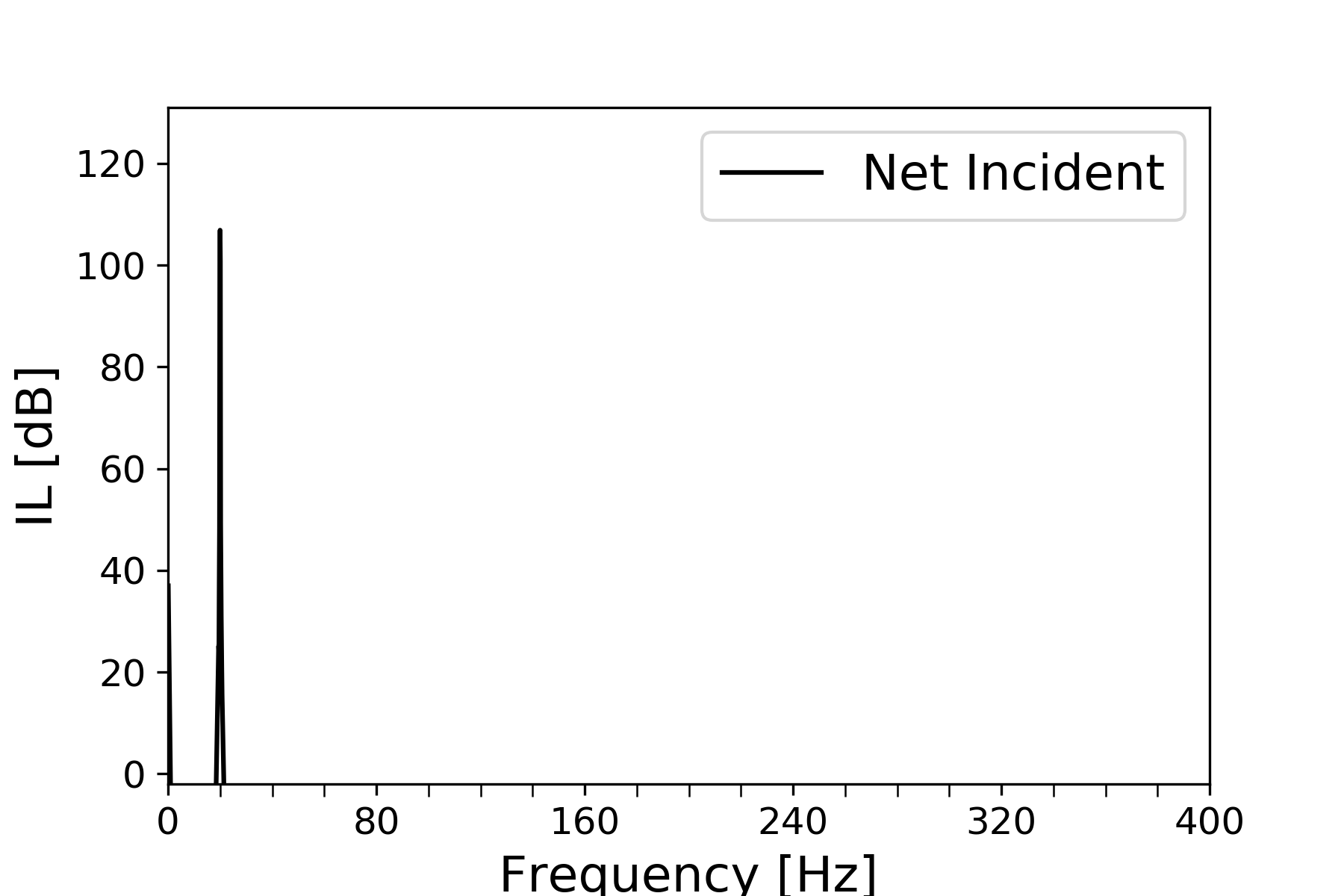}}
\subfigure[]{\includegraphics[width=0.48\textwidth]{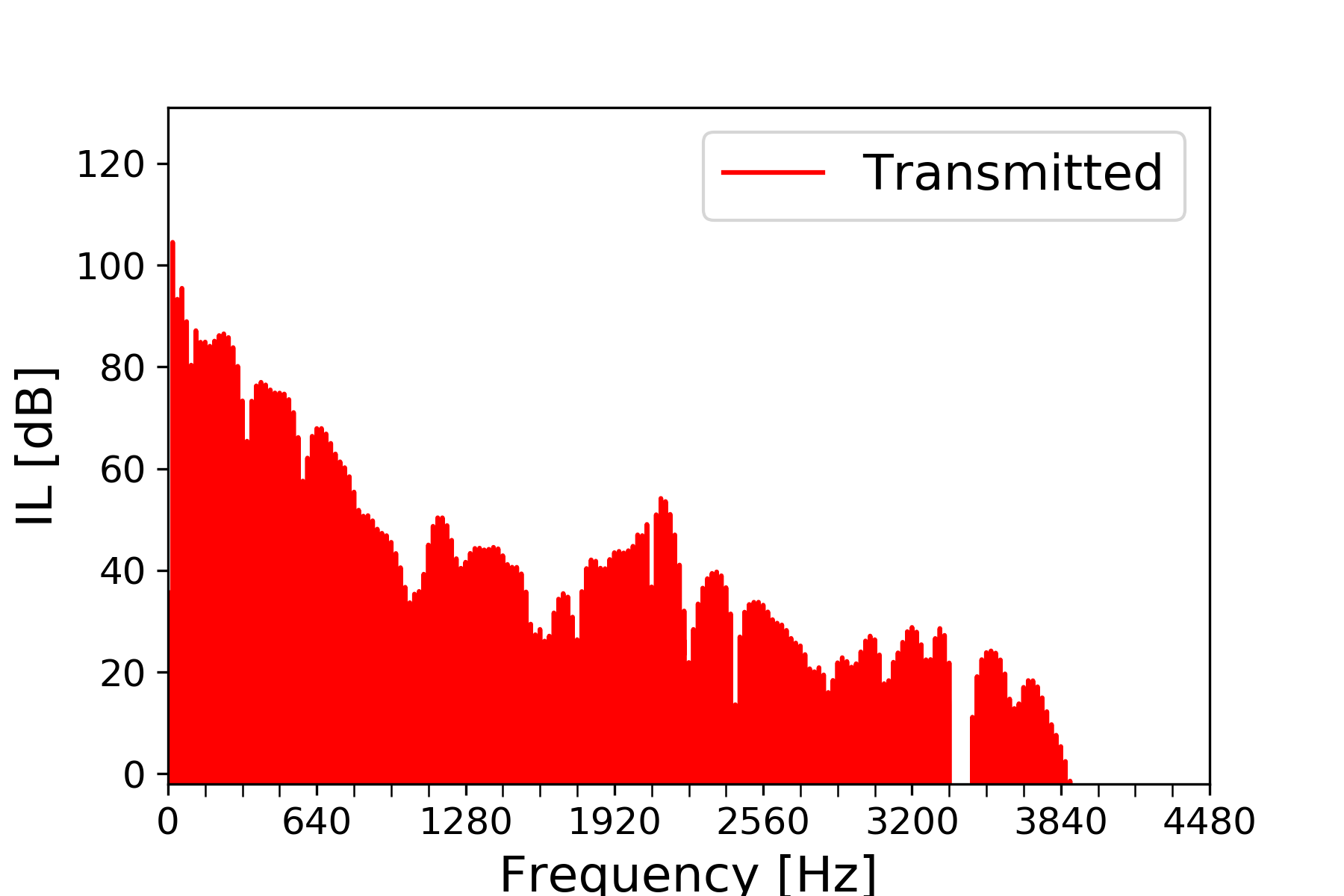}}\\
\subfigure[]{\includegraphics[width=0.48\textwidth]{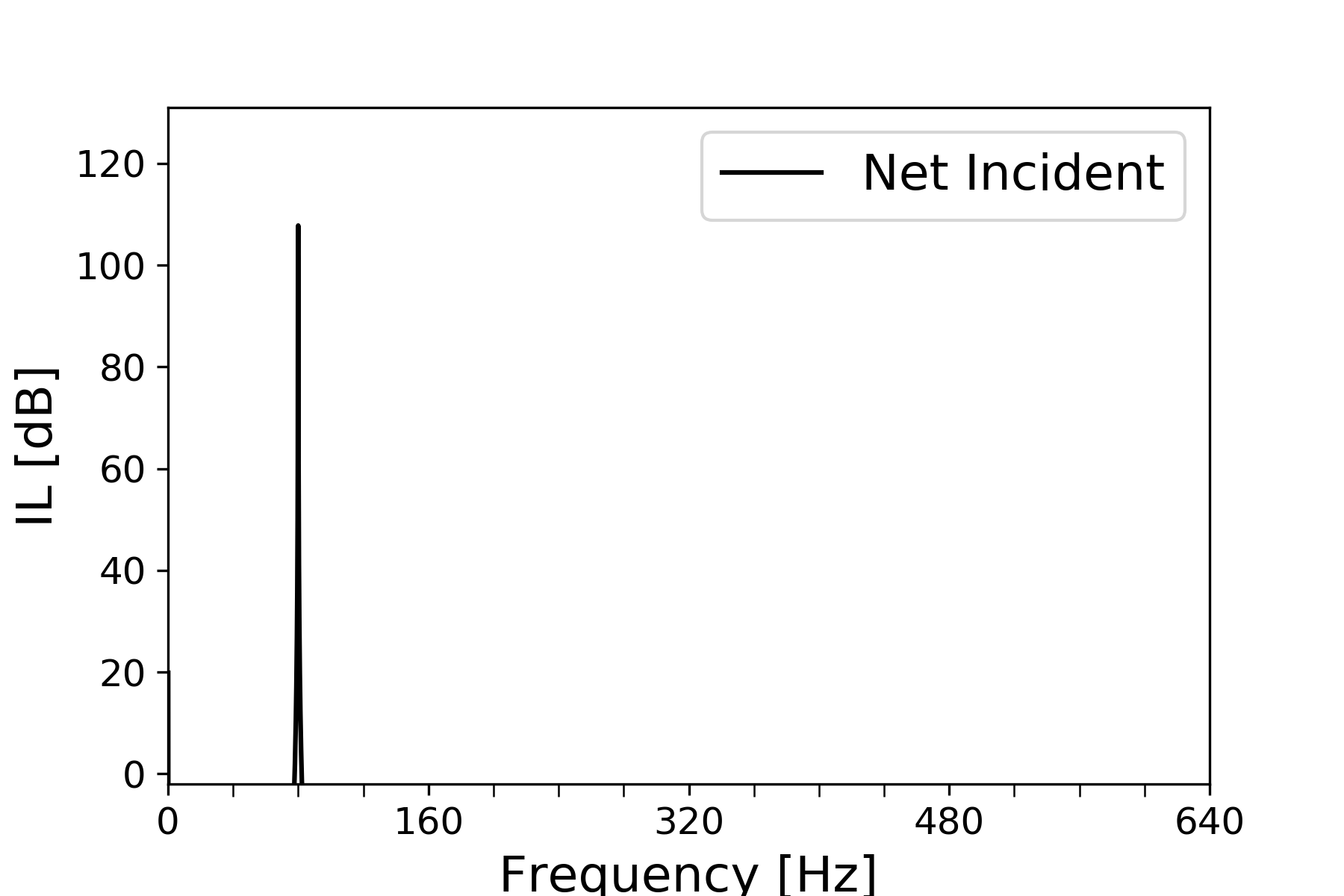}}
\subfigure[]{\includegraphics[width=0.48\textwidth]{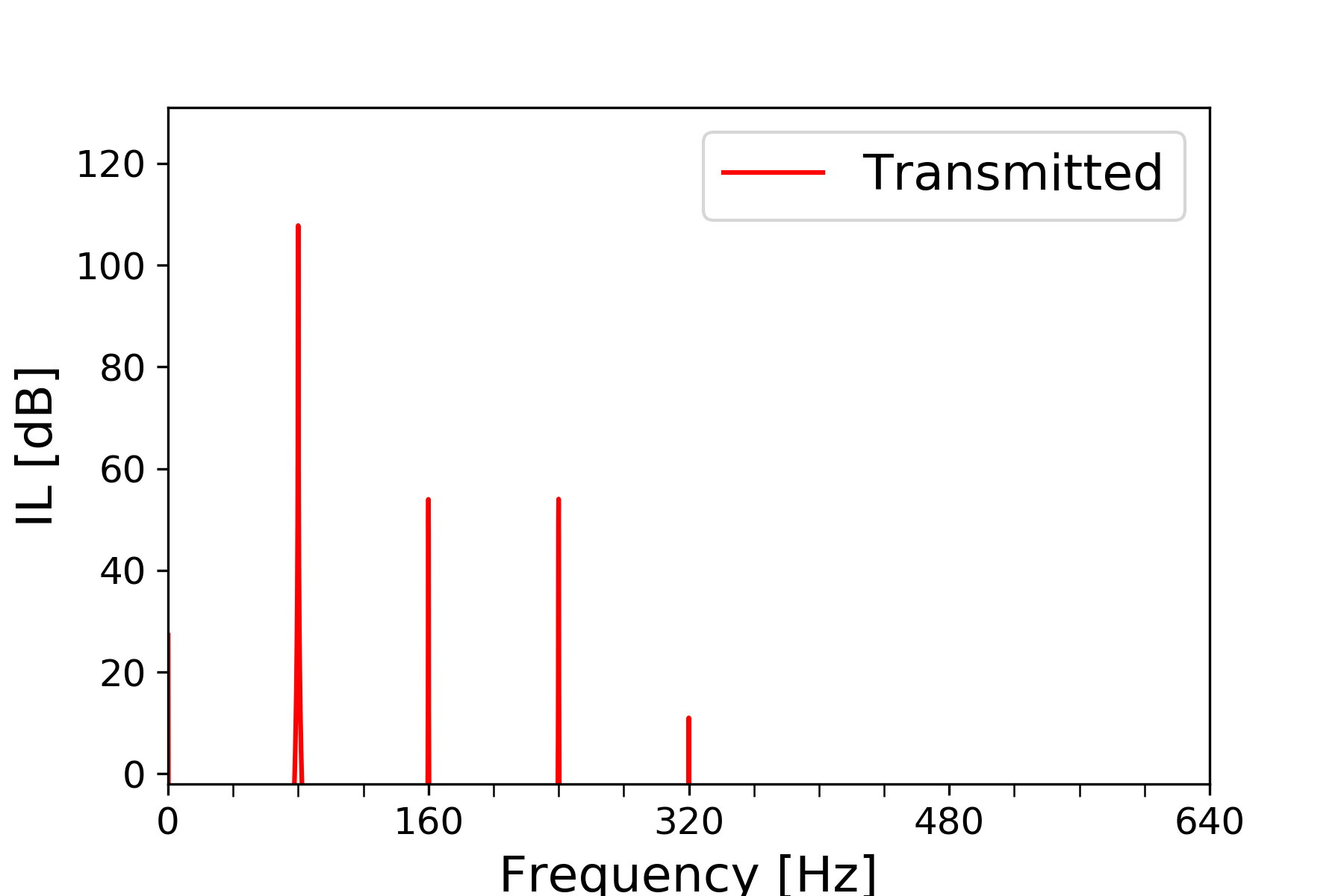}}\\
\subfigure[]{\includegraphics[width=0.48\textwidth]{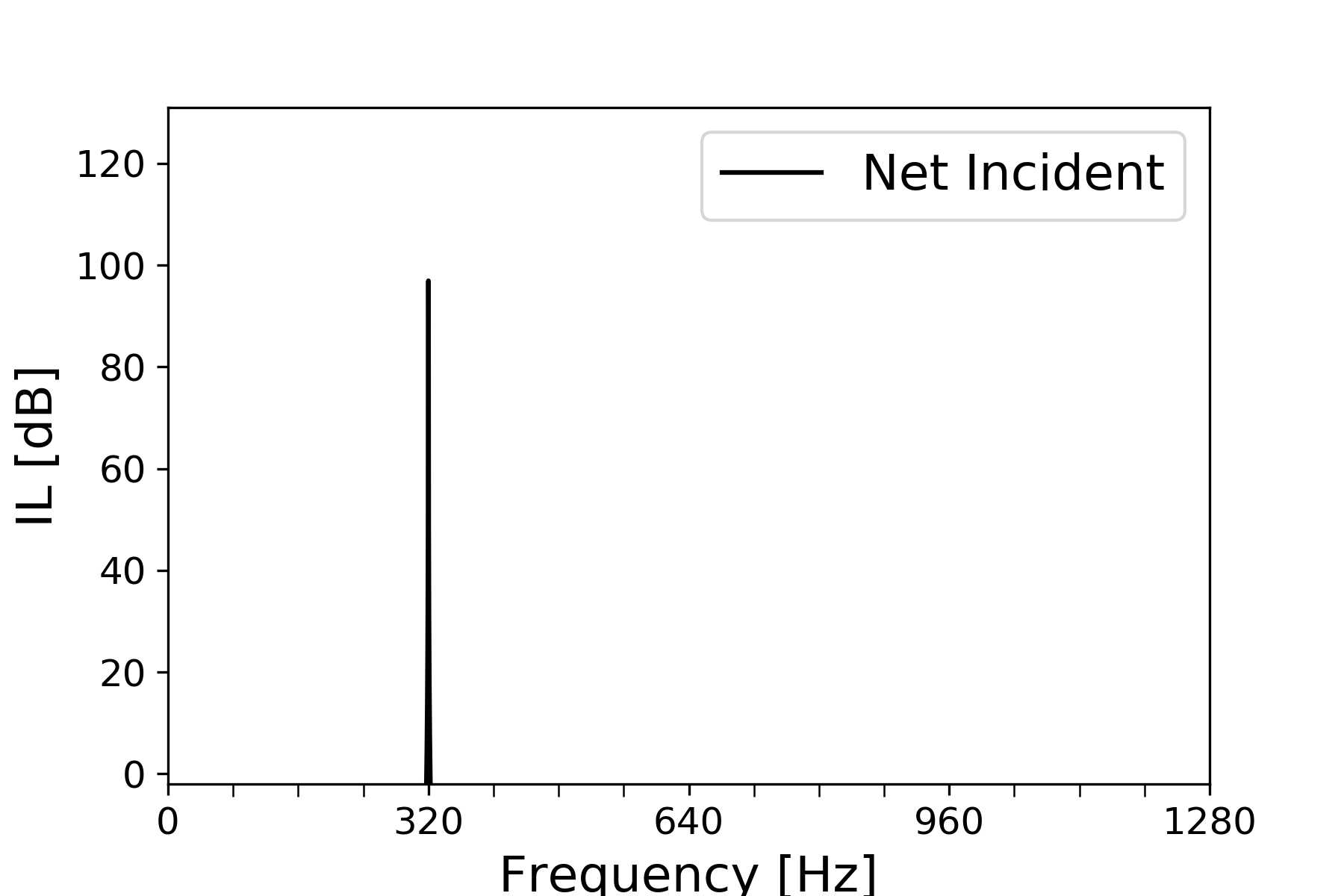}}
\subfigure[]{\includegraphics[width=0.48\textwidth]{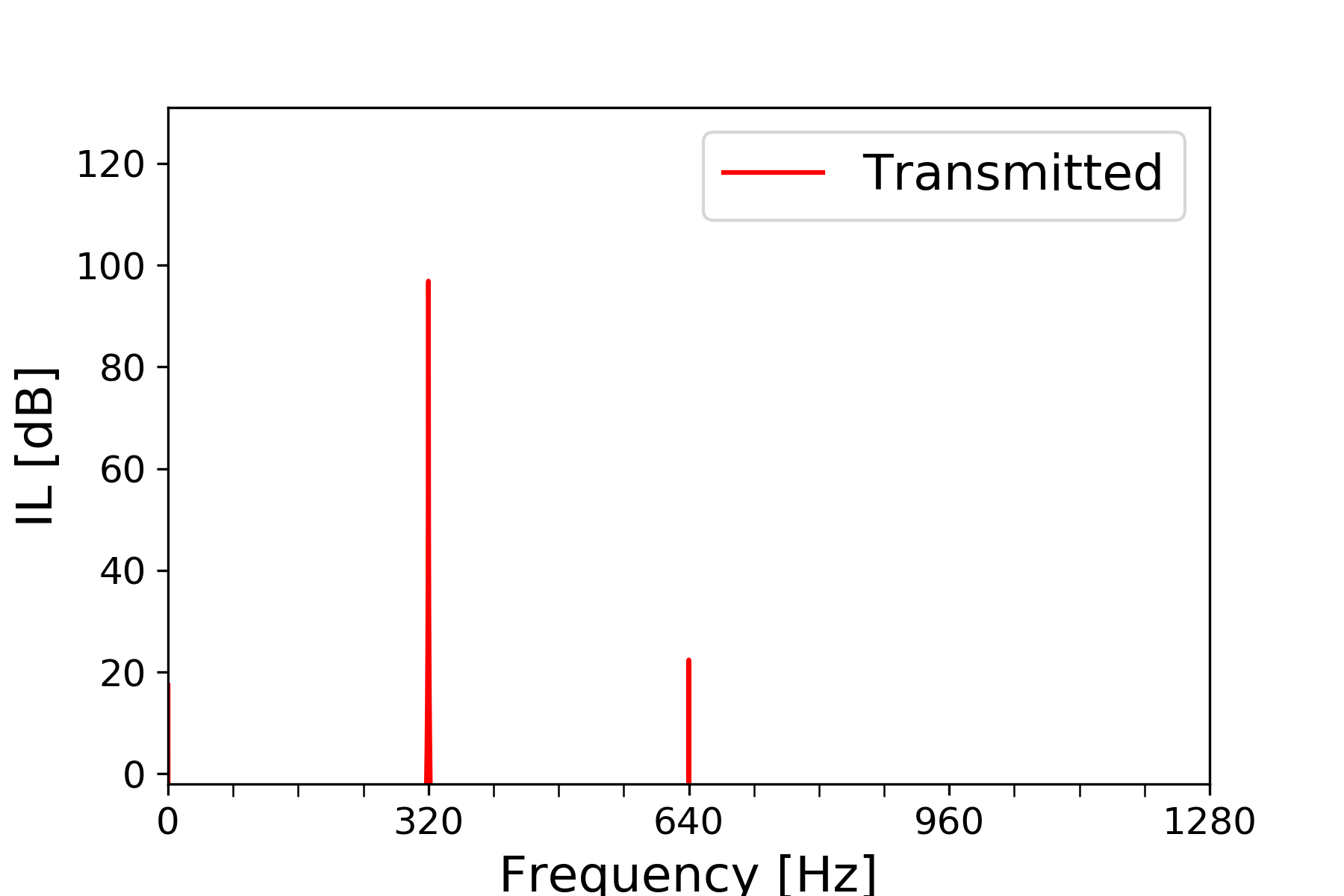}}\\
\caption{Net incident and transmitted sound intensity spectra of M$1$ for a sound source excitation of $10$~Pa amplitude at $20$~Hz, $80$~Hz and $320$~Hz, respectively.}
\label{fig:10Pa_M1}
\end{figure}

\FloatBarrier

\begin{figure}[htb!]
\centering
\includegraphics[width=0.99\textwidth]{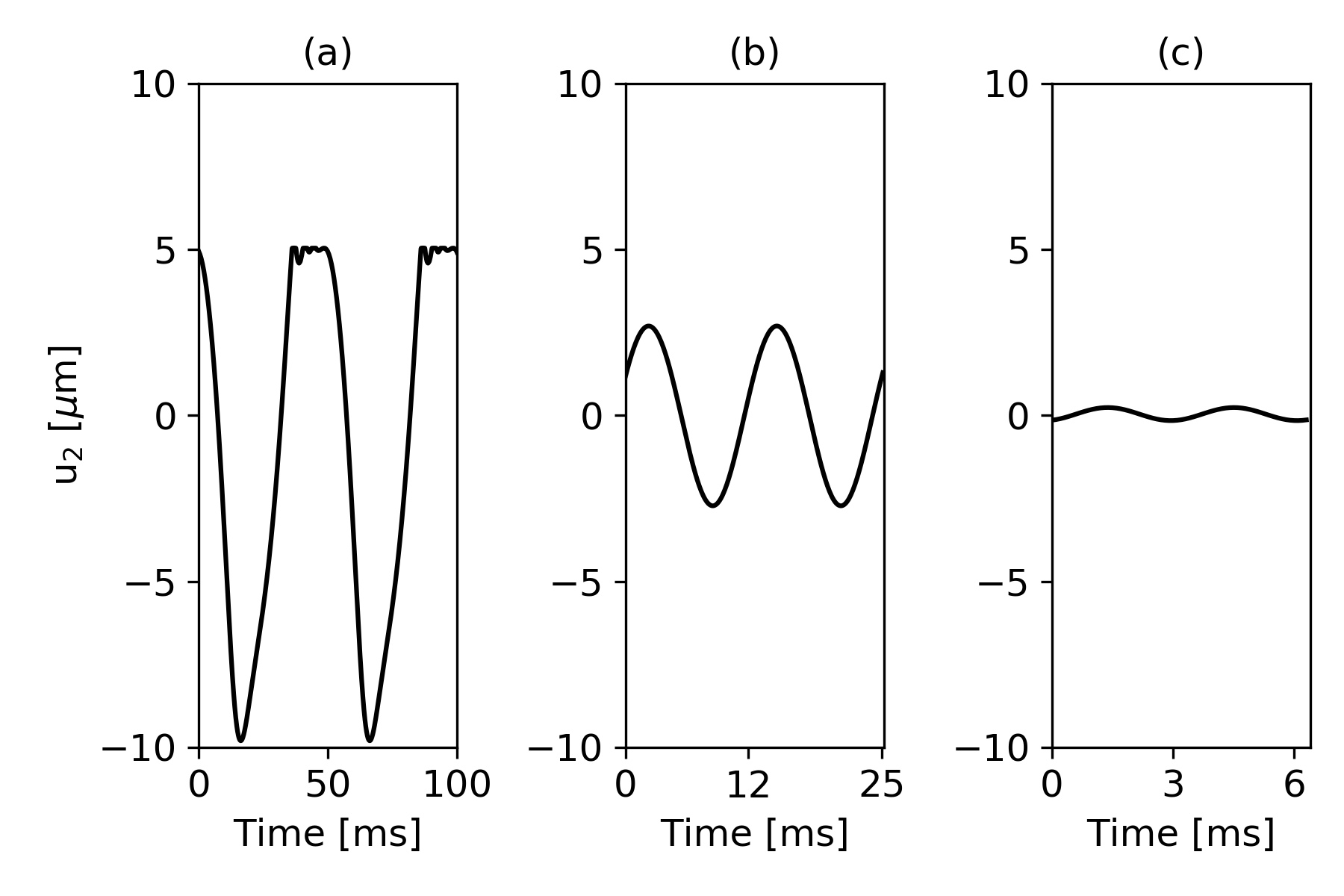}
\caption{Transverse displacement $\left(u_{2}\right)$ of M$1$ for $f_{ex}=$ $20$~Hz, $80$~Hz and $320$~Hz, and $10$~Pa amplitude.}
\label{fig:10Pa_M1_disp}
\end{figure}



\begin{figure}[htb!]
\centering
\subfigure[]{\includegraphics[width=0.49\textwidth]{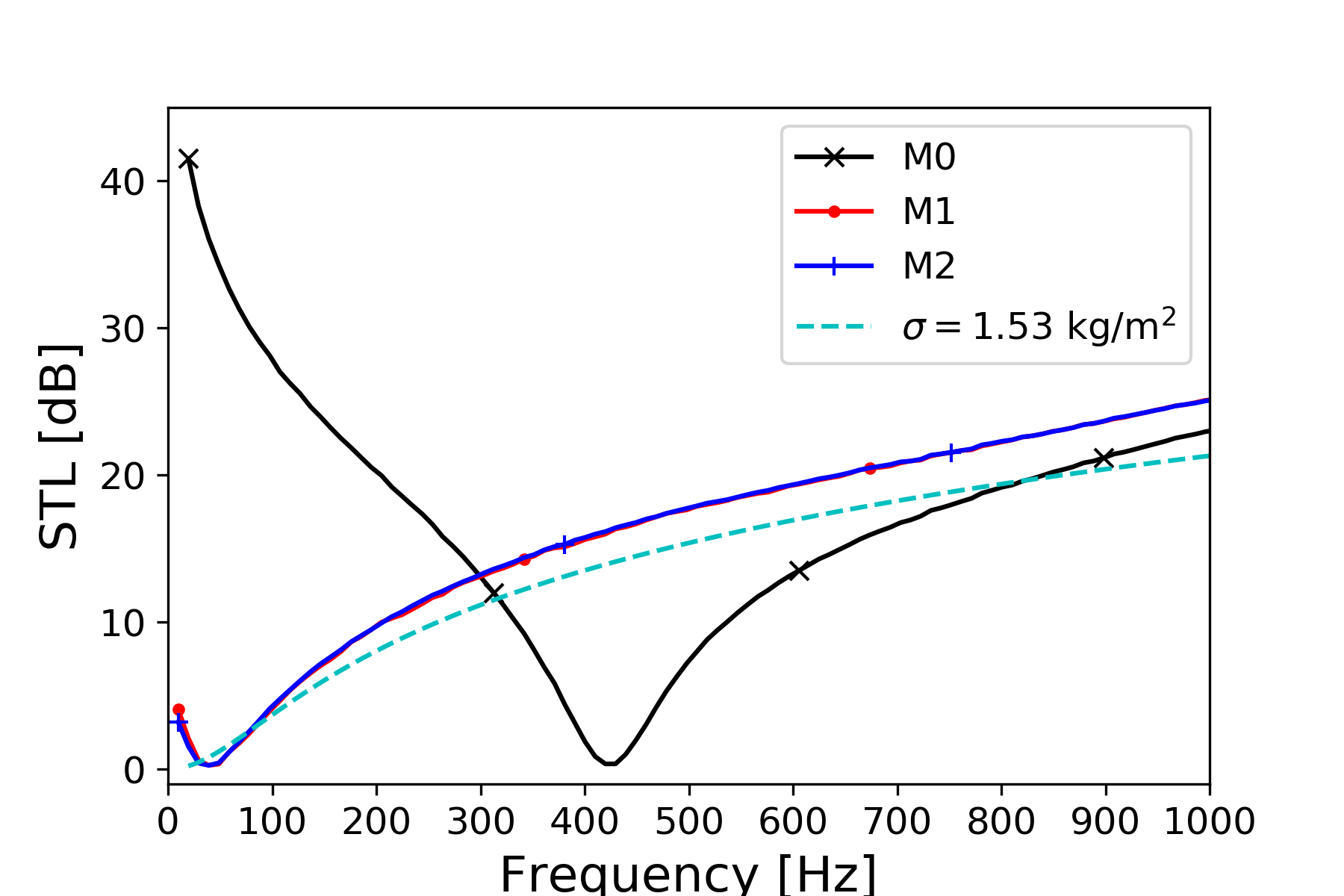}}
\subfigure[]{\includegraphics[width=0.49\textwidth]{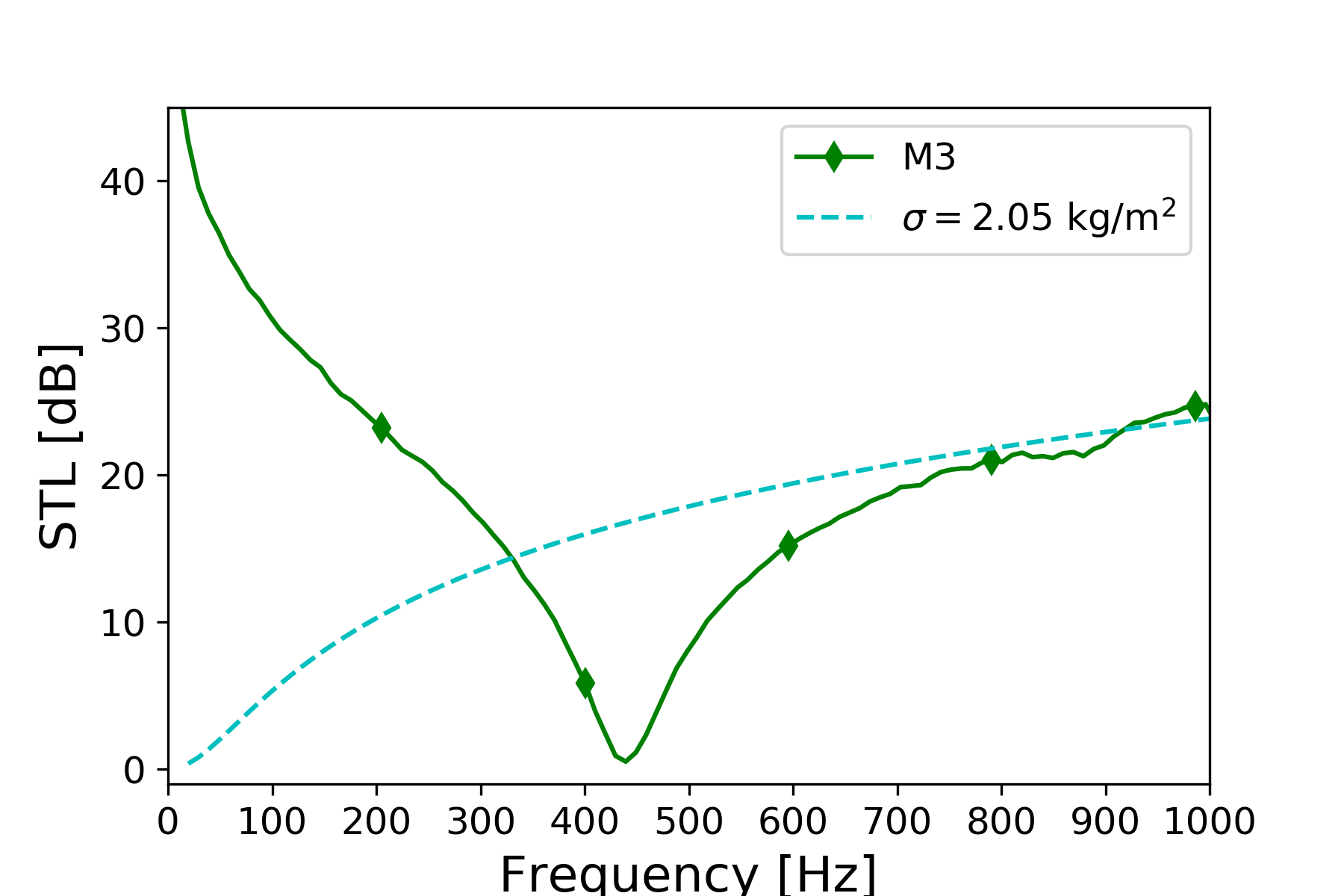}}
\caption{(a) STL for M$0$, M$1$ and M$2$ for a lowpass ($0-1000$~Hz) Gaussian white noise excitation with zero mean and standard deviation of $10$~Pa, and (b) STL for M$3$. The corresponding equivalent limp panel STL characteristics were also added.}
\label{fig:stlBroadbandWhNoiseM0M1M2M3-meas2}
\end{figure}

\begin{figure}[htb!]
\centering
\subfigure[]{\includegraphics[width=0.49\textwidth]{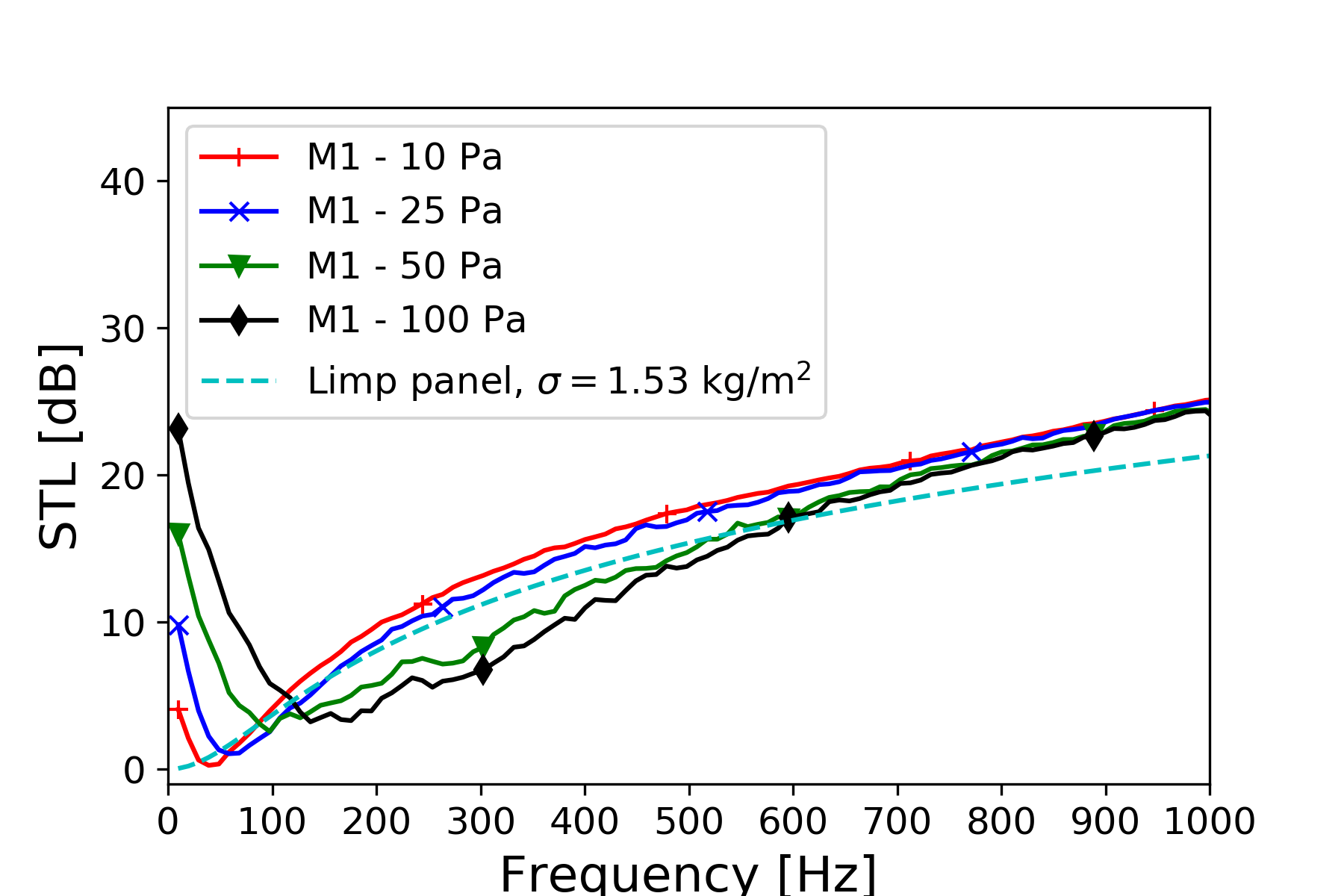}}
\subfigure[]{\includegraphics[width=0.49\textwidth]{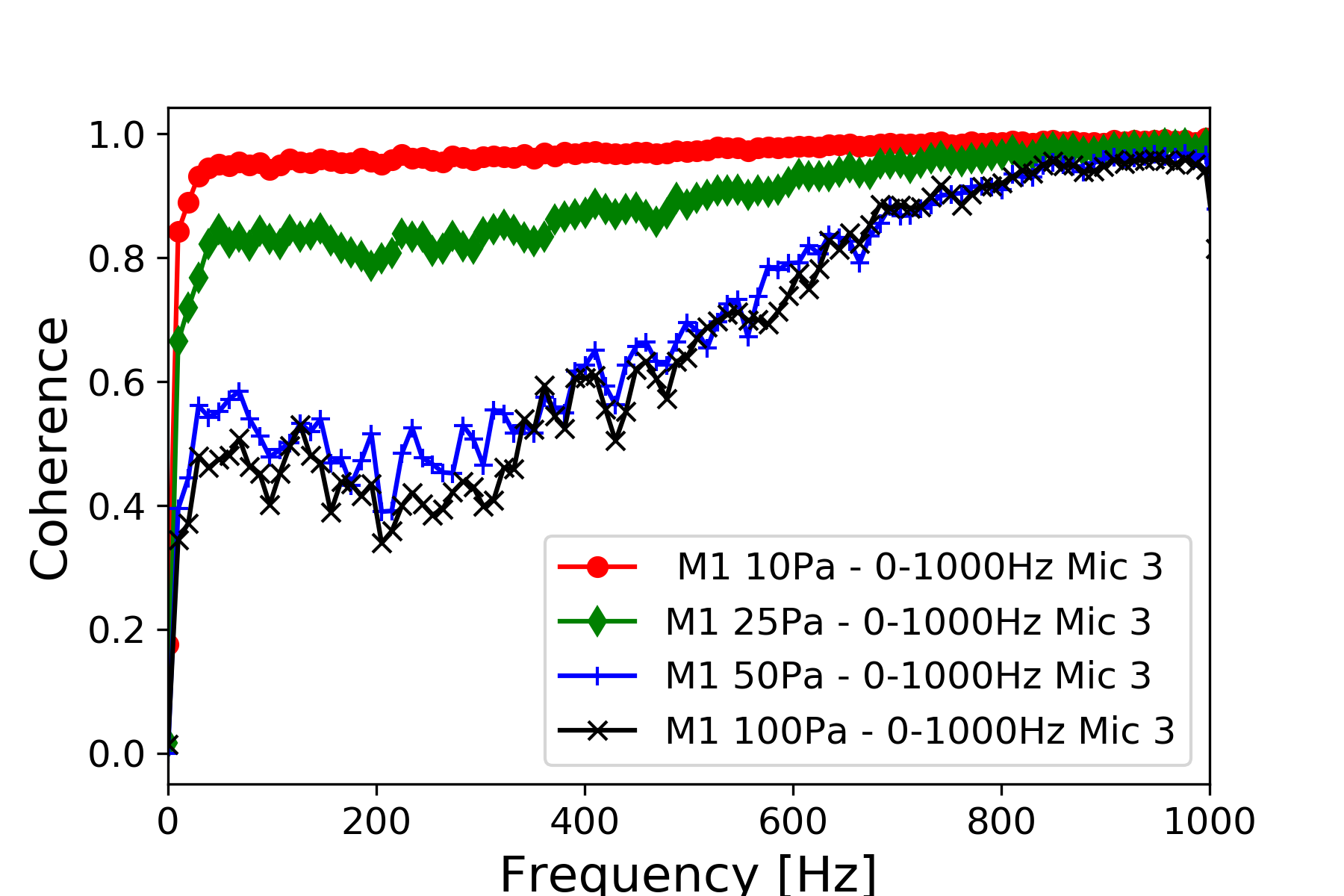}}
\caption{(a) Comparison of STL for M$1$ for a lowpass ($0-1000$~Hz) Gaussian white noise excitation with zero mean and standard deviations of $10$, $25$, $50$ and $100$~Pa, respectively, and (b) the corresponding coherence plots of downstream microphone m$3$ with respect to the input sound source signal.}
\label{fig:stlBroadbandWhNoiseM1AmpEffect}
\end{figure}


\begin{figure}[htb!]
\centering
\subfigure[]{\includegraphics[width=0.48\textwidth]{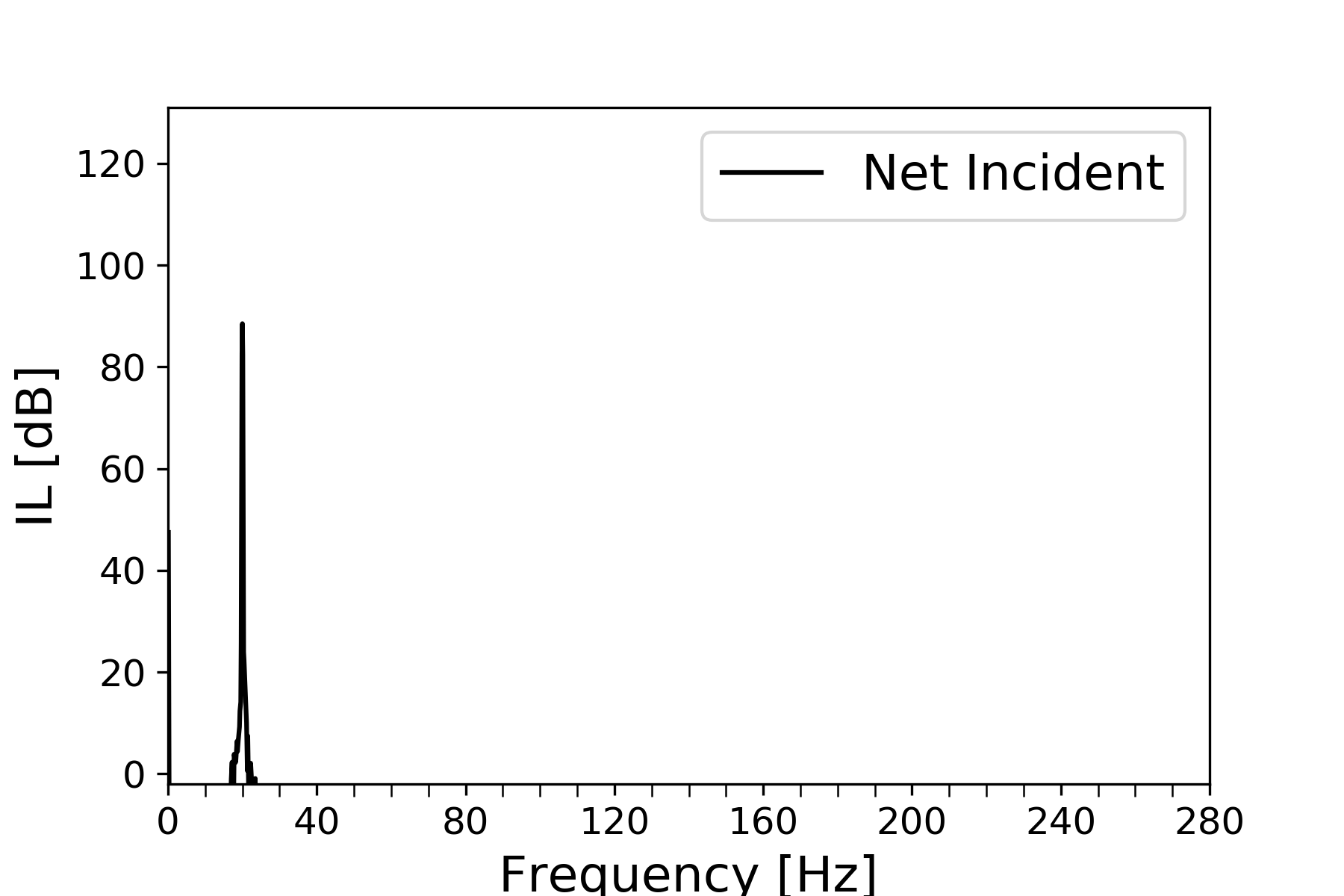}}
\subfigure[]{\includegraphics[width=0.48\textwidth]{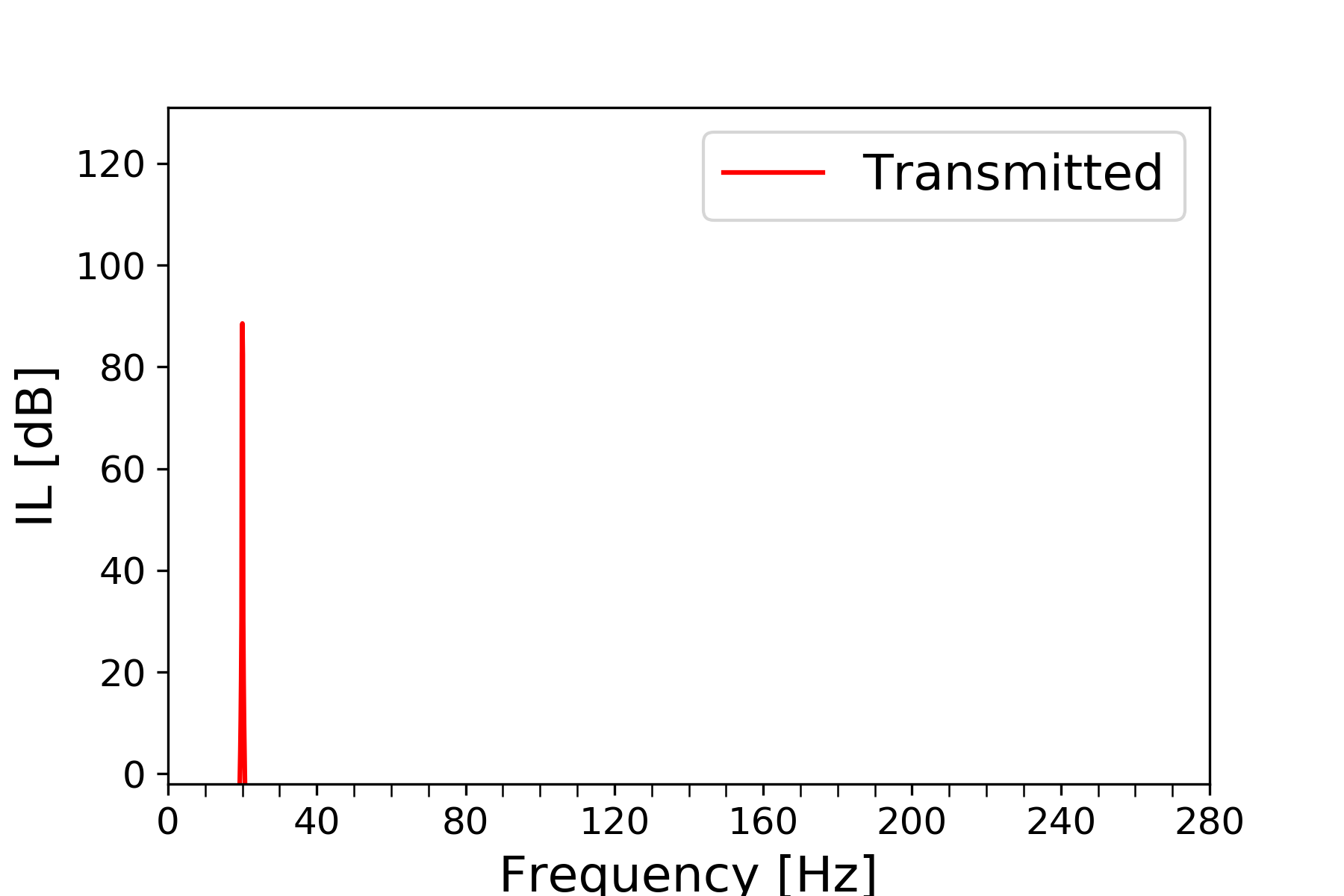}}\\
\subfigure[]{\includegraphics[width=0.48\textwidth]{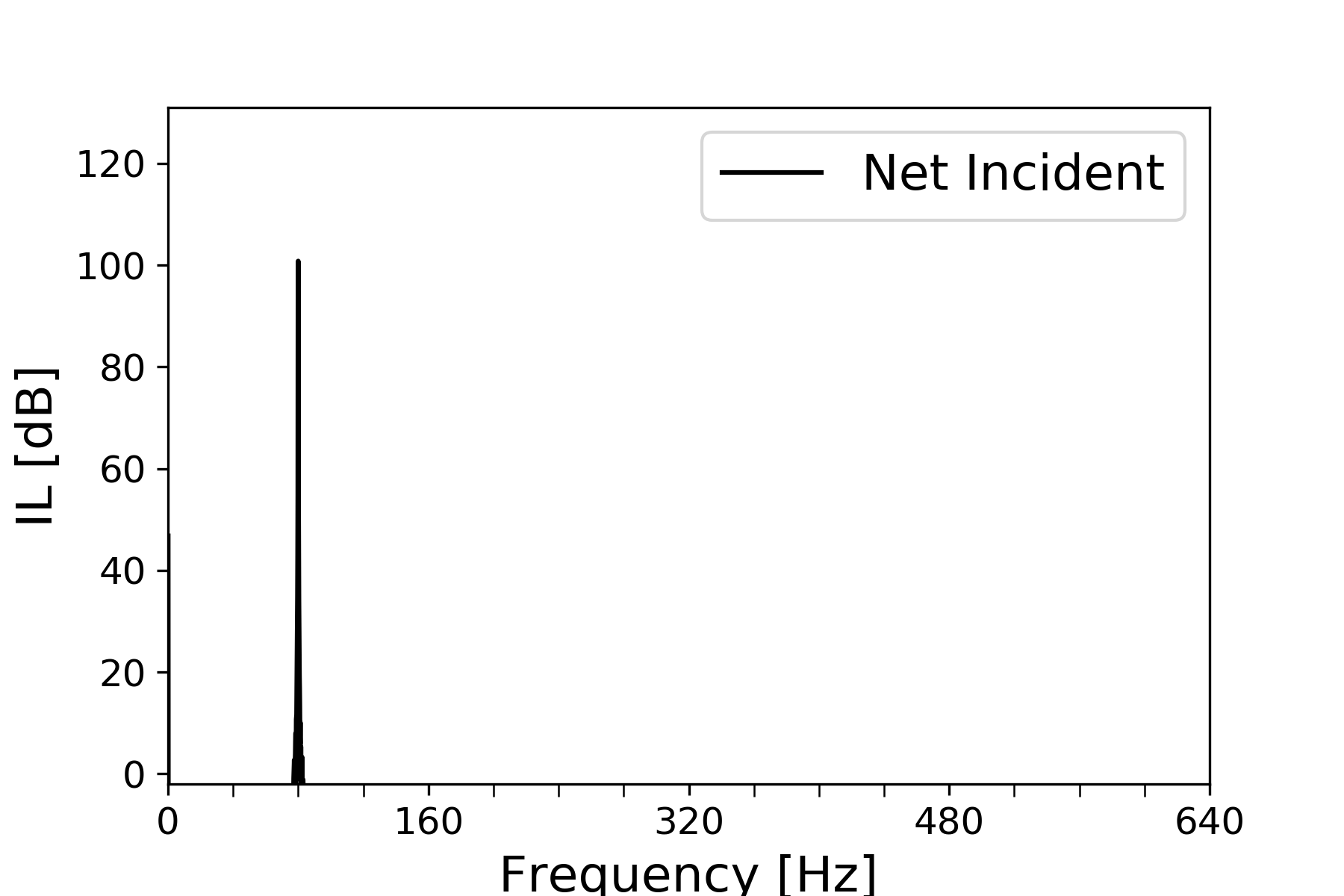}}
\subfigure[]{\includegraphics[width=0.48\textwidth]{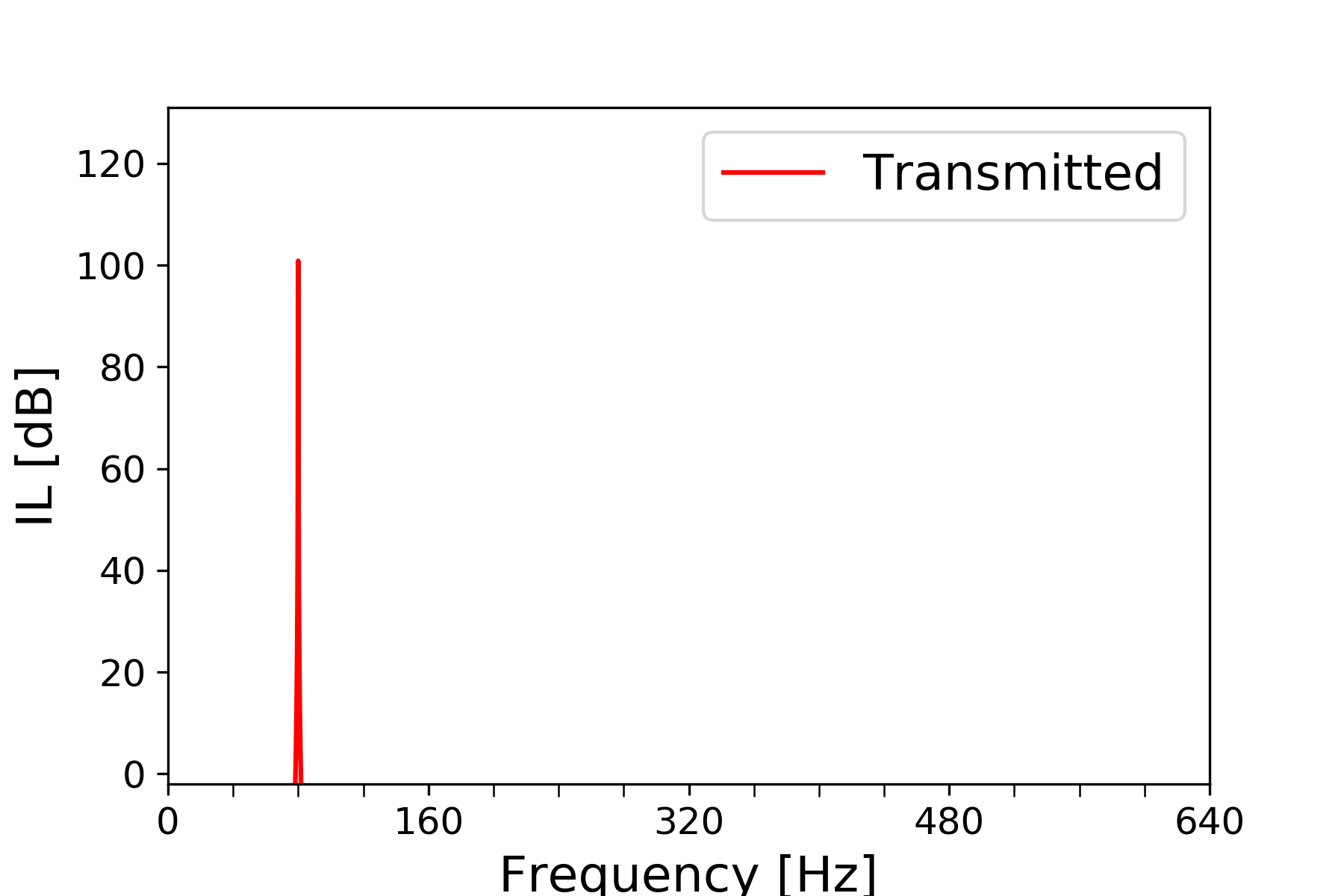}}\\
\subfigure[]{\includegraphics[width=0.48\textwidth]{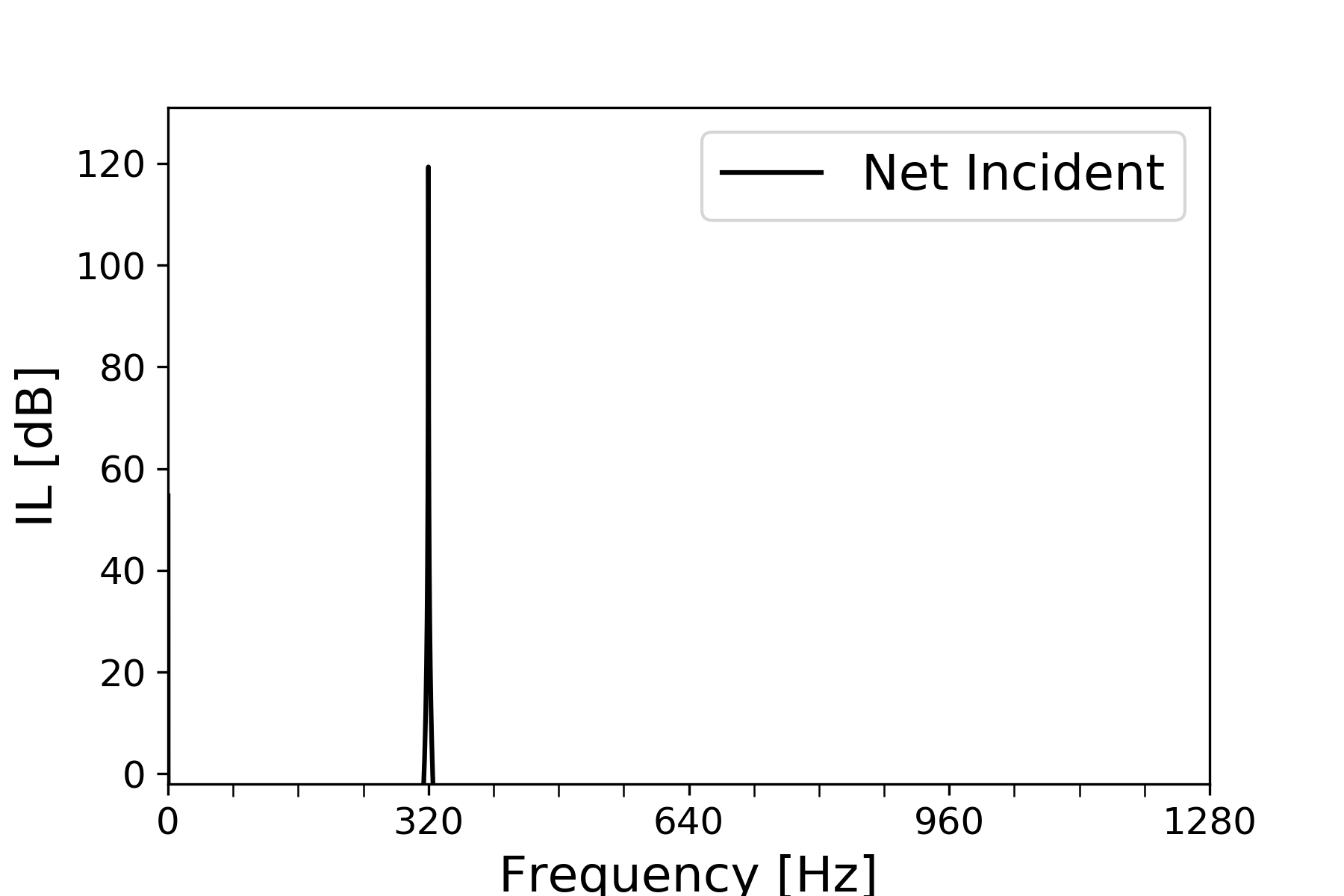}}
\subfigure[]{\includegraphics[width=0.48\textwidth]{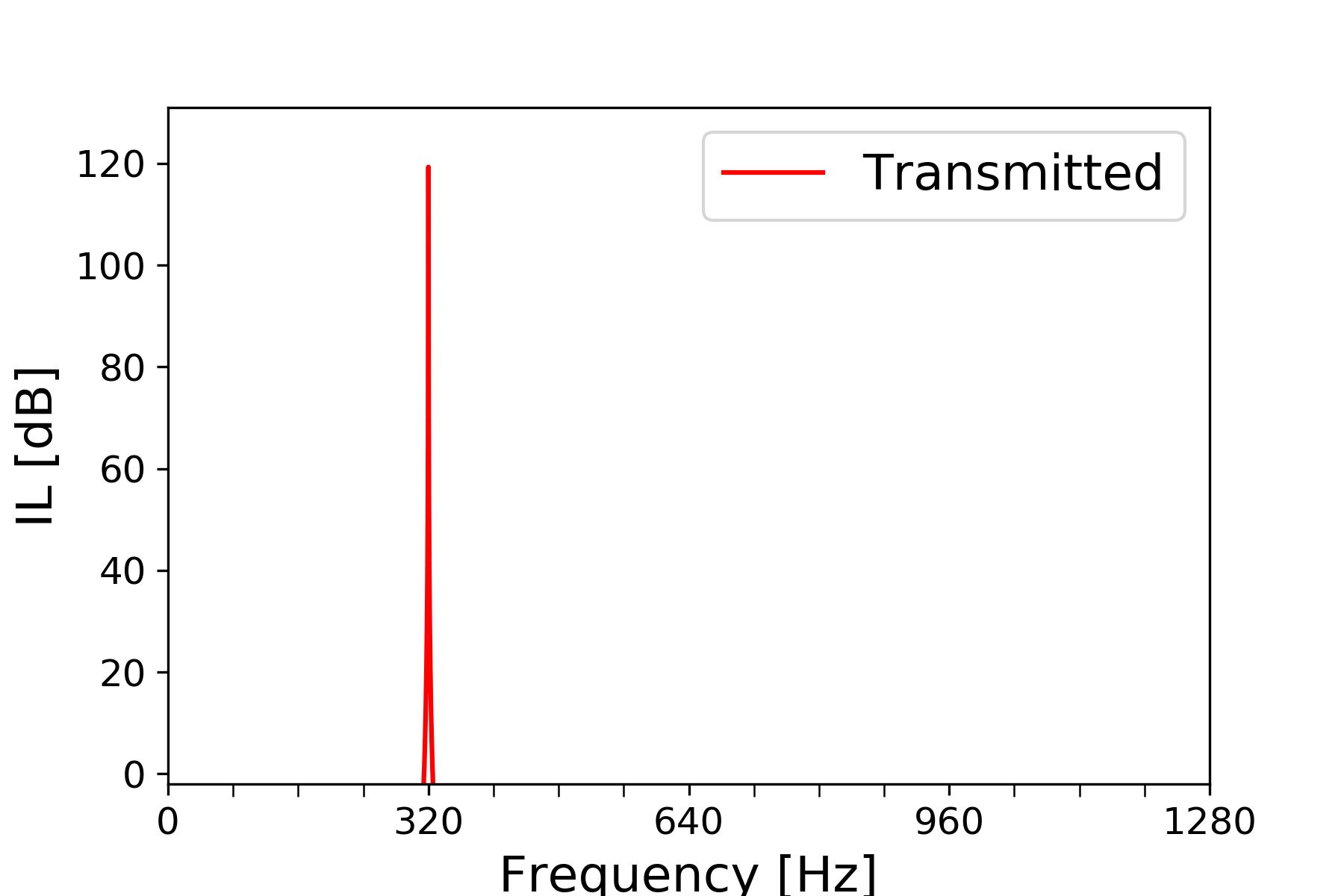}}
\caption{Net incident and transmitted sound intensity spectra of M$0$ considering only the linear terms in the strain-displacement relation for a sound source excitation of $100$~Pa amplitude at $20$~Hz, $80$~Hz and $320$~Hz, respectively.}
\label{fig:100Pa_M0Linear}
\end{figure}

\newpage
\appendix
\section{Sound intensity calculations on the incident and transmitted sides of the segmented plate}\label{sec:soundIntensityApp}

Complex sound pressures obtained from the 'virtual' microphones at the the four locations (Fig.~\ref{fig:soundIntensitySchematic}), i.e., two each in the upstream and downstream ducts are represented by $P_{1}$, $P_{2}$, $P_{3}$ and $P_{4}$, respectively. These pressures can be expressed as a superposition of the positive- and negative- going plane waves in the up- and downstream ducts of the standing wave tube as given in Eq.~\ref{eq:micPressures}\cite{songJASA2000}:
\begin{subequations}\label{eq:micPressures}
\begin{align}
&P_{1}=\left(U_{i}e^{-jkd_{1}}+U_{r}e^{jkd_{1}}\right)e^{j{\omega}t},\\
&P_{2}=\left(U_{i}e^{-jkd_2}+U_{r}e^{jkd_2}\right)e^{j{\omega}t},\\
&P_{3}=\left(D_{i}e^{-jkd_3}+D_{r}e^{jkd_3}\right)e^{j{\omega}t},\\
&P_{4}=\left(D_{i}e^{-jkd_4}+D_{r}e^{jkd_4}\right)e^{j{\omega}t},
\end{align}
\end{subequations}
where $U_{i}$, $U_{r}$ are the positive- and negative- traveling wave amplitudes in the upstream duct, and $D_{i}$, $D_{r}$ are the positive- and negative- traveling wave amplitudes in the downstream duct, respectively. Here, $k$ represents the wave number in the ambient fluid, and the $e^{+j{\omega}t}$ sign convention was adopted, where $\omega=2{\pi}f$ is the circular frequency. The pressure and acoustic particle velocities on either side of the segmented plate can in turn be determined based on a knowledge of the complex amplitude of the positive and negative traveling waves in both the ducts, as given below in Eq.~\ref{eq:PndV}:
\begin{subequations}\label{eq:PndV}
\begin{align}
&P\vert_{x_{1}=0}=U_{i}+U_{r},\\
&V\vert_{x_{1}=0}=\frac{U_{i}-U_{r}}{\rho_{0}c},\\
&P\vert_{x_{1}=t_{p}+t_{s}}=D_{i}e^{-jk\left(t_{p}+t_{s}\right)}+D_{r}e^{jk\left(t_{p}+t_{s}\right)},\\
&V\vert_{x_{1}=t_{p}+t_{s}}=\frac{D_{i}e^{-jk\left(t_{p}+t_{s}\right)}-D_{r}e^{jk\left(t_{p}+t_{s}\right)}}{\rho_{0}c},
\end{align}
\end{subequations}
where $P\vert_{x_{1}=0}$, $P\vert_{x_{1}=t_{p}+t_{s}}$, $V\vert_{x_{1}=0}$ and $V\vert_{x_{1}=t_{p}+t_{s}}$ are the acoustic pressures and the particle velocities on the incident and transmitted sides of the plate, respectively. The sound intensity on the incident $\left(I_{0}\right)$ and transmitted $\left(I_{t}\right)$ sides were then computed by using the following expressions:
\begin{subequations}\label{eq:I0It}
\begin{align}
&I_{0}=\frac{1}{2}\Re\left(P\vert_{x_{1}=0}\overline{V\vert_{x_{1}=0}}\right),\\
&I_{t}=\frac{1}{2}\Re\left(P\vert_{x_{1}=t_{p}+t_{s}}\overline{V\vert_{x_{1}=t_{p}+t_{s}}}\right),
\end{align}
\end{subequations}
where the over-bar denotes the complex conjugate. Also note that the sound source intensity ($I_{s}$) was evaluated as:
\begin{equation}\label{eq:Is}
I_{s}=\frac{G_{P_{w}}\left(\omega\right)}{2\rho_{0}c},
\end{equation}
where $G_{P_{w}}\left(\omega\right)$ is the power spectral density of the lowpass white noise time-domain signal $P_{w}$, and $\omega=2{\pi}f$ is the circular frequency. Finally, the mean square sound pressure acting on the incident face of the panel was
\begin{equation}\label{eq:Pirms}
P_{i_{rms}}^{2}=\frac{P\vert_{x_{1}=0}\overline{P\vert_{x_{1}=0}}}{2},
\end{equation}
and the corresponding sound pressure level was
\begin{equation}\label{eq:spl}
{\mathrm{SPL}}_{x_{1}=0}=10\log_{10}\frac{P_{i_{rms}}^{2}}{4\times10^{-10}}~\mathrm{dB}~re~20~ \mu\mathrm{Pa}.
\end{equation}

\begin{figure}[h!]
\centering
\setcounter{figure}{0}
 \includegraphics[width=0.9\textwidth]{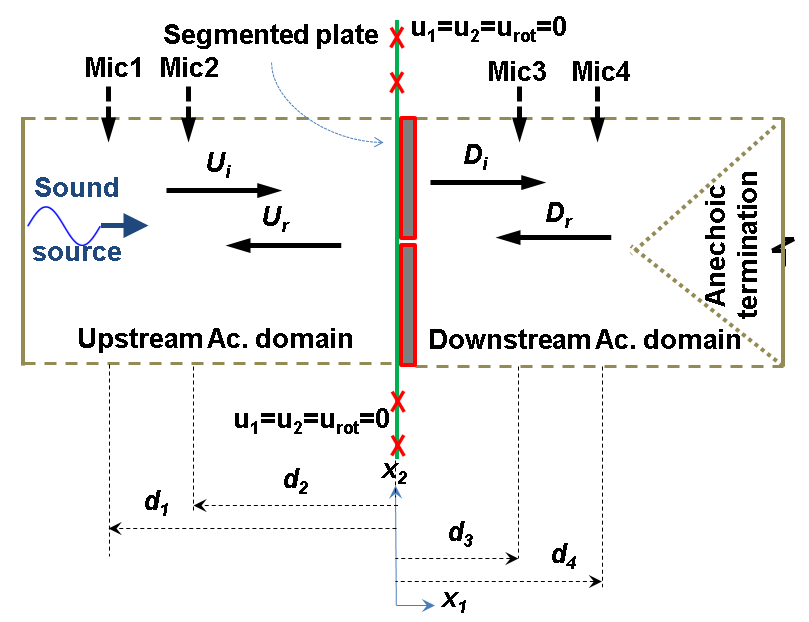}
 \caption{A schematic of the two-dimensional FE model used for acoustical characterization of the segmented plates models with the distance of the microphones from origin.}
\label{fig:soundIntensitySchematic}
\end{figure}
\newpage
\section{Analytical estimation of displacement $u_{1}$ at which the plates come in contact applicable to models M$1$, M$2$ and M$3$}\label{sec:u1Analytical}
From Fig.~\ref{fig:u1Analytical}:
\begin{subequations}\label{eq:u1Analytical1}
\begin{align}
&\angle EAB + \angle BCD = \frac{\pi}{2},\\
&\angle BCD + \angle CDB = \frac{\pi}{2}.
\end{align}
\end{subequations}
From Eq.~(\ref{eq:u1Analytical1}), $\angle EAB = \angle CDB$. Therefore,
\begin{equation}\label{eq:u1Analytical2}
\tan\left(\angle EAB\right)  = \tan\left(\angle CDB\right),
\end{equation}
which translates to:
\begin{equation}\label{eq:u1Analytical4}
\frac{\abs{u_{1}}}{L_{p}/2}=\frac{d_{cl}/2}{t_{p}},
\end{equation}
or $\abs{u_{1}}=\frac{d_{cl}L_{p}}{4t_{p}}$.

\begin{figure}[h!]
\centering
\setcounter{figure}{0}
 \includegraphics[width=0.5\textwidth]{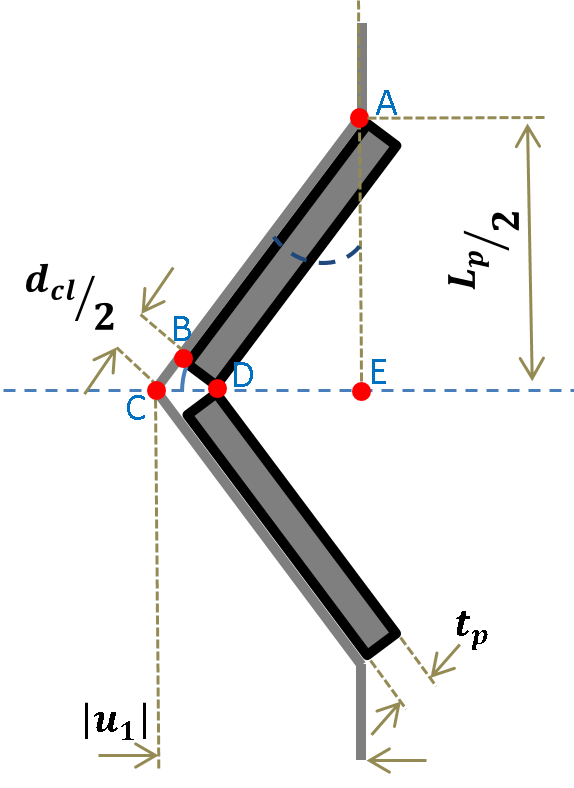}
 \caption{A schematic of the configuration when segmented plates come in contact.}
\label{fig:u1Analytical}
\end{figure}
\clearpage
\bibliography{all_ref_phd}
\end{document}